\documentclass[fleqn,usenatbib]{mnras}

\usepackage{newtxtext,newtxmath}

\usepackage[T1]{fontenc}

\DeclareRobustCommand{\VAN}[3]{#2}
\let\VANthebibliography\thebibliography
\def\thebibliography{\DeclareRobustCommand{\VAN}[3]{##3}\VANthebibliography}

\usepackage{graphicx}	
\usepackage{amsmath}	

\def\msun{\hbox{M$_\odot$}}

\def\t4{\hbox{t$_{\rm 4}$}}

\def\cm3{\hbox{cm$^{-3}$}}
\def\two{\,{\sc ii}}
\def\three{\,{\sc iii}}

\title[High-z clumps in JWST's era]{Star formation at the smallest scales; A JWST study of the clump populations in SMACS0723}

\author[Claeyssens et al.]{
Adélaïde Claeyssens$^{1}$\thanks{E-mail: adelaide.claeyssens@astro.su.se},
Angela Adamo$^{1}$,
Johan Richard$^{2}$,
Guillaume Mahler$^{3,4}$,
Matteo Messa$^{1,5}$,
\newauthor 
\hspace{1pt} 
and Miroslava Dessauges-Zavadsky$^{5}$
\\
$^{1}$The Oskar Klein Centre, Department of Astronomy, Stockholm University, AlbaNova, SE-10691 Stockholm, Sweden\\
$^{2}$Univ Lyon, Univ Lyon1, Ens de Lyon, CNRS, Centre de Recherche Astrophysique de Lyon UMR5574, 69230, Saint-Genis-Laval, FR\\
$^{3}$Centre for Extragalactic Astronomy, Durham University, South Road, Durham DH1 3LE, UK\\
$^{4}$Institute for Computational Cosmology, Durham University, South Road, Durham DH1 3LE, UK\\
$^{5}$ D\'epartement d'Astronomie, Université de Genève, Chemin Pegasi 51, 1290 Versoix, Switzerland\\
}

\date{Accepted 2022 December 12. Received 2022 December 12; in original form 2022 August 23}

\pubyear{2022}

\begin{document}
\label{firstpage}
\pagerange{\pageref{firstpage}--\pageref{lastpage}}
\maketitle

\begin{abstract}

We present the clump populations detected in 18 lensed galaxies at redshifts 1 to 8.5 within the lensing cluster field SMACS0723. The recent JWST Early Release Observations of this poorly known region of the sky have revealed numerous point-like sources within and surrounding their host galaxies, undetected in the shallower HST images. We use JWST multiband photometry and the lensing model of this galaxy cluster to estimate the intrinsic sizes and magnitudes of the stellar clumps. We derive optical restframe effective radii from $<$10 to hundreds pc and masses ranging from $\sim10^5$ to $10^9$ \msun, overlapping with massive star clusters in the local universe. Clump ages range from 1 Myr to  1 Gyr. We compare the crossing time to the age of the clumps and determine that between 45 and 60 \% of the detected clumps are consistent with being gravitationally bound. On average, the dearth of Gyr old clumps suggests that the dissolution time scales are shorter than 1 Gyr. We see a significant increase in the luminosity (mass) surface density of the clumps with redshift. Clumps in { reionisation era galaxies} have stellar densities higher than star clusters in the local universe. We zoom in into single galaxies at redshift $<6$ and find for two galaxies, the Sparkler and the Firework, that their star clusters/clumps show distinctive colour distributions and location surrounding their host galaxy that are compatible with { being accredited or formed during} merger events. { The ages of some of the compact clusters are between 1 and 4 Gyr, e.g., globular cluster precursors formed around 9-12 Gyr ago.} Our study, conducted on a small sample of galaxies, shows the potential of JWST observations for understanding the conditions under which star clusters form in rapidly evolving galaxies.

\end{abstract}




\section{Introduction}


Early deep field observations with the Hubble Space Telescope (HST) revealed that galaxy morphologies rapidly evolve into more irregular and clumpy appearance as a function of increasing redshift \citep{abraham1996, brinchmann1998}. New JWST observations confirm the morphological evolution already traced with HST studies, i.e., galaxy appearance between redshift 2 and 7 is dominated by clumpy structures and only in a small fraction by mergers \citep{treu2022}. 

A long standing effort in the community has focused in understanding the link between clump formation and evolution, and galaxy growth. 
Initial HST studies focused on the physical properties of the clump populations within field galaxies up to redshift $\sim3$, concluded that these stellar clumps are more massive counterparts of local giant star-forming regions, with comparable sizes of $\sim$1 kpc, thus, overall denser stellar entities than typically observed in the local universe \citep{elmegreen2006,wisn2012}. Based on colour properties and the location of clumps within galaxies, it has also been speculated that these clumps survive longer within their disk galaxies and therefore migrate toward the or being responsible for proto-bulges \citep{elmegreen2009, FS2011}. Systematic studies of clump populations in the FUV-optical rest-frames of deep surveys consistently reported that clumps significantly contribute to the total FUV light of galaxies, supporting their young ages, as well as the presence of redder clumps, consistent with old or extincted stellar populations \citep{guo2012, guo2015}. Clumps have been reported as regions of elevated star formation rate (SFR) and specific SFR, as mini-starburst entities within their host galaxies \citep{zanella2015, Iani2021}. Considering the volume densities of galaxies with UV light dominated by multiple clumps, \citet{guo2015} and \citet{shibuya2016} reported that the fraction of galaxies with clumpy structures peaks at cosmic noon, suggesting that clump formation is closely related to the physical conditions under which star formation takes place in galaxies. 

Different approaches have been undertaken to further improve our understanding of star formation operating at sub-galactic scales in rapidly evolving galaxies. 
Numerous studies of local-analogues of main-sequence galaxies at redshift 1--2 have been carried out. The advantage of this approach is the access to several observing facilities sampling different gas phases and their dynamical conditions as well as clump physical properties at hundreds parsec physical resolution. These studies have focused on highly star-forming disk galaxies \citep{fisher2017a, fisher2017b} or starburst systems, with properties similar to those of Lyman $\alpha$ and/or Lyman break galaxies \citep{messa2019}. These local analogues host clump populations that dominate their UV light and have masses and sizes encompassing the range observed in high-redshift main sequence galaxies at cosmic noon \citep{messa2019}. The host galaxies show galactic dynamics dominated by random motion and turbulent thick disks, suggesting that clumps are produced by fragmentation in gravitationally unstable disks \citep{fisher2017b}. These results align with the continuous progress made by numerical simulations. Initial low-resolution simulations of isolated disk galaxies \citep[e.g.,][]{bournaud2007, bournaud2009, Agertz2009} formed massive kiloparsec-scale clumps from fragmentation and collapse taking place in thick, and gravitational unstable, gas-rich disks. In these simulations, clumps survived long enough to migrate and contribute to the formation of bulge structures in galaxies \citep[e.g.,][]{elmegreen2008, ceverino2010}. 

The fate and survival of clumps in high-redshift galaxies has been since long debated. On one side, fundamental constraints have been produced by studies of stellar clumps in gravitationally lensed galaxies \citep{jones2010, livermore2012, Livermore2015, adamo2013, Johnson2017, Cava2018, Vanzella2017b, Vanzella2017a, vanzella2022, mestric2022}. Once the magnification, which can reach factors between 2 and $>$30 at the centre of the lensing cluster, is accounted for, few to several clumps are detected within these star-forming galaxies with sizes from a few parsecs to hundred parsecs and masses ranging between $10^5$ and $10^8$ \msun\, \citep[e.g.,][]{DZ2017, tamburello2017}. Resolution, hence, plays an important role in defining the main physical properties of stellar clumps, both observationally \citep[e.g.,][]{DZ2017, Cava2018}, and numerically \citep{tamburello2015}. On the other side, increasing particle numbers and physical resolution, as well as inclusion of stellar feedback, in numerical approaches has led to question initial results leading to bulge formation via clump migration. Clump survival timescales appear clearly linked to feedback prescriptions and implementations \citep{oklopcic2017,dekel2022} as well as gas fractions in disks, with cosmological simulations producing short-lived clumps while isolated galaxy simulations find longer-lived clumps \citep[e.g.,][]{fensch2021}.

From the observational perspective, clump ages, important to determine clump survival, have been difficult to pin down \citep[e.g.,][]{Cava2018}. Initial HST studies of clumps in lensed galaxies at redshift $\lesssim 2$, find age ranges between a few and tens of Myr up to a few hundreds of Myr \citep[e.g.,][]{adamo2013, Johnson2017, Messa2022, vanzella2022}. The main limitation has been so far the lack of homogeneous coverage of the UV-optical rest-frame of galaxies at redshift $> 2$ \citep[e.g., see][who included ground-based data, although limiting, to improve age and mass estimates of clumps]{mestric2022}, a key phase for star formation and galaxy evolution. 

Stellar clumps have long being suggested to host massive star clusters, which would survive long after clumps dissolve and possibly contribute to the globular cluster population surrounding spiral galaxies in the local universe \citep{shapiro2010, adamo2013}. Indeed, for the highest magnification regions and with HST resolution,  it has been possible to resolve physical scales down to tens of parsec or better, where clump sizes start to overlap to star cluster ones \citep[e.g.,][]{Vanzella2017b,Vanzella2017a, Johnson2017, welch2022}. Star clusters form and evolve within their host galaxies across cosmic times \citep[e.g][]{adamo2020a}. State-of-art cosmological simulations follow cluster formation and evolution during galaxy assembly like the Milky Way and M31 \citep[e.g.,][among the latest]{grudic2022, reinacampos2022}. These simulations agree that the peak formation of massive star clusters (masses $>10^5$ \msun) is around redshift 3 \citep{reinacampos2019} in these Milky-Way-like progenitors. The feedback produced by the numerous massive stars residing in star clusters might have played a significant role toward the reionisation of the universe \citep{ma2021, vanzella2020} and could be detected in JWST observations with the aid of gravitational lensing \citep{renzini2017, sameie2022}. Very little is actually known from direct observations about star cluster formation and evolution in the high-redshift universe. Pioneering work conducted by \cite{Vanzella2017b, Vanzella2017a, vanzella2020, vanzella2022} would suggest that star clusters are indeed detectable and produce extreme feedback within their host galaxies. 

JWST observations of several gravitationally lensed regions will be a game-changer for the field. Initial studies already give an idea of the potentialities of having access to high-spatial resolution optical rest-frames of lensed galaxies at redshift higher than 1. For example,  \citet{mowla2022} and \cite{vanzella2022b} report detections of cluster candidates and discuss the implications for massive star cluster formation.

In this work we use the first imaging observation acquired with the JWST of the galaxy cluster SMACS J0723.3--7323 (hereafter, SMACS0723, $z = 0.388$, Fig.~\ref{fig:im_cluster}). This region has been previously observed with HST as part of the RELICS program by \citet{Coe2019}. We move beyond the initial single--object studies by \citet{mowla2022} and \cite{vanzella2022b}, by extending our analysis to the clump populations of galaxies in the redshift range 1 to 8. This statistical approach enables us to look at clump formation and evolution across the most important phases of galaxy growth and address key questions related to their survival timescales, as well as the capability of detecting and resolving them into their star cluster components. 

The manuscript is organised as following. In Sect.~\ref{sec:data}, we present the observational data used in this work. The lensing model and galaxy selection are described in Sect.~\ref{sec:lens_model} and \ref{sec:galaxies}. In Sect.~\ref{sec:cl_analysis} we present the method we use to perform photometry and size measurements, the conversion to de--lensed quantities, and the spectral energy distribution (SED) analysis. The results obtained by studying clumps as a population are presented and discussed in Sect.~\ref{sec:results}, while in Sect.~\ref{sec:single_gal} we focus on the physical properties of clumps in particularly interesting galaxies. In Sect.~\ref{sec:starclusters} we discuss the physical properties of clumps in the reionisation era. Final remarks and conclusions are gathered in Sect.~\ref{sec:conclusion}.

\section{Observations and data reduction}
\label{sec:data}
\subsection{JWST data}
\label{sec:jwst}

\subsubsection{NIRCam observations}
In this work, we utilise the first JWST imaging data targeting the lensing cluster SMACS0723. These observations have been obtained as part of the Early Release Observations (ERO) program (ID 2736; PI: Pontoppidan, \citet{ERO}) and released on the July 13th and retrieved from the Mikulski Archive for Space Telescopes (MAST). The cluster has been imaged using the Near-Infrared Camera (NIRCam; \citet{Rieke2005}) F090W, F150W, and F200W short wavelength (SW) and F277W, F356W, F444W long wavelength (LW) filters, covering an observed wavelength range of $\lambda_{\rm obs} = 0.8 - 5 \mu$m, for a total of 12.5 hours of integration time.  We  focus on the portion of data that covers the centre of the cluster, that is entirely contained in the NIRCam module B. 
{ The HST and JWST processed imaging data for SMACS0723 
are publicly accessible\footnote{\url{https://s3.amazonaws.com/grizli-v2/JwstMosaics/v4/index.html}}. They have been processed  using the \emph{grizli} pipeline\ \citep{Grizli}, which
co-adds all calibrated single exposures in each filter,
and aligns all stacked images to the GAIA DR3 catalogues \citep{Gaia_EDR3}.
The NIRCam SW images are drizzled to $0.02''$ pixels, while the HST final images and the NIRCam LW images are drizzled to $0.04''$ pixels. 

The \emph{grizli} v4 images use \texttt{jwst\_0942.pmap} NIRCam calibrations
first made available July 29. Improved relative calibrations for each NIRCam module have been available and independently tested with other approaches \citep{boyer2022} showing agreement consistent to within $<5\%$ of the most recent
\texttt{jwst\_0995.pmap} calibrations in each filter and detector. We applied these flux corrections to our final photometry\footnote{\url{https://zenodo.org/record/7143382\#.Y1-ucS0w2jg}}, accounting that our targets are acquired in module B, making our photometric measurements consistent with the latest calibration release. 
We use AB magnitude system throughout our analysis.}

\subsubsection{NIRSpec observations}

Additionally, we retrieve from MAST the reduced NIRSpec 2D and 1D spectra for all targets observed within the NIRCam module B. We identify the main spectral features ([OII], [OIII], Balmer or Paschen lines.) for 10 galaxies in the redshift range $1<z<9$, including one multiple system (S7) from \citet{Mahler22}. The spectroscopic redshift and the main identified lines from NIRSpec are provided in Table~\ref{tab:sample}.

\subsection{HST data}
\label{sec:hst}
This cluster is part of the RELICS program (\citet{Coe2019}), and therefore was observed with HST in 7 filters F435W, F606W, F814W, F105W, F125W, F140W, and F160W using the Advance Camera for Survey (ACS) and Wide Field Camera 3 (WFC3). On average, between 0.5 and 2 orbits have been used per filter, covering a wavelength range between 0.4 and 2 microns. These observations are, however, significantly shallower than the JWST ones of 2 to 3 mag at comparable wavelengths (see Table~\ref{tab:table_observations}).

 The HST data have been reduced in a similar fashion as the NIRCam data (Brammer et al. in prep) and aligned on the JWST observations. The 7 science frames have a pixel scale of 0.04''. 
\begin{table*} 
 
\begin{tabular}{llllll} 
Filter & Exp Time & ${mag}_{\rm {lim}}$  & $\rm E(B-V)_{\rm gal}$ &  PSF FWHM\\ 
 &   [s] & AB mag  & & [arcsec]\\ 

\hline 
\hline 

HST F435W & 2,233 &  27.0 & 0.776 & 0.120 \\ 
HST F606W & 2,289 &  27.5 &  0.502 & 0.120\\ 
HST F814W & 2,529 &   27.2 &  0.315 & 0.120\\ 
JWST F090W & 7,537  & 29.2  & 0.265  & 0.050  \\ 
HST F105W & 6,058 &  26.7  &  0.203 & 0.160\\ 
HST F125W & 3,248 &  26.0  &  0.151 & 0.180\\ 
HST F140W & 5,233 &  26.2  &   0.125 & 0.190\\ 
JWST F150W & 7,537  &  29.1  & 0.112 & 0.060 \\ 
HST F160W & 4,523 &  26.4 &  0.102 & 0.190\\ 
JWST F200W & 7,537  &  29.1  & 0.074 & 0.068 \\ 
JWST F277W & 7,537 &  29.1   & 0.048 & 0.110\\ 
JWST F356W & 7,537 &  29.1   & 0.034 & 0.130\\ 
JWST F444W & 7,537 &  29.0  & 0.027 & 0.140\\ 
\end{tabular} 
\caption{Calibration details of the HST and JWST images. The magnitude limits reported in the third column have been measured in a PSF-like aperture in each filter. The HST filter limiting magnitudes are consistent (within 0.2 mag) with the values reported by \citet{Coe2019}.  The fourth column gives the galactic extinction we have removed in each filter. The last column gives the PSF FWHM (measured on a stack of 3 isolated and not saturated stars).}
\label{tab:table_observations} 
\end{table*}

\section{Lensing model}
\label{sec:lens_model}

We briefly summarise the methodology and discuss relevant aspects of the lens model used in this work \footnote{The lens model used in this work is available here: \url{https://github.com/guillaumemahler/SMACS0723-mahler2022 }}. We refer the reader to \citet{Mahler22} for the details.  A more in-depth discussion of the adopted lensing algorithm to construct the model of the cluster mass distribution is given in \citet{Kneib1996,Richard2010,Verdugo11}.

We adopt a parametric approach using \textsc{Lenstool} \citep{Jullo2007} to model the galaxy cluster mass distribution surrounding our targets as a combination of dual pseudo-isothermal ellipsoids (dPIEs; \citealt{Eliasdottir07}). Using a Monte Carlo Markov Chain (MCMC) method we estimate the mass model parameters and their uncertainties. These dPIE clumps are combined to map the dark matter (DM) at the cluster scale and to model the cluster mass distribution, while galaxy-scale DM potentials are used to describe galaxy-scale substructure. To reduce the overall parameter space, we scale the parameters of each galaxy using a reference value with a constant mass–luminosity scaling relation \citep{Limousin2007}. We construct a galaxy cluster catalogue using the red sequence technique \citep{Gladders2005}, where we select galaxies
that have similar colours in the HST filter F606W–F814W colour versus F814W-band magnitude diagram. In this method, the central galaxy of the cluster is modelled separately. To allow for a better estimation of the lensing magnification, we  remove from the galaxy catalogue one cluster member responsible for the main perturbation of the lensed object galaxy S2.2 and model it separately. 

In addition, as discussed in \cite{Mahler22}, the model includes one additional large-scale dark matter clump associated with the smooth extended profile of the intra-cluster light (ICL) west to the core. 
Our final lens model of SMACS\,J0723 includes one cluster-scale DM halo parameterized as dPIE profile. We constrain the cluster lens model using 21 multiply-imaged lensed systems with five systems with spectroscopic constraints (see Table~2 in \cite{Mahler22} for the exact coordinates).
The resulting model presents a good rms of 0\farcs3. The goodness of the model assures a reduced statistical uncertainty for the magnification globally. We note that this is not an indication of the systematic uncertainty and further analysis beyond the scope of this paper would be necessary to estimate it.

\section{Galaxy selection}
\label{sec:galaxies}
\begin{figure*}
	\includegraphics[width=18cm]{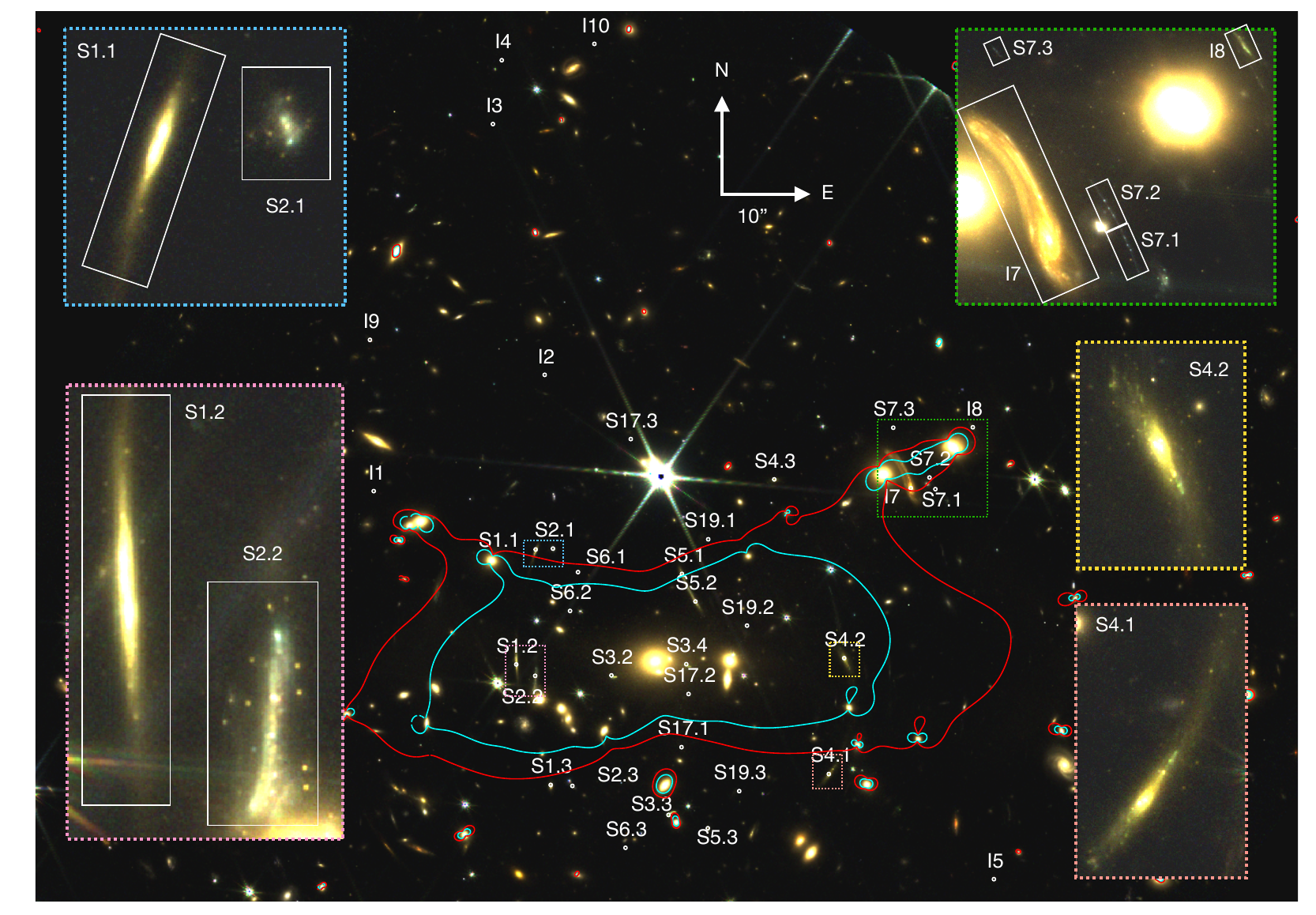}
    \caption{ JWST/NIRCam colour image (F090W, F150W, F200W) of the SMACS0723 cluster. The blue and red lines represent the critical lines at $z=2$ and $z=5$, respectively. The selected galaxies are shown in white (cf Table~\ref{tab:sample}). The 5 boxes zoom in on the 5 more extended sources detected in the cluster. }
    \label{fig:im_cluster}
\end{figure*}

In this work, we focus on studying stellar clumps within galaxies with a robust redshift measurement and/or a high lensing magnification factor (i.e. to get high spatial resolution on these images). A significant fraction of targeted galaxies has been selected from the multiple systems identified and used by \citealt{Mahler22} to build the lens model. Among them, 4 have a spectroscopic redshift value measured from MUSE data (systems S1, S2, S3 and S5) and one measured from the NIRSpec data (system S7, see \citealt{Mahler22}). The remaining systems have a geometric redshift value obtained from the lens model optimisation (cf. Sect.~\ref{sec:lens_model}). Each galaxy image has been visually inspected on NIRCam F150W and F200W filters (reference filters for this study, see Sect.~\ref{sec:cl_identification}) to confirm that at least 1 stellar clump (compact source surrounded by diffuse extended emission) is clearly  detected in both filters. Of the 21 multiple systems identified by \citealt{Mahler22}, we rejected 12 systems with poorly constrained redshift estimations from the lens model (with a 1$\sigma$ uncertainty $>$0.3)  and/or very faint/diffuse (S/N$<$3) detections in the F150W and F200W NIRCam images. In addition to the multiple systems, we include 9 galaxies located outside the multiple images area, but for which we were able to measure a robust spectroscopic redshift from the available NIRSpec data. The final sample is, therefore, composed of 18 galaxies producing 36 images presented in Table~\ref{tab:sample}. The selected images are shown in Fig.~\ref{fig:im_cluster}.

\begin{table*}
\begin{tabular}{lllllllll}
ID & $\alpha$ & $\delta$ & $z_{\rm spec}$ & $z_{\rm model}$ & Main spectral features & $\mu$ & Reference & \# clumps\\
   & [deg] & [deg] & \\
\hline 
\hline 
S1.1 & 07:23:21.77 & -73:27:03.89 &  1.449 & & MUSE - [OII] & 5.4$^{+0.5}_{-0.4}$ & \citep{Mahler22} & 7\\[0.05cm]
S1.2 & 07:23:22.31 & -73:27:17.42 &  1.449 & & MUSE - [OII]  & 9.8$^{+1.5}_{-1.1}$ & \citep{Mahler22} & 12 \\[0.05cm]
S1.3 & 07:23:21.36 & -73:27:31.63 &  1.449 & & MUSE - [OII] & 5.6$^{+0.4}_{-0.4}$& \citep{Mahler22} & 1 \\[0.05cm]
S2.1 & 07:23:21.30 & -73:27:03.78 &  1.378 & & MUSE - [OII] & 5.1$^{+0.5}_{-0.4}$ & \citep{Mahler22}& 8 \\[0.05cm]
S2.2 & 07:23:21.79 & -73:27:18.77 &  1.378 & & MUSE - [OII] & 9.6$^{+1.5}_{-1.2}$& \citep{Mahler22} & 27 \\[0.05cm]
S2.3 & 07:23:20.76 & -73:27:31.75 &  1.378 & & MUSE - [OII] & 5.1$^{+0.4}_{-0.4}$ & \citep{Mahler22}& 6 \\[0.05cm]
S3.2 & 07:23:19.68 & -73:27:18.72 &  1.991 & & MUSE - [CIII] & 3.0$^{+0.5}_{-0.2}$ & \citep{Mahler22} & 5 \\[0.05cm]
S3.3 & 07:23:18.11 & -73:27:35.17 &  1.991 & & MUSE - [CIII] & 6.9$^{+1.0}_{-0.6}$ & \citep{Mahler22} & 5 \\[0.05cm]
S3.4 & 07:23:17.61 & -73:27:17.40 &  1.991 & & MUSE - [CIII] & 1.8$^{+0.3}_{-0.2}$ & \citep{Mahler22} & 6 \\[0.05cm]
S4.1 & 07:23:13.68 & -73:27:30.35 &  &2.19$^{+0.05}_{-0.08}$ & & 7.9$^{+0.6}_{-0.6}$ & \citep{Mahler22}& 15 \\[0.05cm]
S4.2 & 07:23:13.26 & -73:27:16.69 &  & 2.19$^{+0.05}_{-0.08}$ & & 13.9$^{+1.8}_{-1.3}$ & \citep{Mahler22} & 27 \\[0.05cm]
S4.3 & 07:23:15.19 & -73:26:55.62 &  &2.19$^{+0.05}_{-0.08}$ & & 4.5$^{+0.4}_{-0.3}$ & \citep{Mahler22} & 4 \\[0.05cm]
S5.1 & 07:23:17.73 & -73:27:06.77 &  1.425 & & MUSE - [OII] & 19.0$^{+2.7}_{-2.5}$ & \citep{Mahler22} & 11 \\[0.05cm]
S5.2 & 07:23:17.36 & -73:27:10.01 &  1.425 & & MUSE - [OII] & 18.8$^{+2.7}_{-1.3}$& \citep{Mahler22} & 10 \\[0.05cm]
S5.3 & 07:23:17.02 & -73:27:36.74 &  1.425 & & MUSE - [OII] &  3.3$^{+0.2}_{-0.2}$ & \citep{Mahler22} & 4 \\[0.05cm]
S6.1 & 07:23:20.60 & -73:27:06.55 &  & 1.67$^{+0.03}_{-0.02}$ & & 13.8$^{+2.1}_{-2.0}$ & \citep{Mahler22} & 2 \\[0.05cm]
S6.2 & 07:23:20.82 & -73:27:11.12 &  & 1.67$^{+0.03}_{-0.02}$ & & 14.2$^{+1.1}_{-1.0}$& \citep{Mahler22} & 1 \\[0.05cm]
S6.3 & 07:23:19.29 & -73:27:39.05 &  & 1.67$^{+0.03}_{-0.02}$ & & 3.2$^{+0.2}_{-0.2}$ & \citep{Mahler22} & 1 \\[0.05cm]
S7.1 & 07:23:10.74 & -73:26:56.76 &  5.173 & & NIRSpec - [OIII] Hbeta Halpha & 26.4$^{+17.2}_{-3.5}$ & This work  & 5 \\[0.05cm]
S7.2 & 07:23:10.91 & -73:26:55.39 &  5.173 & & NIRSpec - [OIII] Hbeta Halpha & 22.2$^{+22.6}_{-5.7}$ & This work & 5 \\[0.05cm]
\S7.3 & 07:23:11.91 & -73:26:49.52 &  5.173 & & NIRSpec - from S7.1 and S7.2 &5.4$^{+1.0}_{-0.4}$ & This work & 2 \\[0.05cm]
S17.1 & 07:23:17.75 & -73:27:27.18 &  &2.10$^{+0.10}_{-0.06}$ & & 17.5$^{+2.5}_{-1.8}$ & \citep{Mahler22} & 1 \\[0.05cm]
S17.2 & 07:23:17.55 & -73:27:20.92 &  & 2.10$^{+0.10}_{-0.06}$ & & 7.9$^{+0.9}_{-0.5}$& \citep{Mahler22} & 1 \\[0.05cm]
S17.3 & 07:23:19.15 & -73:26:50.87 &  & 2.10$^{+0.10}_{-0.06}$ & & 2.6$^{+0.1}_{-0.1}$ & \citep{Mahler22} & 1 \\[0.05cm]
S19.1 & 07:23:17.01 & -73:27:02.68 & & 1.35$^{+0.03}_{-0.02}$ & & 5.8$^{+0.8}_{-0.4}$ &  \citep{Mahler22} & 1 \\[0.05cm]
S19.2 & 07:23:15.94 & -73:27:12.85 & & 1.35$^{+0.03}_{-0.02}$ & &  8.1$^{+0.9}_{-0.7}$ & \citep{Mahler22} & 1 \\[0.05cm]
S19.3 & 07:23:16.15 & -73:27:32.36 & & 1.35$^{+0.03}_{-0.02}$ & &  4.2$^{+0.3}_{-0.3}$ & \citep{Mahler22} & 1 \\[0.05cm]
I1 & 07:23:26.24 & -73:26:56.99 &  8.500 & & NIRSpec - [OIII] Hbeta &  4.0$^{+0.3}_{-0.3}$ & This work & 2 \\[0.05cm]
I2 & 07:23:21.52 & -73:26:43.30 &  6.380 & & NIRSpec - [OIII] Hbeta Halpha & 2.2$^{+0.1}_{-0.1}$  & This work  & 5 \\[0.05cm]
I3 & 07:23:22.95 & -73:26:13.74 &  3.715 & & NIRSpec - [OIII] Hbeta Halpha &  1.4$^{+0.1}_{-0.1}$ &This work  & 1 \\[0.05cm]
I4 & 07:23:22.70 & -73:26:06.22 &  7.663 & & NIRSpec - [OIII] Hbeta & 1.3$^{+0.1}_{-0.1}$ &  This work  & 4 \\[0.05cm]
I5 & 07:23:09.12 & -73:27:42.73 &  5.280 & & NIRSpec - [OIII] Hbeta [OII] Halpha & 3.3$^{+0.3}_{-0.3}$ & This work  & 4 \\[0.05cm]
I7 & 07:23:11.42 & -73:26:56.64 &  1.160 & & NIRSpec - CaT, Paschen lines & 6.4$^{+2.6}_{-0.5}$ & This work  & 28 \\[0.05cm]
I8 & 07:23:09.71 & -73:26:49.47 &  2.120 & & NIRSpec - Halpha [SIII] HeI & 10.0$^{+1.0}_{-5.7}$ & This work  & 4 \\[0.05cm]
I9 & 07:23:26.34 & -73:26:39.17 &  2.742 & & NIRSpec - Halpha [NII] [SII] & 1.7$^{+0.1}_{-0.1}$ &This work  & 4 \\[0.05cm]
I10 & 07:23:20.15 & -73:26:04.29 &  7.661 & & NIRSpec - [OIII] Hbeta & 1.4$^{+0.1}_{-0.1}$ & This work  & 2 \\[0.05cm]
\hline
\end{tabular}

\caption{ Galaxy sample used for clump identification. From left to right: identification (S$=$multiple system from \citealt{Mahler22}, I$=$additional individual image), right ascension and declination, spectroscopic redshift, Lenstool-predicted redshift, spectrograph and main spectral features, magnification measured at the centre of the image of the galaxy (notice it can change significantly as a function of position in the image), reference, and the number of clumps detected in each image.}
\label{tab:sample}
\end{table*}

\begin{figure*}
	\includegraphics[width=18cm]{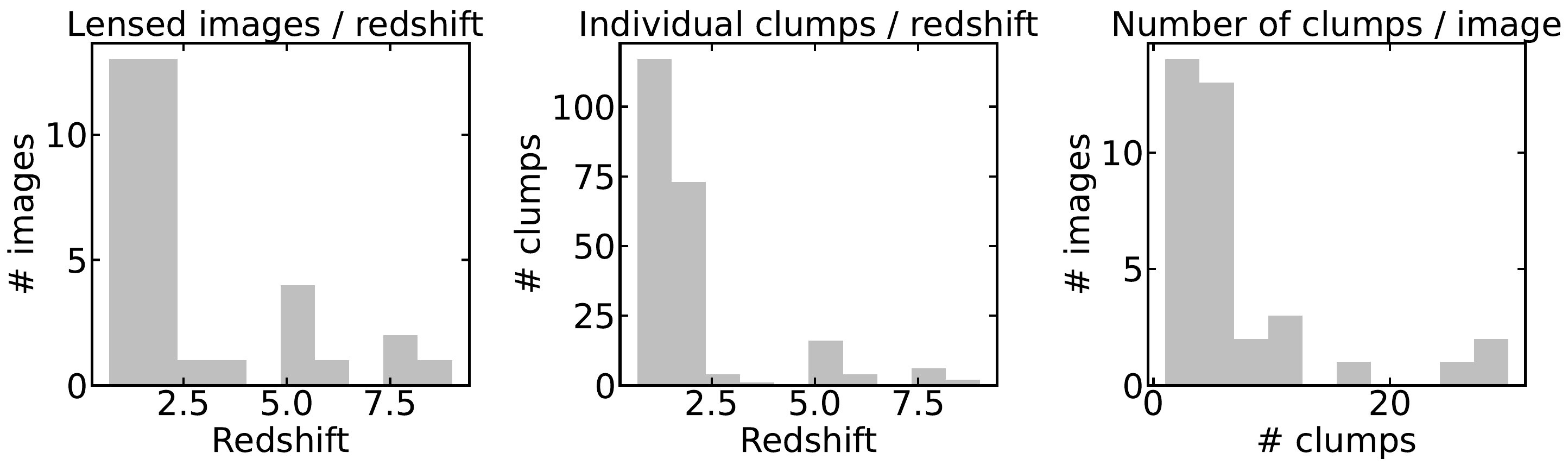}
    \caption{ From left to right: Galaxy images versus  redshift, individual clumps  versus redshift, and number of clumps per galaxy image distributions. The full sample of 223 clumps has been used to produce these distributions.}
    \label{fig:histo_sample}
\end{figure*}

\begin{figure}
	\includegraphics[width=\columnwidth]{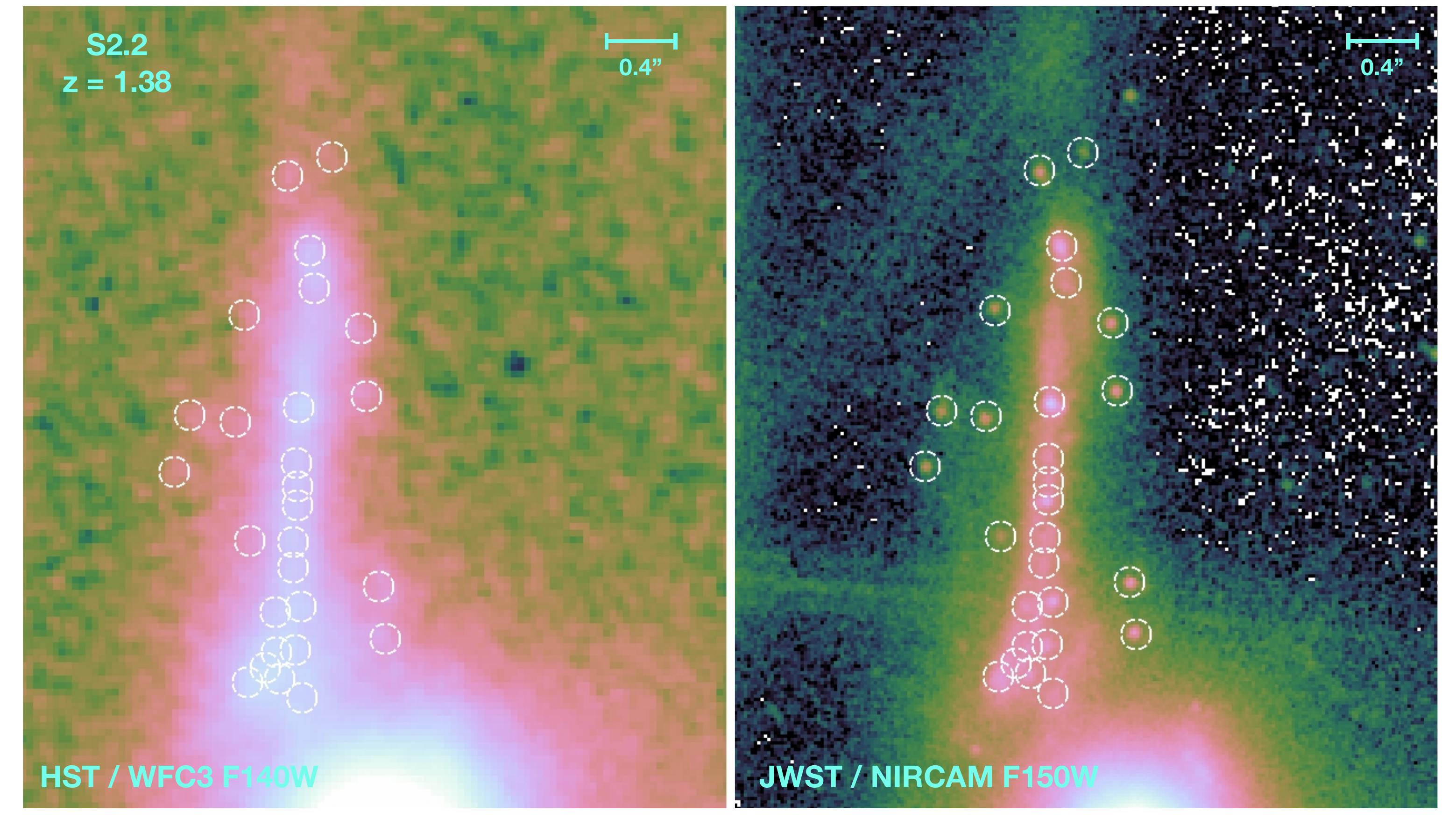}    \caption{  Comparison between the HST/WFC3 F140W image (left, 5880 \AA\, rest-frame) and JWST/NIRCam F150W image (right, 6300 \AA\, rest-frame) of the Sparkler (here listed as S2.2). The two images are produced with the same scales (AB magnitudes) and colour bars. The white circles show the positions of all the clumps detected in the NIRCam data. Almost no clumps are visible in the current HST observations.}
    \label{fig:fig_HST_JWST}
\end{figure}

\section{Clump photometry and spectral energy distribution analysis}
\label{sec:cl_analysis}

\subsection{Clumps identification}
\label{sec:cl_identification}
Thanks to the high resolution and sensitivity of the NIRCam images, we detect a large quantity of clumpy structures within the selected galaxies. We first visually identify clumps on a colour image composed of the six NIRCam filters (Blue: F090W+F150W, Green: F200W+F277W and Red: F356W+F444W, \citet{ERO}). Clump candidates are selected as compact systems that stand out of the diffuse light of the host galaxy. The use of colour images has been indispensable in order to discriminate which compact clump belongs to the targeted galaxy and avoid inclusion of interlopers, such as globular clusters belonging to the galaxy cluster or other point-like background and foreground sources (an example is presented in Fig.~\ref{fig:selection_clumps}). Critical lines have been used to separate clump candidates belonging to different images of the same galaxy.
In a second step, we produce high contrast median smoothed images of the F150W and F200W filters (using a 30 pixels side median filter) and we visually confirm each clump detection in the F150W and F200W NIRCam filters and add any missing one. These filters resulted to be the most sensitive to clump detection and analysis and they are, hence, used as reference filters in this study.

The methodology adopted above is limited by contrast, i.e. bright clumps will easily dominate and be detected against fainter clumps. To overcome this problem, we add a third iteration of detection. The latter is performed on the F150W residual images produced by the procedure described in Sect.~\ref{sec:cl_model} { (see Figs.~\ref{fig:residuals1} to \ref{fig:residuals6})}. After the light of the bright clumps has been fitted and removed, we inspect the residual images for remaining compact sources that have been missed. In total, 10 clumps have been added to the final catalogue.

We notice that our extraction method has also included faint clump candidates with detection below the 3$\sigma$ even in the reference frames (about 25 in total). To ensure that we recover reliable physical properties for the clumps,  we also apply as a condition a minimum   $3\sigma$ detection in the two NIRCam reference bands (see next section). Our final sample consists of 223 clumps. The galaxy images and clump redshift distributions are presented in Fig.~\ref{fig:histo_sample}. In most of the cases, we detect less than 5 clumps per image (cf. Fig.~\ref{fig:histo_sample}). However, three galaxies host more than 20 clumps (S2.2 has 27 clumps at $z=1.3779$; S4.2 has 27 clumps at $z=2.19$, I7 has 28 clumps at $z=1.16$). For the large majority of the clumps we do not have detection in the HST images both because HST observations are significantly shallower and/or resolution is significantly worse (see, for example, Fig.~\ref{fig:fig_HST_JWST}). This result alone, is a clear motivation for why JWST observations are paramount.  

We show the final clump selection in the galaxy S2.2, named the "Sparkler" by \cite{mowla2022}, in Fig.~\ref{fig:fig_HST_JWST} using both the HST/F140W, on the left, and the NIRCam/F150W frame, on the right, as background images. At a first glance, it is immediately apparent that the power of the higher spatial resolution and sensitivity of the JWST observations, at comparable wavelengths as HST, is enabling us to detect fainter point-like sources, which are smeared out and lost in the diffuse light gathered by the HST NIR detector. Our catalogue comprises all the candidate clumps analysed in \cite{mowla2022} (except their clump 9, which we excluded this clump of our analysis because his colour is very similar to galaxy cluster globular clusters), but triplicate the final number of clumps, improving the detection significantly along and within the main galaxy body.

\subsection{Clumps modeling}
\label{sec:cl_model}

Since stellar clumps will appear with different sizes and resolutions depending on the magnification and distortion, fixed aperture photometry will produce significant contamination and might introduce systematics. We, therefore, simultaneously fit for clump size and fluxes, using the method initially developed by \citet{messa2019} and later improved and adapted to take into account lensing models as presented in \citet{Messa2022}. 

We determine clump sizes and fluxes in the two reference observed frames, the JWST/NIRCam F150W and F200W filters. We assume that clumps can be modelled with an elliptical 2D Gaussian profile convolved with the instrumental PSF in the image plane and a spatially-varying local background emission, that we want to remove. The model grid is composed of a 2D Gaussian profile convolved with the instrumental point-spread function (PSF) (the latter size will depend on the standard deviation and ellipticity) and a first-degree polynomial function to take into account possible background emission. The degree polynomial function is described by three free parameters. The 2D Gaussian profile is parameterised by the clump centre ($\rm x_0$ and $\rm y_0$), the minor axes standard deviation ($x_{\rm std}$), the ellipticity\footnote{the ellipticity is the axes ratios, such that $x_{\rm std}*\epsilon = y_{\rm std}$} ($\epsilon$), the positional angle ($\theta$, describing the orientation of the ellipses), and the flux ($f$). We model the PSF of each filter included in the analysis from a stack of 3 bright and non-saturated stars, detected within the field of view of the galaxy cluster region. The HST or JWST image cutouts of each star are combined to create a  $1\arcsec\times1\arcsec$ averaged image of the PSF. In Sect.~\ref{sec:comp_psf} of the Appendix, we compare this {empirical} PSF with the publicly available webbpsf model, showing that the agreement is very good. We will use the empirical PSF for the remaining of the analysis as it is the closest representation of the true PSF of the observed data.  

Following the method by \citet{Messa2022}, we determine an analytical expression of the PSF shape in each band, by fitting the resulting PSF image with an analytical function described by the combination of a Moffat, Gaussian, and 4th-degree power-law profiles. This fit provides a good description of the PSF up to a radius of $\sim$0.5\arcsec, which is significantly larger than the physical region used to fit the clump light distribution. This size also corresponds to the physical scale to which aperture correction is estimated (i.e. in a similar fashion to PSF photometry). We report the {measured} PSF FWHM of each filter in Table~\ref{tab:table_observations}.

The spatial fit is performed on a cutout image (9$\times$9 pixels) centred on each clump, using a least-squared approach via the python package \texttt{lmfit} by \citet{Newville2021}. We tested different cutout sizes{. For example, 7$\times$7, worked equally well for a large fraction of clumps, but not for the extended ones. The 9$\times$9 box size performed best in most of the cases and produced the least residuals. We notice that when two clumps are separated by less than 4 pixels in the F150W images (i.e. 0.08") we fitted the two clumps simultaneously using a 13$\times$13 pixel cutout image. This simultaneous fit on a larger box size produced better results than fitting the two nearby clumps individually. }  

During testing, the clump sizes determined by independently fitting the F150W and the F200W images have been found consistent (see Fig.~\ref{fig:comp_sizes} in Appendix \ref{sec:comp_sizes}). Therefore, we choose the resulting 2D Gaussian parameters produced by fitting clumps in the F150W frames as reference for extracting the fluxes in the other bands. 

 Once the Gaussian shape ( $x_{\rm std}$, $\epsilon$ and  $\theta$ parameters) is fixed, we fit the clump cutouts on the other bands by treating the flux $f$ and the local background as free parameters. We also allow  the position of the centre ($\rm x_0$, $\rm y_0$) to vary by 1 pixel maximum, to account for shifts due to different pixel scales among the data. This assumption ensures that the clumps have intrinsically the same sizes and morphology in all the filters and reduce the risk of including flux originating in areas surrounding the clumps. For 41 clumps located in very crowded regions, we fixed the position at the values measured in the reference filter F150W. This approach was necessary to avoid contamination.  We  fit all the HST/ACS HST/WFC3 and JWST/NIRCam available (13 filters, cf Sect.~\ref{sec:data}).

According to the testing performed by \citet{messa2019, Messa2022}, we assume 0.4 pixel as the minimum resolvable Gaussian axis standard deviation (std) in the F150W and F200W frames. If a clump is only resolved along the shear direction of magnification (i.e., the minor axis std is $\leq 0.4$ pixel but the major axis std is $> 0.4$ pixel) we estimate the size based only on the major axis value, assuming the clump is intrinsically circular in the source plane.  If a clump is not resolved (i.e. the two axes std are $\leq 0.4$ pixel), the size is reported as upper limit at 0.4 pixel).

The photometric error in each filter accounts not only for the Poissonian noise, but also for the uncertanties in the size measurement due to the underlying local background. Applying a bootstrapping approach, we repeat 100 times the size and flux measurements in the F150W reference filter by randomly sampling the standard deviation of the local sky under the clump. The best size parameters of each of the 100 realisations are then used to estimate the fluxes in all the other filters. The final error associated to the size measurement is given by the 68\% confidence interval on each side of the best-value measured on the original F150W image. The flux errors are estimated by summing in quadrature the Poissonian error from the best-fitted model and the standard deviation of the hundred flux measurements in each filter.

For each filter, we also determine the intrinsic magnitude limit ($mag_{\rm lim}$, reported in Table~\ref{tab:table_observations}) corresponding to the minimum flux of a PSF aperture measured in an empty region of the sky. This minimum flux is converted to an AB magnitude for each filter and used as upper limit when performing SED analyses (see Sect.~\ref{sec:sed}). We use these upper-limit values during the SED fitting analysis, for clumps that are detected with a S/N lower than 2 in a given filter.

In Fig.~\ref{fig:im_residuals}, using the Sparkler as a showcase (see Sect.~\ref{sec:single_gal}), we plot the clump positions in the observed F150W image plane on the left panel, the best-fitted model of each clump in the central panel, and the residual image after the clump light has been subtracted on the right. The residuals at the location of the clumps are minimal, while the diffuse light of the galaxy is clearly visible in the last panel. Appendix \ref{sec:clump_models} includes similar figures for all the targeted galaxies. Overall, the residuals show the quality of our analysis approach in reproducing the size and flux distribution of each clump, without resulting in over or under-substractions of the clump light on the diffuse light of the galaxy. We list in Tables~\ref{tab:table_clumps_phot1} to~\ref{tab:table_clumps_phot4} the JWST/NIRCam observed magnitudes of each clumps. 

\subsection{Unlensed clumps properties}
\label{sec:unlensedclump}
At this stage, the resulting clump catalogue includes all the 223 clumps that have detection in both reference JWST bands, F150W and F200W, their IDs, galaxy redshift, positions, standard deviation and ellipticity ($x_{\rm std}$ and $\epsilon$, which provide the major and minor axes of the ellipsoidal), position angle, fluxes in all the bands as measured in the image plane of the galaxies. Uncertainties associated to all the measured quantities are also included.

In order to estimate the intrinsic properties of the clumps, we use the galaxy cluster lens model (Sect.~\ref{sec:lens_model}) to produce magnification maps of each galaxy in the image plane.  At the location of the clump within the galaxy, we measure the median value of the magnification map in the region enclosed within the ellipsoidal describing the shape of the clump (using the best-fit parameters $x_{\rm std}$ and $\epsilon$ and $\alpha$). We refer to these median values as $\mu$ and we list them in Tables~\ref{tab:table_clumps1} to \ref{tab:table_clumps4} for each clump.

The uncertainties introduced by the magnification are composed of two components. First, we measured the standard deviation of
the magnification values within each clump region which represents the potential spatial variation of magnification across the size of the clumps ($\delta \mu_1$, this value is particularly high for the clumps located very close to the critical lines). 

Second, we randomly produce 100 magnification maps for each clump, selected from the lenstool MCMC posterior distributions of the lens model. We measure the magnification of each clump on the 100 maps and estimate the 68\% confidence interval on each side of the best value to obtain asymmetrical errors on the magnification ($\delta \mu_{2-}$ and $\delta \mu_{2+}$ for lower and upper error-bars).

The final  lower and upper magnification uncertainties of each clump are given by $\delta \mu_{+,-}= \sqrt{ \delta \mu_1 ^2 + \delta \mu_{2+,-} ^2}$. These values are included in Table~\ref{tab:table_clumps1} and associated to their respective $\mu$ values. 

Intrinsic fluxes are derived by dividing the {observed} fluxes by the magnification value. The intrinsic fluxes  are converted into AB absolute magnitudes after correcting from the reddening introduced by the Milky Way \citep{SF2011} in each filter (these corrections are listed in Table~\ref{tab:table_observations}).

To derive the intrinsic effective radius, $R_{\rm eff}$, we first estimate the radius of the circle having the same area of the ellipses describing the morphology of the clump, i.e., $R_{\rm cir} = \sqrt{x_{\rm std}\times y_{\rm std}}$. We assume that $R_{\rm cir}$ is the standard deviation of a 2D circular Gaussian, and derive the observed PSF-deconvolved effective radius as $R_{\rm eff,obs}=R_{\rm cir}\times \sqrt{2 ln(2)}$. 

We consider three cases when measuring the intrinsic effective radius. If the clump is resolved along both axes in the image plane, the intrinsic effective radius, $R_{\rm eff}$, is obtained by dividing $R_{\rm eff, obs}$ by $\sqrt{\mu}$. If the clump is not resolved in one direction we divide $R_{\rm eff, obs}$ directly by the magnification along the shear direction ($\mu_t$). The underlying assumption in the latter case is that the clump is intrinsically circular and its radius is resolved in the tangential direction \citep[e.g.,][]{Vanzella2017c}. The same assumption (circular shape) is made in the third case, i.e., if the clump is unresolved in both direction. In this latter case we use $\mu_t$ to derive the size upper-limit.
Finally the physical sizes in parsecs are derived considering the pixel size of the F150W images (0.02'') and the angular diameter distance of each galaxy.
We propagate the magnification uncertainty both in the intrinsic flux and effective radius errors.

We list in Tables~\ref{tab:table_clumps1} to \ref{tab:table_clumps4} the IDs, redshift, positions, magnification and uncertainties, intrinsic flux in the reference filter F150W, intrinsic $R_{\rm eff}$, and uncertainties associated to the measured quantities. 

\subsection{Broadband SED fitting}
\label{sec:sed}
The final photometric catalogue is then used as input to perform the SED analysis of the clumps and estimate ages, masses and extinction of the clumps. We use the Yggdrasil stellar population library (\citet{Zackrisson2011}), based on Starburst99 Padova-AGB tracks \citep{leitherer1999, VL2005}, generated with a universal \citet{Kroupa2001} initial mass function (IMF) in the interval 0.1-100 $\rm M_{\odot}$. The Yggdrasil models also include contribution from nebular continuum and emission line estimated with Cloudy \citep{ferland1998}. In this work, we use spectral evolutionary models with a gas covering fraction $\rm f_{cov}=0.5$ (i.e., they assume that 50 \% of Lyman continuum photons are escaping the star-forming region). We perform 12 different runs. We fit the clump SEDs using four metallicity values ($Z=0.02$, $Z=0.008$, $Z=0.004$ and $Z=0.0004$) and three different star formation histories (SFHs). Solar metallicity models are not used to fit clumps/clusters in galaxies at z$>$3, as their morphological appearance suggests they are low mass systems more similar to dwarf galaxies, thus less metal rich.

The first SFH assumes instantaneous burst (hereafter, IB), i.e., the stars form in a single event and then they evolve with time. These models are typically used to analyse young star clusters in local galaxies \citep[see review by][]{adamo2020a}. In the second SFH, we assume constant star formation for a time span of 10 Myr followed by stars aging with time (hereafter, 10Myr). This model is more realistically representing star formation taking place in giant star-forming regions of a few 100 pc in the local universe \citep[see][for a discussion regarding timescale for star formation in star-forming regions]{adamo2013}.  For the third SFH, we use constant star formation over 100 Myr, followed by stars evolving with time (hereafter, 100Myr). This model would better describe star formation in regions of galaxies that are constantly replenished by gas and sustain star formation over a long time range in a relatively small region of the galaxy (e.g., circum--nuclear starbursts within disk galaxies). 

We also include attenuation in the form of the \citet{Calzetti2000} attenuation law. We do not use the differential expression formulated by \citet{Calzetti2000}, but we apply the same reddening to both stellar and nebular emission, assuming that stars and gas are well mixed. The attenuation is applied to the model spectra before convolution with the filter throughput in the range $\rm E(B-V)=0$ to 1 mag, in steps of 0.05 mag. We allow the extinction to vary freely when fitting clumps within the galaxy host. While we fix the extinction to 0 if clumps/clusters are located outside their host galaxies (the latter are outlined in the Table~\ref{tab:table_clumps1}). This assumption helps to minimise the age-extinction degeneracy that affect optical broadbands SED analyses.   

We do not use the filter containing Lyman-$\alpha$ emission for galaxies at $z>6$, as strong unknown inter-galactic medium absorption will attenuate the flux transmission at this wavelength range.
We perform the SED fitting in flux units. We redshift model spectra at each clump redshift and then convolve with the filter shapes (using the pyphot library). 
A minimum reduced $\chi^2$ analysis is performed to determine the best match between model and observed SED.

\begin{figure*}	\includegraphics[width=18cm]{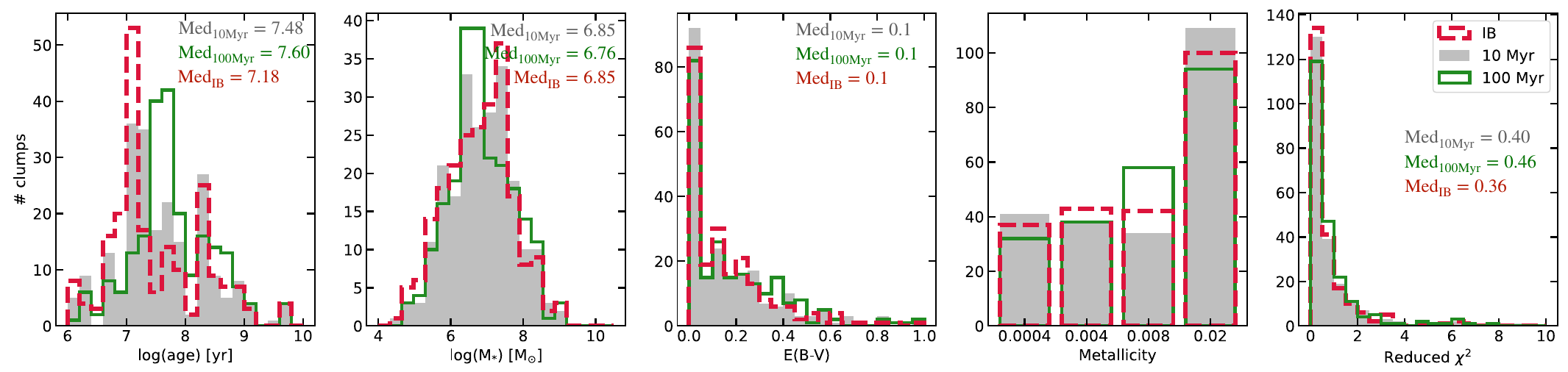}
    \caption{Clump physical properties and resulting reduced $\chi^2$ distributions, and median values obtained  from the different SED models. We show the best-fit solution obtained with the three different SFH:  instantaneous burst (red) or continuum star formation during 10 Myr (grey), 100 Myr (green). See Sect.~\ref{sec:sed} for more detailed discussion.}
    \label{fig:comp_SED}
\end{figure*}

We present in Fig.~\ref{fig:comp_SED}, the best-fit results and $\chi^2$ distributions for the three different SFH.
They are all remarkably similar in terms of mass, metallicity and $\rm E(B-V)$ parameters.
More noticeable are the changes in the resulting ages of the clumps when we use different SFH assumptions (left panels). Overall, we observe that as we use models where star formation proceeds for a longer time (from IB to 100Myr) the ages of the clumps get on average older but the range in age does not change significantly. These trends have already been remarked in the literature \citep{adamo2013, Messa2022}, who find similar behaviour and dependencies.

For the remaining of the analysis we use the clump properties obtained by fitting clump SEDs with the 10Myr SFH. 
We motivate the choice of this SFH as it better reflects what we know of star formation taking place in giant star-forming regions in local starburst galaxies \citep{bastian2006, sirressi2022}, assuming that clumps in high-redshift galaxies form in a similar fashion to those in the local universe. In the remaining of the paper, we will remark whether we reach different conclusions using different SFH assumptions.

We include in our analysis only clumps with at least detection in 4 bands with magnitude values brighter than the $\rm mag_{\rm lim}$ and with S/N$>$2 (in other filters than F150W and F200W reference data, in the latter we impose a S/N$>$3 for selection, see Section~\ref{sec:cl_identification}). During the fit we use upper limits as values to exclude families of solutions that would predict fluxes two times brighter than these limits. To estimate the uncertainties on the SED fitting results, we randomly draw a set of 100 magnitudes for each clump using the 1$\sigma$ uncertainty in each filter and run a new SED fitting analysis on each set. Error bars correspond to the 68\% confidence interval on each side of the best-model value measured on the observed data. 

In total, we recover physical parameters for the 221 clumps.

\begin{figure*}
	\includegraphics[width=18cm]{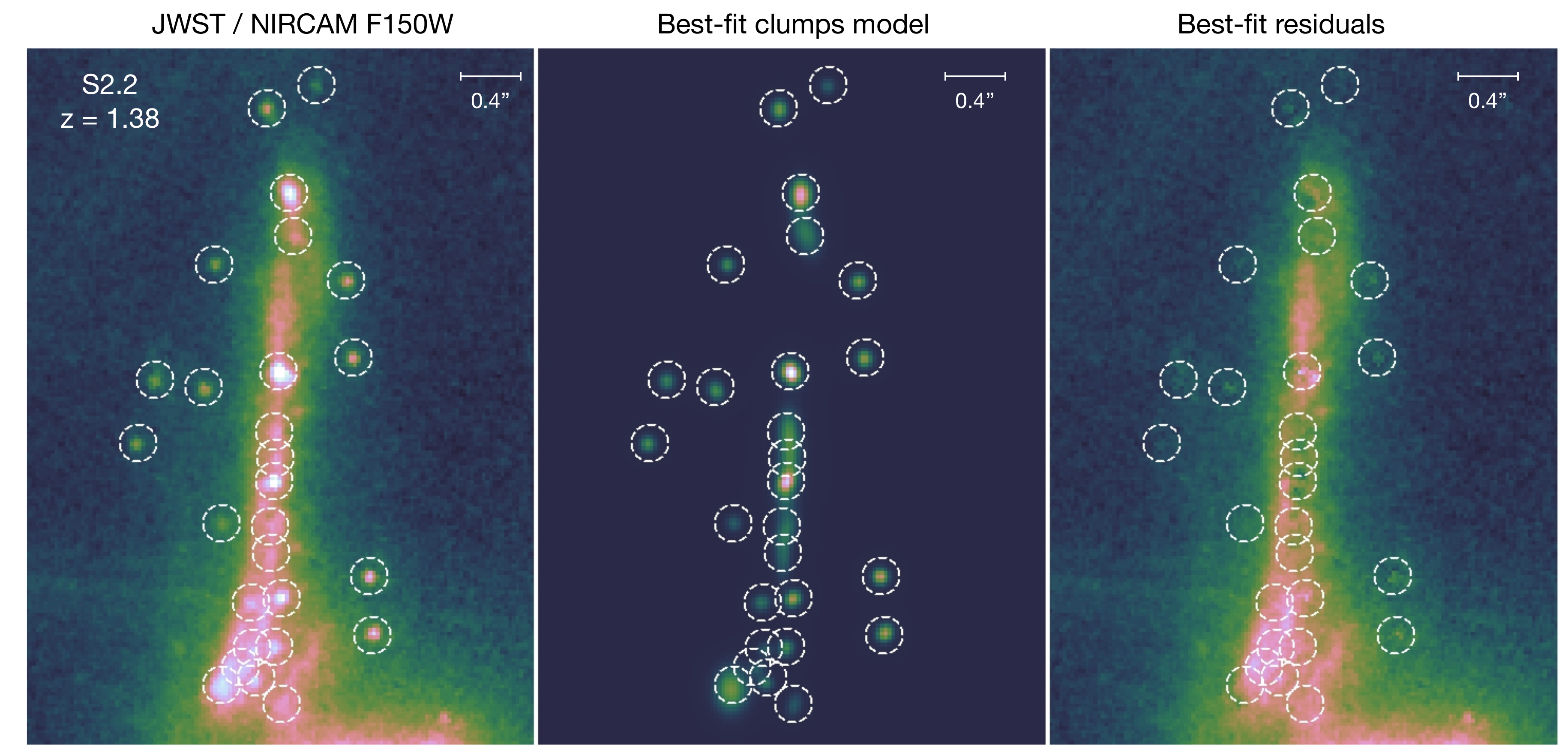}
    \caption{ Example of measurements of clump sizes and fluxes for the Sparkler galaxy. From left to right: F150W observations, best-fit clump model, best-fit residual images. The three images are produced with the same scales and colour bar. The white circles indicate the position of all the clumps. The same fits and residuals of the other galaxies are presented in Fig.~\ref{fig:residuals1} to ~\ref{fig:residuals6}.}
    \label{fig:im_residuals}
\end{figure*}

\section{Results}
\label{sec:results} 
\subsection{The effect of magnification on derived clump physical properties}

\begin{figure*}
    \includegraphics[width=18cm]{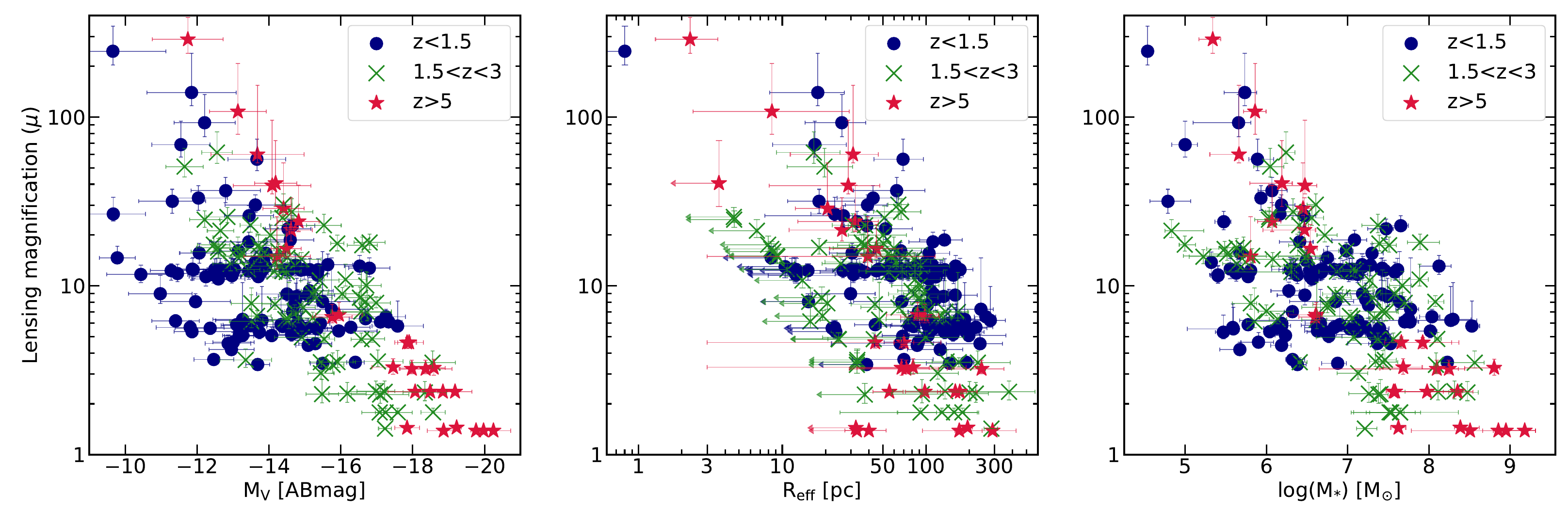}
    \caption{{\bf Left}: Lensing magnification as a function of the absolute V band magnitude of the clumps. {\bf Middle}: Lensing magnification as a function of the clump effective radius. The size upper limits are indicated with an arrow. { \bf Right}: Lensing magnification as a function of the clump stellar mass. The clumps with the highest derived stellar masses are, as expected, the ones with the lowest lensing magnification values. In the three panels the clumps are colour-coded by redshift bins.}
    \label{fig:MU_VS_masse}
\end{figure*}

In Fig.~\ref{fig:MU_VS_masse}, we plot the average magnification, at the position of each clump, versus the clump intrinsic pseudo-$V$ band absolute magnitude, M$_V$  (measured in the filter which overlaps with 5500 \AA\, rest-frame),  the $R_{\rm eff}$, and the stellar mass of the clumps. The sample is divided in redshift bins which bracket clump properties at the cosmic dawn ($z>5$, filled stars), cosmic noon ($1.5\leq z \leq3$, crosses), and cosmic afternoon ($z<1.5$, filled circles). The recovered range of magnifications and their distribution in the clump populations detected in the SMACS0723 region are similar to the range reported for the region of MACSJ0416 by \citet{mestric2022} and in other two Frontier Field (FF) clusters, Abell 2744 (A2744) and Abell 370 (A370), for which the clump populations are currently under analysis (Claeyssens et al., in prep.). The highest magnifications ($\sim 100$) are achieved in proximity of the critical lines and are also the values with the largest size uncertainties. The recovered M$_V$ ranges from $-10$ to $-20$ ABmag, while masses go from $\sim10^5$ to $10^9$ \msun. We found that 20\% of the clumps are unresolved, i.e., we can only estimates upper limits which range between 10 and 50 pc. These systems line-up forming a sequence on the central panel of Fig.~\ref{fig:MU_VS_masse}. We notice a small cloud of clumps at low magnification, these systems are either clumps in counter images of multiple systems (e.g. blending of several single clumps), or are detected in single image galaxies at the highest redshift. The lack of clumps in the lower-left (compact sources but faint and low mass with relatively low magnification) and upper-right corners (extended, thus, low surface brightness objects) are possibly due to completeness limits \citep[see][for a discussion on completeness effects in lensed regions]{Messa2022}.  

As already reported by \citet{mestric2022}, we see a continuity in the distribution of clump sizes which go from typical giant star-forming regions (a few 100 pc) to star clusters ($\sim 10$ pc or smaller but with very high uncertainties). This behaviour is linked to the hierarchical and fractal properties of star formation that operates in these high-redshift galaxies in a similar fashion, but under more extreme conditions, than in local galaxies. Clumps are not star-forming units, but simply correlated regions of star formation, that appear compact because of resolution effects.   

It is interesting to compare these recovered physical properties with the luminosities of the most massive star clusters observed in the local universe. \citet{whitmore2010}  report for the Antennae system absolute magnitudes reaching $\sim-13$ Vega mag\footnote{AB and Vega mag are comparable in V band to better than 5 \%}, corresponding to stellar masses of $\sim10^7$ \msun. The spatial resolution achieved with HST at the distance of the Antennae is about 7 pc. The most luminous star clusters in the HiPEEC sample of merging galaxies \citep{adamo2020b} and in Haro 11 \citep{sirressi2022} reach M$_V\sim-16$ mag and masses of $\sim 10^8$ \msun. In the latter sample, these values are recovered over physical resolutions of about 30 pc. We clearly see an overlap between these extreme but rare (in the local universe) cluster formation events and the average properties of highly resolved clumps in high redshift galaxies, suggesting that the latter are truly the sites where globular cluster progenitors form.

\subsection{Clump luminosity-size relation as a function of redshift}

Numerous works, studying clumps within galaxies, use the  luminosity (or SFR) versus size plot as a tool to compare physical conditions of clumps as a function of redshift. \citet{Livermore2015}, compiling several literature studies that included star-forming regions in typical main-sequence galaxies of the local universe as well as clumps in lensed galaxies with redshift $<$3, concluded that higher surface brightness clumps are found in increasingly higher redshift galaxies. This trend has been linked to the fact that clumps form out of fragmentation in galaxies that have high gas fractions  for increasing redshift. 

\begin{figure*}
    \includegraphics[width=18cm]{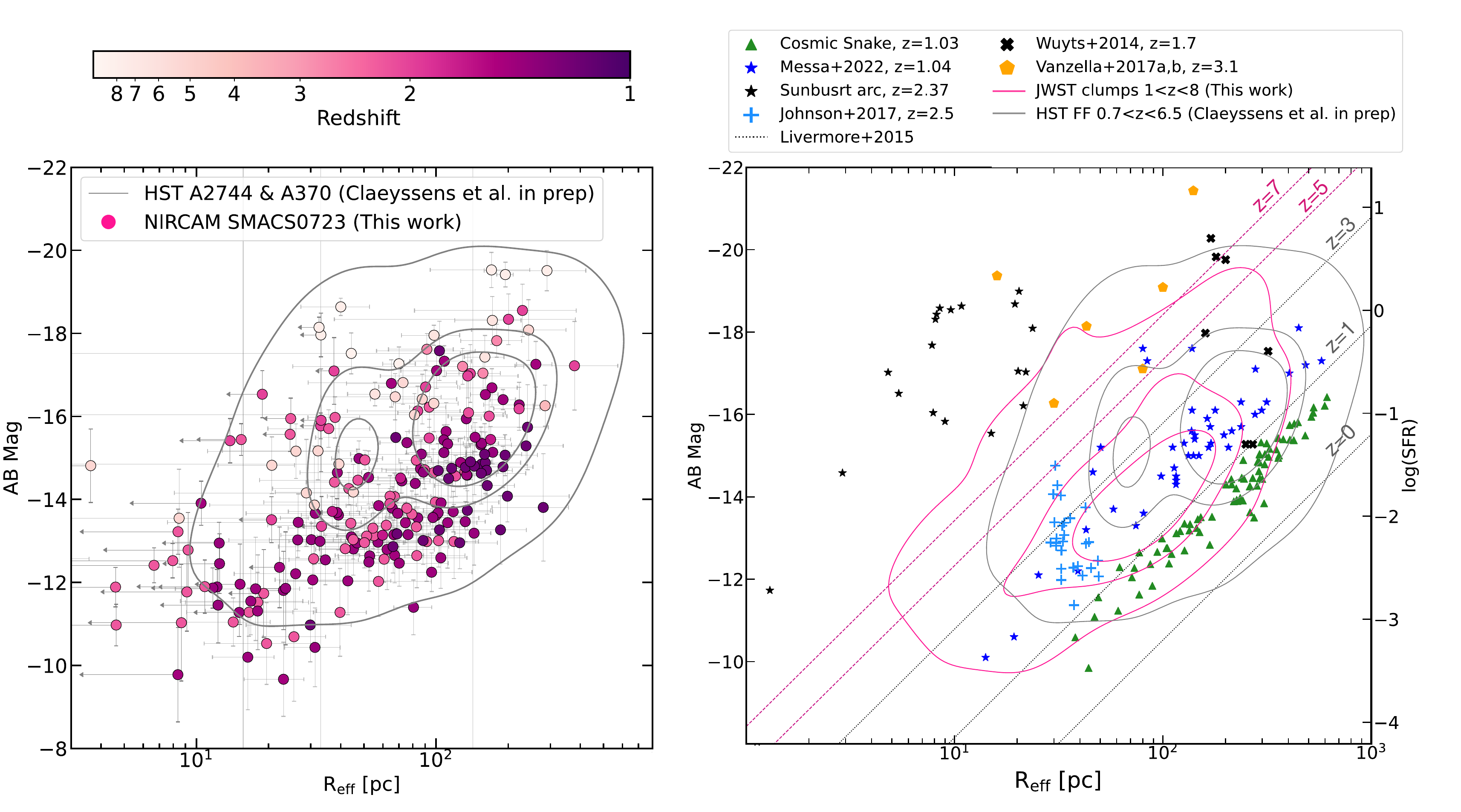}
    \caption{ {\bf Left:} Clump effective radius, $R_{\rm eff}$, plotted versus absolute magnitudes, M$_{\rm F150W}$ for clumps with $z>4$ and M$_{\rm F090W}$ for clumps with $z<4$ to cover the rest-frame UV. The points are colour-coded by the redshift. The grey contours, referring to the right y-axes, enclose 95\%, 50\% and 25\% of the sample of FUV clumps detected in two other lensing clusters (A2744 and A370, both FF clusters) analysed by Claeyssens et al. (in prep.) from HST data. For the latter sample, sizes and fluxes are measured in rest-frame FUV wavelength using the same method presented here. {\bf Right:} The luminosity (or SFR) versus size relation including data from the literature. The left y-axes shows AB magnitudes. They correspond to F150W of F090W magnitudes (depending on redshift) for the JWST clumps of this work (pink contours enclosing 95\%, 50\% and 25\% of the sample, notice the rest-frame wavelengths does not correspond to FUV mag, see left panel). We include the FUV magnitudes for the HST FF sample by Claeyssens et al (in prep.) visualised again with grey contours, and literature data for which we use the reported radii and FUV magnitudes (Cosmic Snake by \citealt{Cava2018}; A521-sys1 by \citealt{Messa2022}; Sunburst arc by \citealt{vanzella2022}; the SDSS J1110+6459 system by \citealt{Johnson2017}; several clumps studied by \citealt{Vanzella2017b, Vanzella2017a}). The clump sample by \citet{Wuyts2014} is plotted referring to the right-side y-axes. i.e. the SFR. The grey and pink lines on the right panel represent the relation between SFR and sizes measured by \citet{Livermore2015} at $z=0$, 1 and 3, and $z=5$ and 7, respectively. The sample used by Livermore and collaborators is omitted for clarity but is represented here by relations.}
    \label{fig:Reff_VS_M150}
\end{figure*}

It is common to find in the literature the FUV magnitude vs. $R_{\rm eff}$ diagram, where the FUV luminosities are used as proxy for SFR because they are easier to retrieve from HST data. An example of the latest compilation of such diagram can be found in \cite{mestric2022}, who include also clumps detected in the Frontier Field (FF) lensing cluster MACSJ0416. In our analysis, only a small fraction of the JWST detected clumps in the SMACS0723 lensing cluster region have detection in the shallower HST data. For this reason we build the luminosity--size diagram using the F090W or F150W magnitudes for clumps with $z<4$ and $z>4$, respectively, to probe the bluest optical rest-frame wavelengths (1500 to 4000 \AA\ rest-frame depending on the redshift). In the left panel of Fig.~\ref{fig:Reff_VS_M150}, we plot all clumps in our sample having detection in the 2 reference filters (F150W and F200W see Sect.~\ref{sec:unlensedclump}). The clumps are colour-coded based on their redshift. As a comparison we overplot (against the right y-axes) the contours that encompass the 95\%, 50\%, and 25\% of FUV rest-frame magnitudes of the clumps detected in the two Frontier Field lensing clusters, Abell 2744 and Abell 370, analysed by Claeyssens et al. (in prep.) with the same method used in this work. The grey contours cover the same parameter space of the  clump sample published by \cite{mestric2022}.  We observe, thanks to the improved spatial resolution and sensitivity of the JWST data, that smaller physical scales and luminosities are found in the clump sample of SMACS0723. This trend is better visualised by the contours in the right panel 
of Fig.~\ref{fig:Reff_VS_M150}. There, we compile several datasets of lensed clumps at redshift ranges between 1 and 6 with FUV determined sizes and luminosities, along with the clump samples (in magenta contours the SMACS0723 sample, in grey contours the FF one from Claeyssens et al.) showed in the previous plot. The clumps detected in the JWST data are on average smaller and less luminous. We also plot with dotted black lines the constant surface brightness prediction for redshifts 0 to 3 as presented by \citet{Livermore2015} and we extrapolate the relation to redshifts 5 and 7 (dashed lines in magenta), which brackets the redshift ranges reached with the current study. These predictions should be considered as representative of the average clump densities at a given redshift range. Overall, we see that they still hold, suggesting that clumps can reach highers densities at higher redshift. These intrinsic properties do not only facilitate their detection in JWST data, but would also point out an evolution on the average properties of the clumps with redshift. However, larger statistics are needed to further confirm this result.

\subsection{Clump physical properties as determined from the SED analysis}

In Sect.~\ref{sec:sed}, we have introduced and discussed the method used to determine clump physical properties, the assumption on SFH and metallicity, and the resulting degeneracies. We use here the outputs produced with the reference model, namely assuming constant SFH over a time-frame of 10 Myr to discuss the general physical properties of the clump population detected in the region of SMACS0723.  

\begin{figure*}
    \includegraphics[width=18cm]{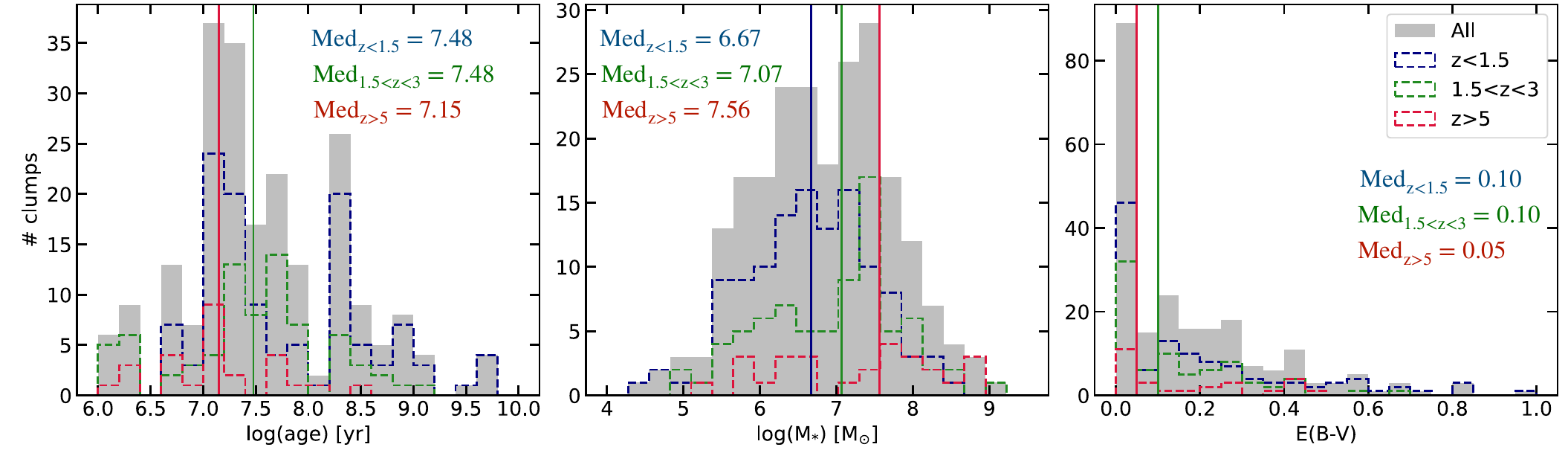}
    \caption{ SED fitting results for the 221 clumps. From left to right: clumps stellar masses, stellar ages and colour excess $\rm E(B-V)$. The grey histograms show the distribution of the total sample, while the blue, green and red distributions show the values measured in three different redshift ranges: $ z<1.5$, $1.5<z<3$ and $z>5$, respectively. The vertical lines represent the median in each redshift bin and the values are given in each panel. The distributions have been obtained from measurements derived by  using the chosen reference model (assuming a continuous stellar formation for 10 Myr).}
    \label{fig:histo_SED}
\end{figure*}

In Fig.~\ref{fig:histo_SED}, we show clump ages, masses and extinction, broken down in different redshift ranges centred and around the cosmic noon. The age range for the three sub--samples are similar. We notice that median ages for clumps at redshift $z>5$ are younger (10 Myr) versus the median age in the redshift range $1.5<z<3$ and for the bin at $z<1.5$ (30 Myr). These medians are all very close to the first broad peak in the age distribution of the clump sample that encompasses the 10-30 Myr. Another secondary but pronounced peak is noticeable in the age range of  300-500 Myr.

Mass distributions have similar ranges but different medians. Clumps at redshift below 1.5 have median masses close to $5\times10^6$ \msun, but they appear increasingly more massive at redshift $1.5<z<3$ (median close to $10^7$ \msun), reaching $5\times10^7$ \msun\, in the highest redshift bin.

Finally, the majority of the clumps have extinction values E($B-V$) smaller than 0.4 mag, with median values that become smaller for increasing redshift (0.10, 0.05, 0.05 mag, respectively in the three redshift bins), suggesting clumps at higher redshift have lower extinction.

\begin{figure}
    \includegraphics[width=9cm]{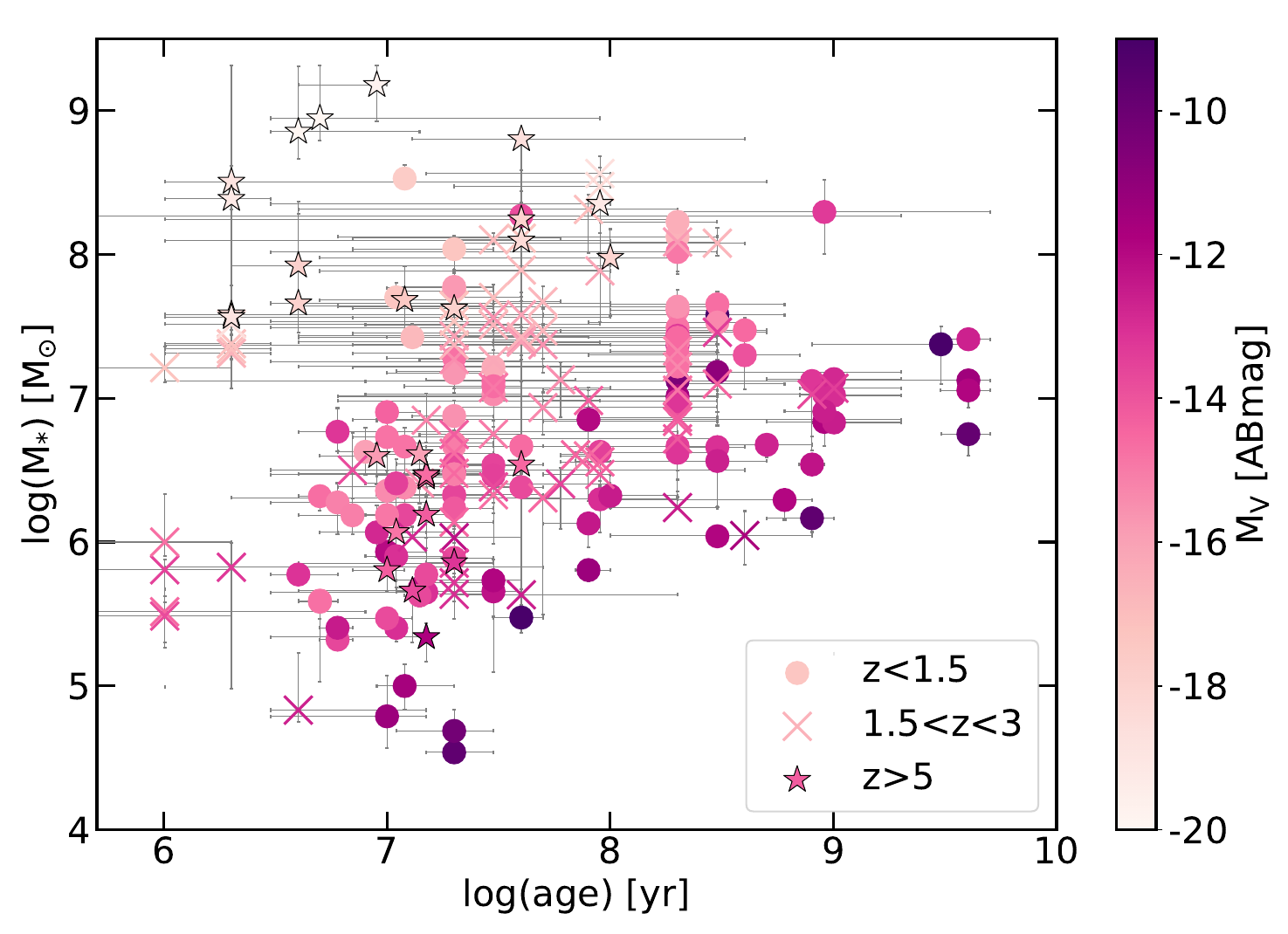}    
    \caption{ Clump stellar masses as a function of their ages. The clumps are colour-coded by their absolute V band magnitude and the different markers represent the redshift bins.}
    \label{fig:EBV_VS_age}
\end{figure}

In Fig.~\ref{fig:EBV_VS_age}, we show the age--mass diagram, typically used for star cluster population analyses, but adapted here to show clump physical properties. Different symbols encode the redshift bin of the clumps (see inset in the figure), while the colour represents the M$_V$ of the clumps. As expected, luminosity in optical increases with mass, with the most massive clumps being also the most luminous. At the low mass (luminosity) range, the lack of clumps is mainly driven by the faintness of the sources, i.e., they fall below detection. It is interesting to notice that, for the entire age range, clumps in the highest redshift bin ($z>5$) are also the most massive. This trend could be driven by observational bias, i.e., they correspond to the least magnified clumps.

\subsection{Clump dynamical age and size--mass relation }

\begin{figure*}
    \includegraphics[width=8.7cm]{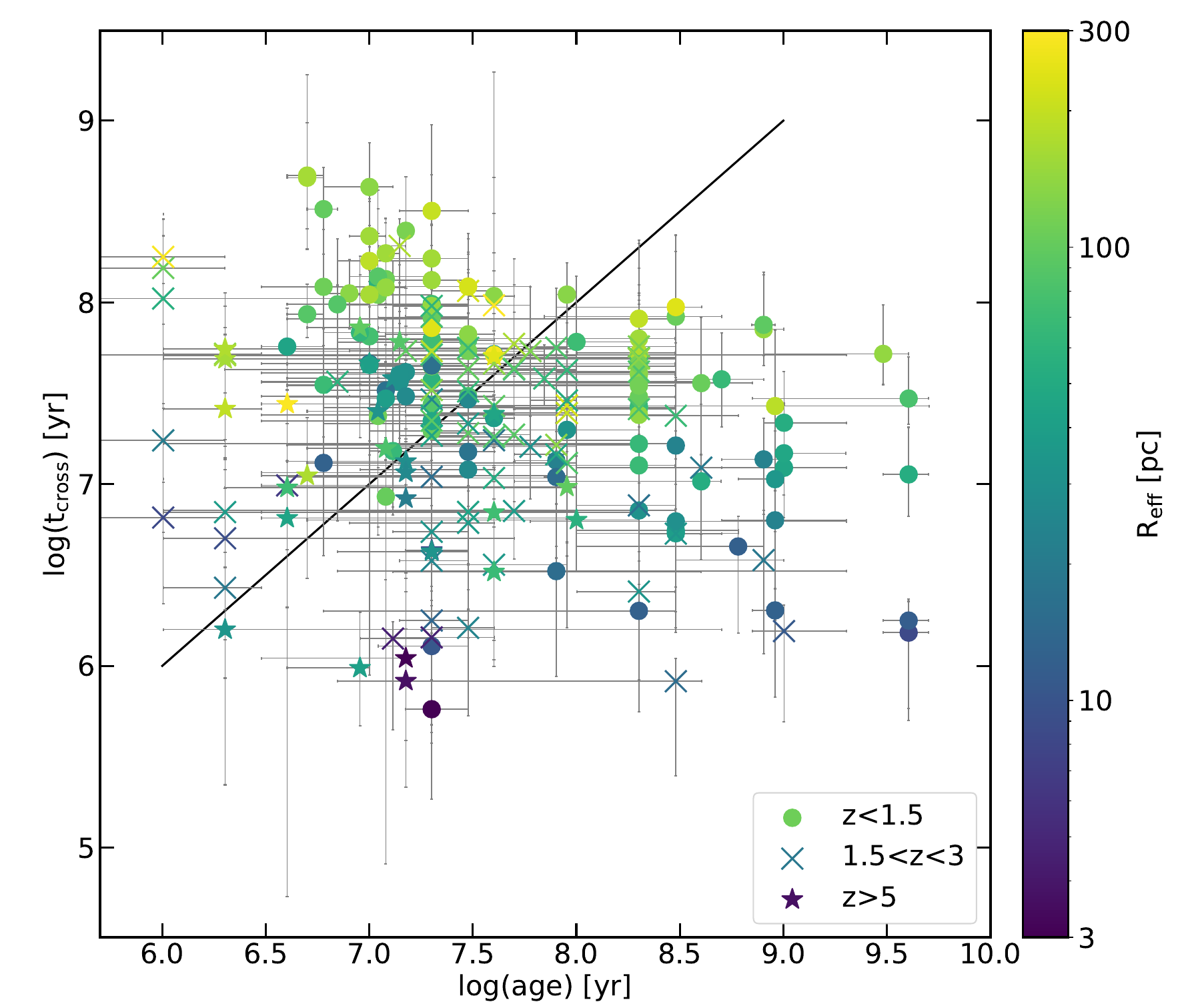}
    \includegraphics[width=7.5cm]{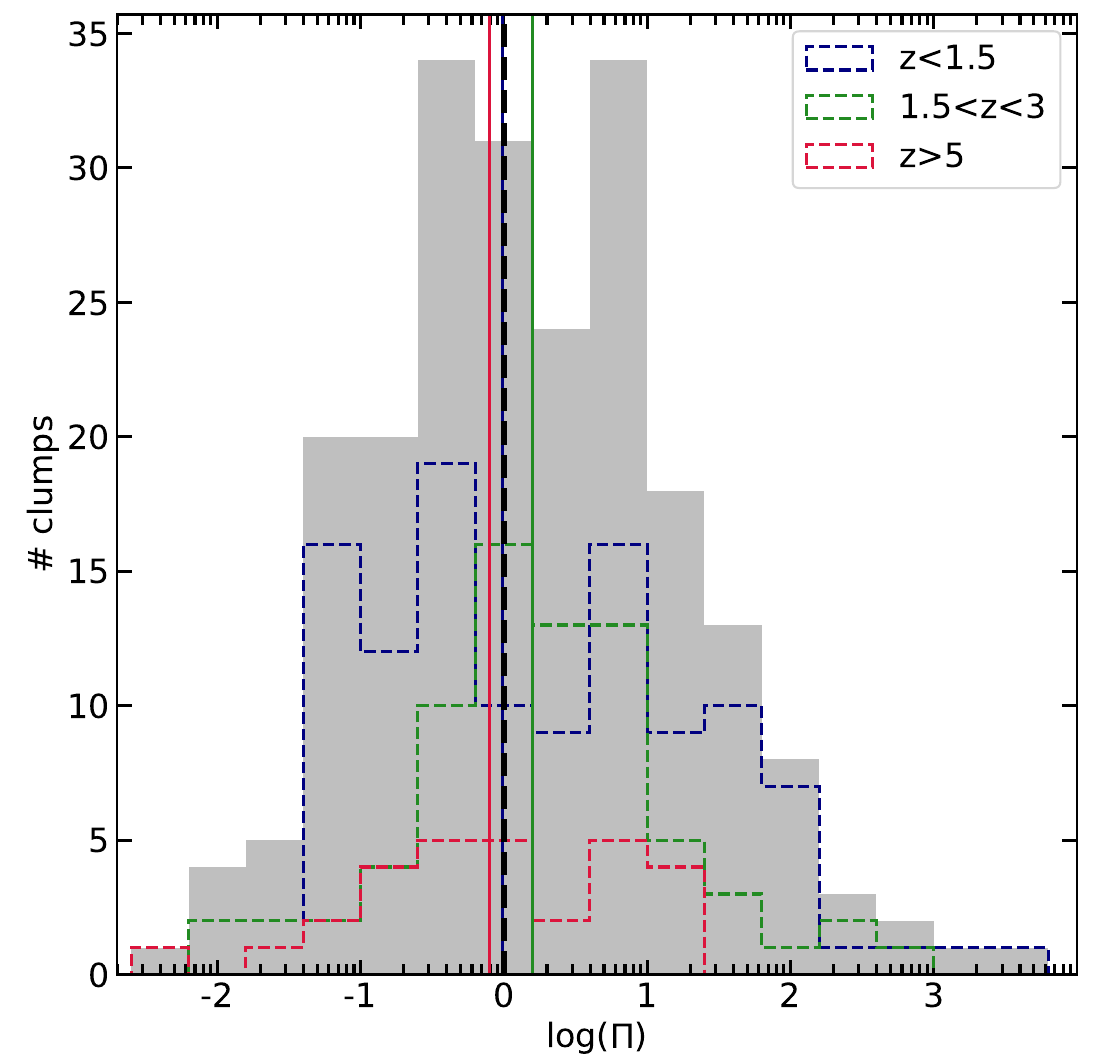}
    \caption{ {\bf Left:} Clump crossing times as a function of clump ages, colour-coded by their R$_{\rm eff}$. These values have been derived using the reference SED model (with a continuous star formation for 10 Myr). The black line shows where the crossing time is equal to the age. Clumps with a crossing time higher than their age (i.e. $\Pi>1$) are expected to be bound. The different markers represent the three redshift bins. {\bf Right:} $\Pi$ value distributions (dynamical age) for the 221 clumps. The grey histogram shows the distribution of the total sample, while the blue, green and red distributions show the values measured in three different redshift ranges: $ z<1.5$, $1.5<z<3$ and $z>5 $, respectively. The vertical lines represent the median values in each redshift bin.}
    
    \label{fig:tcross}
\end{figure*}

In the left panel of Fig.~\ref{fig:tcross}, we compare the crossing time, $\tau_{\rm cross}$ \footnote{As defined by \citet{GPZ2011}, where\\ $\tau_{\rm cross} = 10 \sqrt{R_{\rm eff}^3/GM}$ } to the age of the clumps and find that about 53\% of the clumps have ages older than $\tau_{\rm cross}$.  The fraction changes only slightly for different SFH assumptions, from 45\% in case of IB to 61\% for 100Myr models. 

\begin{figure*}
    \includegraphics[width=8cm]{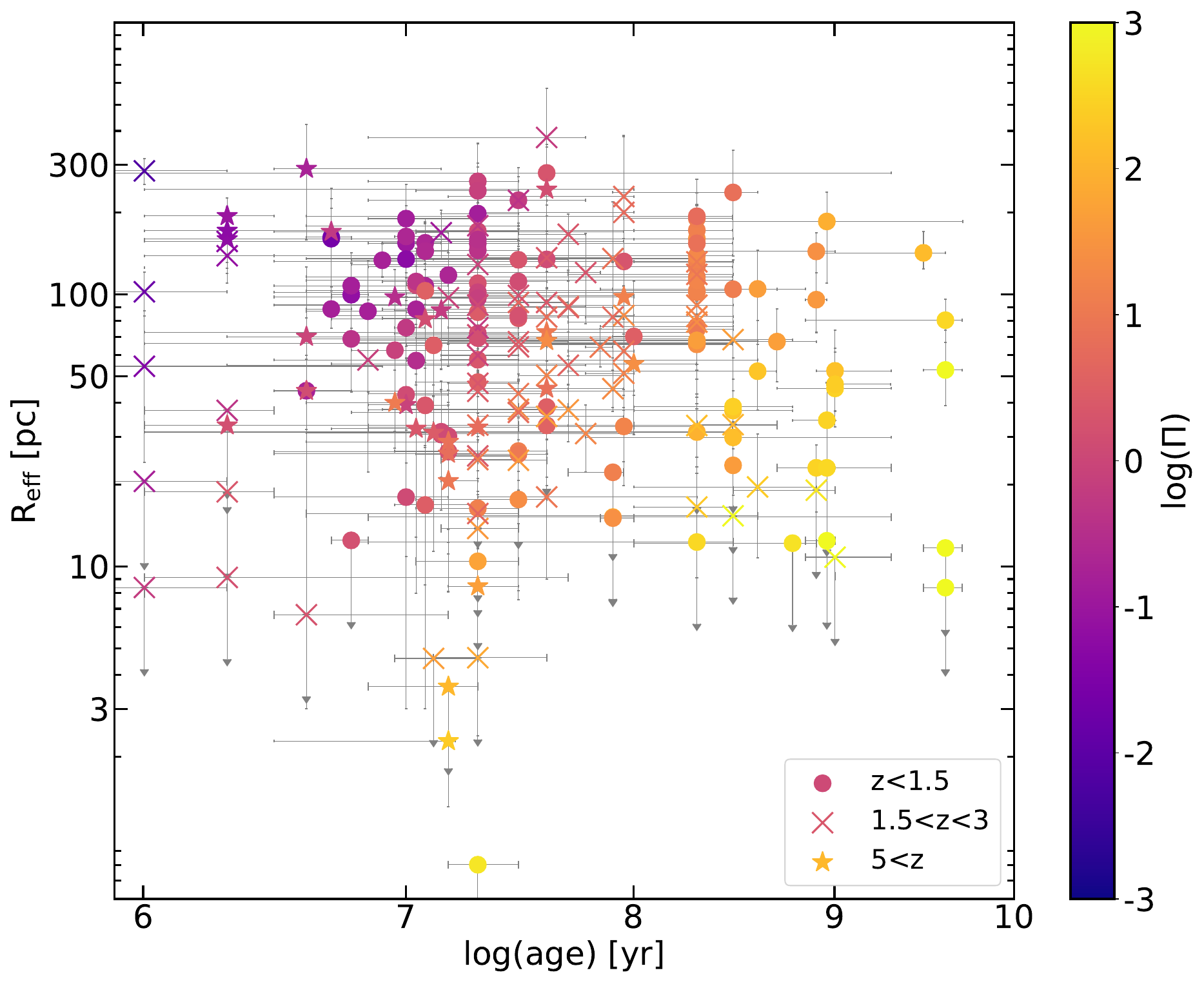}  
    \includegraphics[width=8cm]{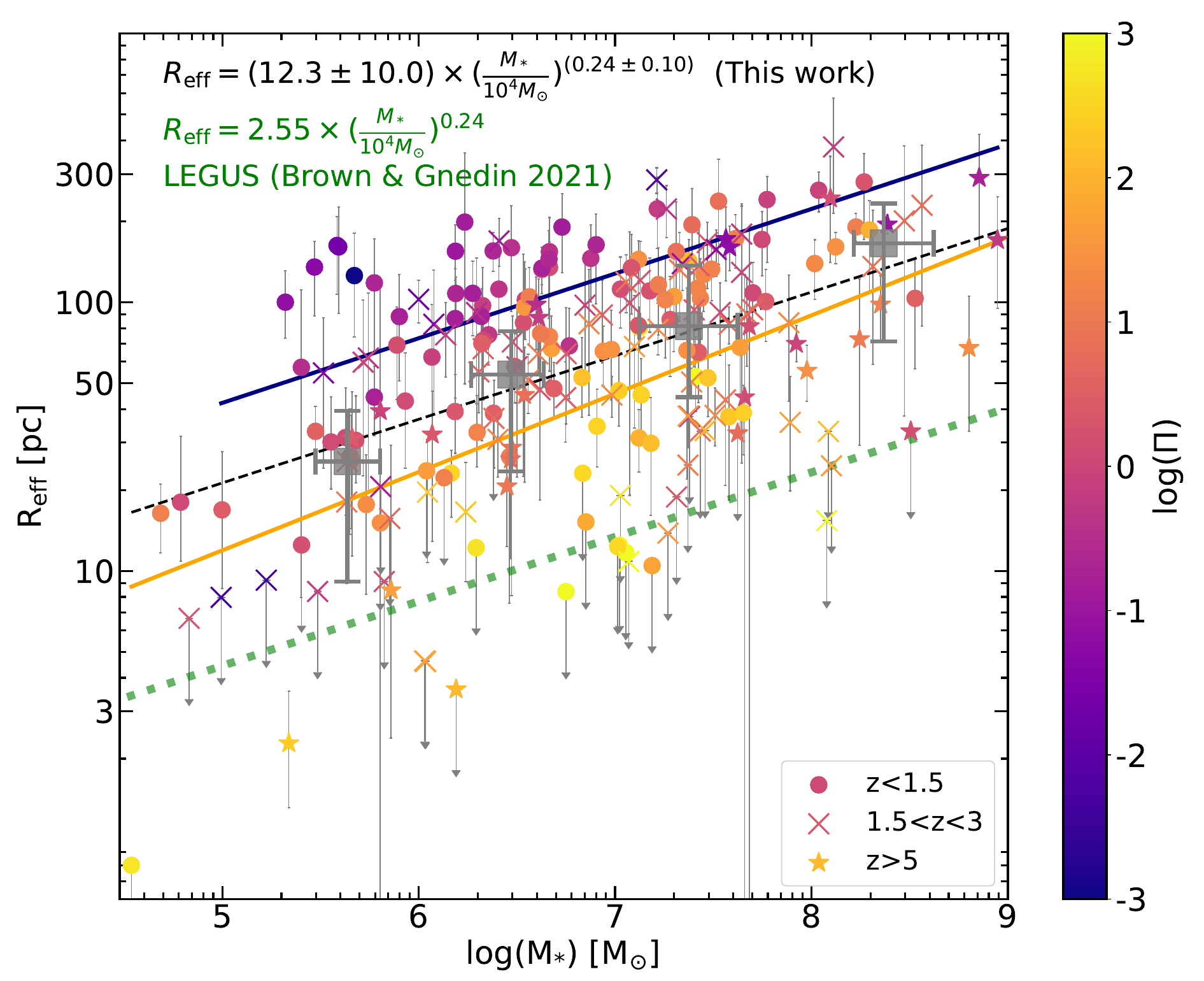}
    \caption{ {\bf Left}: Clump ages as a function of their effective radius.  Points are colour-coded accordingly to the dynamical age of the clump, i.e.,  the ratio of clump ages with respect to crossing time ($\Pi$), using the physical parameters derived with the reference SED model (with a continuous star formation for 10 Myr). The different markers represent the redshift bins. \textbf{Right: } Sizes of the clumps as a function of their stellar masses. The dashed black line represents the fitted relation between size and mass. The relation with the recovered parameter is indicated in black on the top of the plot. The green line represents the relation measured for star clusters by \citet{Brown2021} using the LEGUS sample.    The grey squares indicate the median, 1rst and 3rd quartiles of the clumps in the four mass bins: $M_* < 10^6 \ \rm M_{\odot}$, $10^6 < M_* < 10^7 \ \rm M_{\odot}$, $10^7 < M_* < 10^8 \ \rm M_{\odot}$, and $M_* >10^8 \ \rm M_{\odot}$. The orange and blue lines show the relation fitted only considering clumps with $\rm log(\Pi)>0$ and $\rm log(\Pi)<0$, respectively. }
    \label{fig:Reff_VS_mass}
\end{figure*}

We see as expected that longer crossing time corresponds to larger effective radii.
The ratio between the age of the clump and the crossing time defines its dynamical age, $\Pi$, first introduced by \citet{GPZ2011} for star clusters.   To a first order, stellar structures with ages longer than their crossing time are by definition gravitationally bound and can survive longer times as single entities. On the other hand, systems with $\Pi\leq1$ are considered transient stellar structures in the act of expanding. In the right panel of Fig.~\ref{fig:tcross}, the distributions of $\Pi = \rm age/ \tau_{\rm cross}$ are plotted per redshift bins. Overall, between 45 and 60 \% of the clumps (depending on the assumptions on the SFH) in each redshift bin are consistent with being bound structures, i.e., dynamically evolved and can potentially survive for a significant timescale. By plotting the age vs. the R$_{\rm eff}$ colour-coded by $\Pi$ in the left panel of Fig.~\ref{fig:Reff_VS_mass}, we see that very young clumps, in spite of their size, appear to be only marginally bound. This result is mainly driven by the fact that the definition of $\Pi$ does not work well in systems that are not yet dynamically evolved. As clumps age, first the most compact and slowly the entire sample across the recovered sizes appears dynamically evolved. In short, clumps/clusters with ages older than 100 Myr are consistent with being gravitationally bound. We will discuss below the implications for clump survival times. 

It is well-known that giant molecular clouds (GMCs) in the local and in high--redshift galaxies follow the relation $R \propto M^{0.5}$ as result of the almost constant gas surface density \citep[see][]{larson1981, bolatto2008}, although with different normalisation \citep[surface densities in high-redshift GMCs is significantly higher than in the local universe][]{DZ2019}. On the other hand, young star clusters, formed within these GMCs, show observationally shallower mass--radius relations, with reported slopes ranging around $0.25$ \citep[see][and references therein]{krumholz2019, Brown2021}. Analysing star-forming complexes in the local universe, \citet{bastian2005} found that these regions follow the GMC mass--size relation, i.e. their sizes trace the parent GMC from which stars have formed in the regions, while the stellar masses are scaled down by the star formation efficiency. They also concluded that young star clusters deviate from this relation because they form in the densest cores of GMCs and over a size range that is very narrow. Indeed, such weak dependence between star cluster mass and radius also hints toward the fact that stellar surface densities are significantly higher in more massive clusters. As already discussed above, lensing partially resolves clumps into smaller components, almost reaching sizes and masses of massive star clusters. It is therefore interesting to look whether clumps in our sample trace the sizes of their GMC progenitors. In the right panel of Fig.~\ref{fig:Reff_VS_mass}, we plot the sizes versus the masses of the clumps (different symbols encode different redshift bins). We fit the sample in log-space with the python package \texttt{lmfit} and report the recovered slope and normalisation in the figure, along with the recovered fit and relation for young star clusters in the local universe by \citet{Brown2021}.  The latter relation has been derived over cluster mass ranges of $10^3$ to $10^5$ \msun, and we extrapolate here over the mass range of the detected clumps. We also include median and quartiles by binning our clump distributions. Although a different normalisation, the slope we recover for the clumps is very similar with the Brown \& Gnedin's results.  The fact that the relation is shallower (i.e. with a coefficient $<$0.5) than the one found for GMC suggests that the detected clumps are not the largest coherent structures, but thanks to lensing, we are able to resolve down into the densest stellar structures within these galaxies. Within 1$\sigma$ uncertainty, the fit to the clump populations in the 3 redshift bins produce the same slope, suggesting that the size range in the 3 redshift bins is similar and that we detect similar physical scales.   The symbols in Fig.~\ref{fig:Reff_VS_mass} are colour-coded accordingly to their dynamical age. We notice that clumps with higher $\Pi$ are more compact and sit below the average median values of the distribution. By fitting the distributions of clumps with $\Pi<1$ and $\Pi\geq1$ separately, we get similar slopes ($0.23 \pm 0.15$ and $0.27 \pm 0.12$, respectively) but significantly different normalisations ($38 \pm 18$ for $\Pi<1$ and   $8 \pm 8$ for $\Pi\geq1$). 
In their recent simulations (using a sub-pc scale and including for the first time in a cosmological context and in a grid code, 
the feedback of individual stars), \citet{Calura2022} studied the size-mass relation of high redshift ($z\sim6$) clumps and found a consistent slope ($0.38$) overlapping our result within 1.5 $\sigma$. The shallower size vs. mass relation {(with respect to the GMC one)} in the simulations would point toward a significant role of stellar feedback in the resulting clump size vs. mass relation. When comparing with the sizes and masses of the simulated clump sample, we notice that the sizes overlap quite well with our observed sample, while the masses are lower than the observed ones, resulting in lower normalisation of the size vs. mass relation  than observed for clumps in SMACS0723.

\begin{figure}
    \includegraphics[width=\columnwidth]{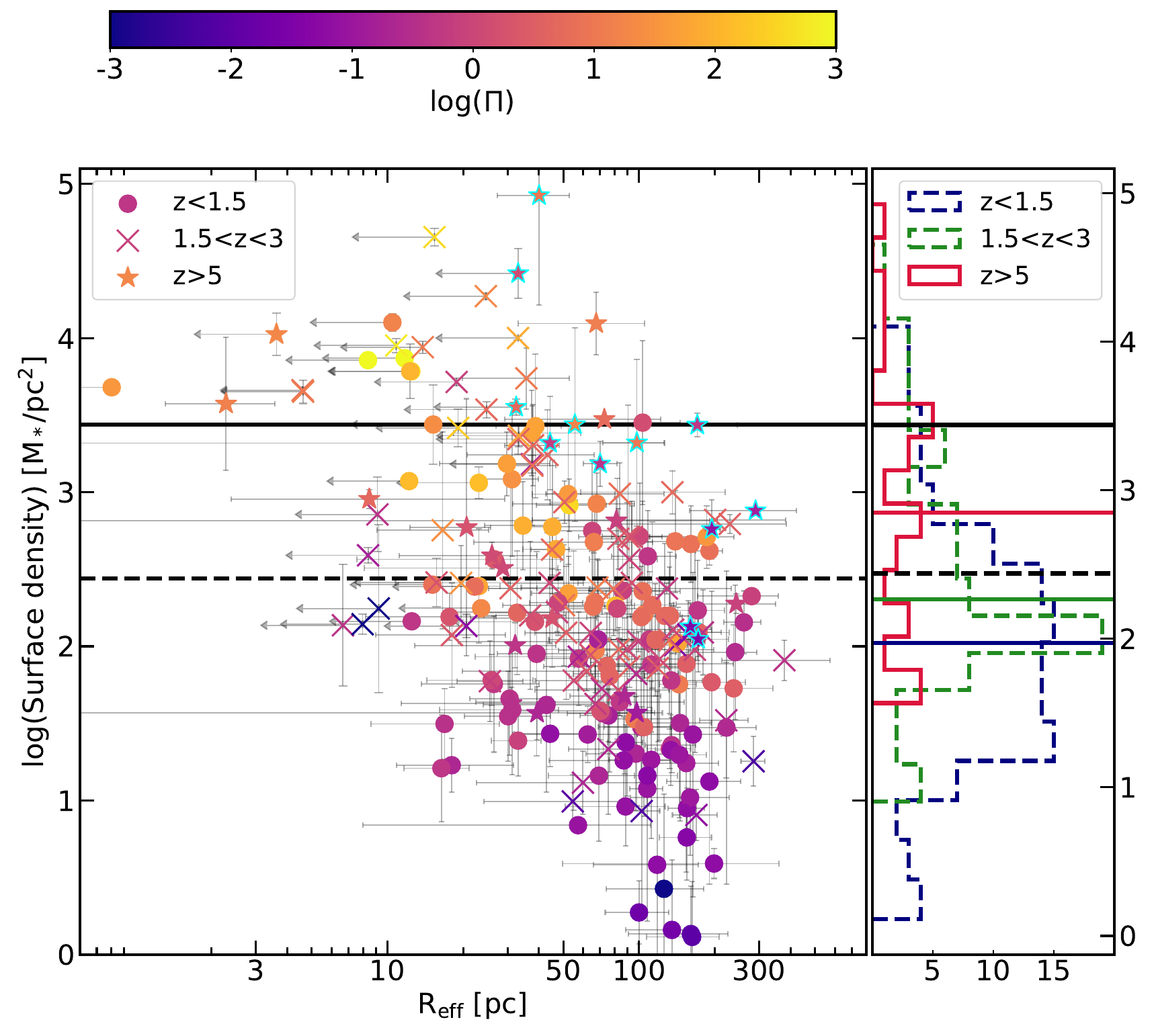}
    \caption{ Clump surface density as a function of their size. The clumps are colour-coded accordingly to their dynamical age ($\Pi$) and the different markers represent the three redshift bins. Arrows represent upper limits on the size. The cyan stars represent the clumps detected in galaxies with z$>6$. The dashed black horizontal line represents the average surface density of a nearby massive cluster, from \citet{Brown2021}. The solid black line represents the typical surface density of a globular cluster with $\rm R_{\rm eff}=3 \ pc $ and $\rm M_{*}=10^{5.2} \ \rm M_{\odot}$.}
    \label{fig:density_VS_Reff}
\end{figure}

 In Fig.~\ref{fig:density_VS_Reff}, we plot the stellar surface density distributions against the sizes of the clumps and include as reference the stellar surface density of a typical $10^{5.2} \ \rm M_{\odot}$ globular cluster with $R_{\rm eff}=3$ pc \citep{brodie2006} and the average surface density of nearby bound star clusters detected in local galaxies \citep{Brown2021}.

We notice that clumps that are potentially bound and can survive for a longer time within their host galaxies are as dense or denser than gravitationally bound star clusters in the local universe. As already noticed in the previous sections, clumps in the higher redshift bin ($z>5$) are among the densest structures in our sample, while clumps forming at redshift lower than 1.5 are the least dense. The fact that we observe on average higher densities at high redshift could be due to completeness effect (i.e. it is more difficult to detect very diffuse clumps at high redshift), however, the lack of very dense clumps in the lower redshift galaxies shows that higher redshift clumps are denser on average. This result confirms already the hypothesis made by \citet{Livermore2015} that clumps forming in higher-redshift universe show higher stellar densities, thus requiring more extreme gas conditions and higher gas fractions in their host galaxies. The highest surface densities values almost reach the maximum  surface density reported for stellar systems ($\ 10^5$ \msun\,pc$^{-2}$) in the local universe and considered as the largest value that can be supported for a stellar population that forms over a short timescale \citep{hopkins2010}.

\section{Tracing recent star formation within galaxies}
\label{sec:single_gal}

\begin{figure*}
    \includegraphics[width=18cm]{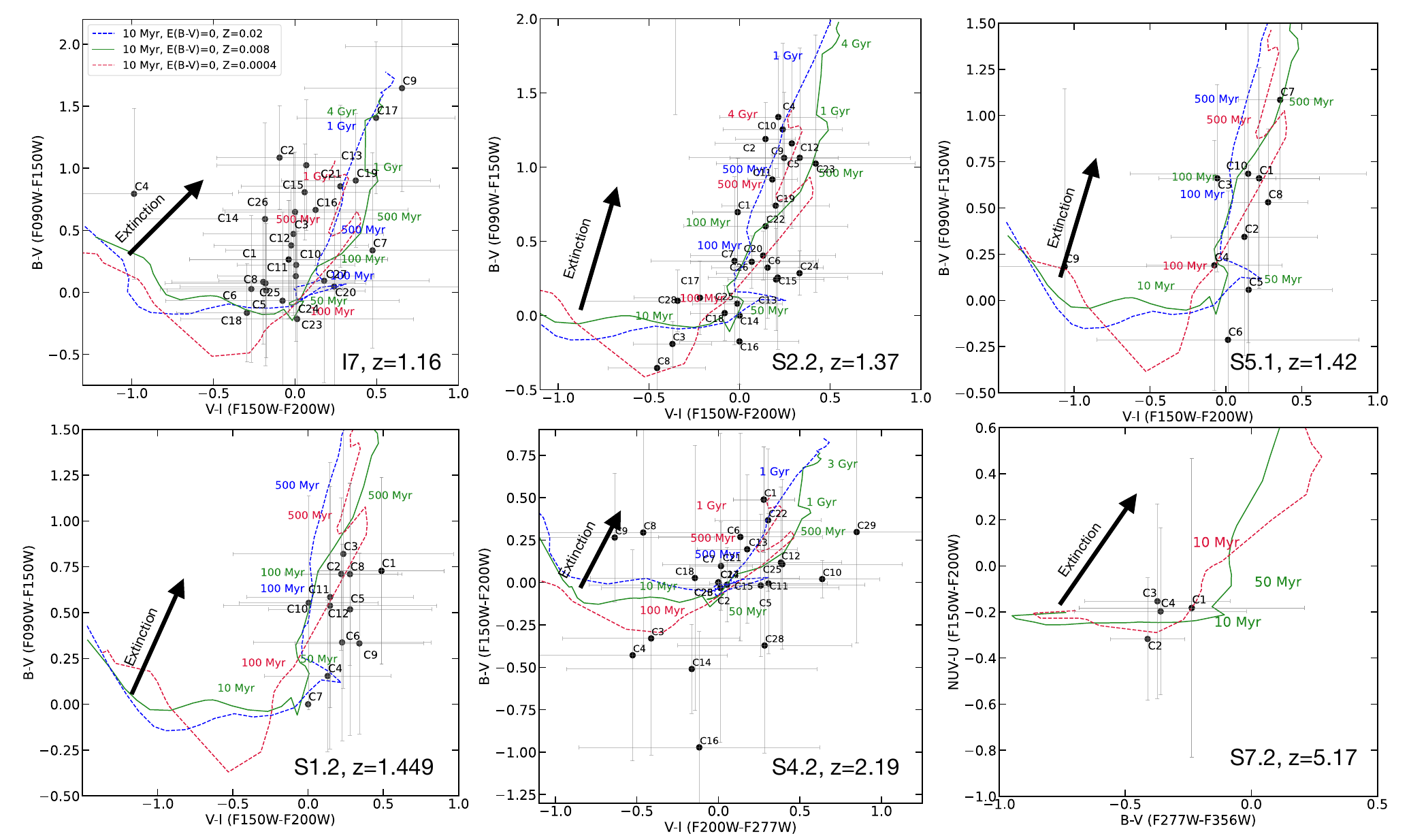}
    \caption{ Optical colour-colour diagrams of six galaxies, I7 (The Beret), S2 (The Sparkler), S5, S1, S4 (The Firework), and S7, ordered as a function of increasing redshift. We include only clumps (represented as black dots) detected in one multiple image (the one where the highest number of clumps is detected). For the highest redshift target we plot the NUV-U versus B-V colours. We show the tracks of the 10 Myr models with $Z=0.02$, $Z=0.008$ and $Z=0.0004$ in blue, green and red lines, respectively with $\rm E(B-V)=0$. The black arrow shows the evolution of the tracks when extinction increases from $\rm E(B-V)=0$ to $\rm E(B-V)=0.3$}.
       \label{fig:CCDs}
\end{figure*}

\begin{figure*}
    \includegraphics[width=19cm]{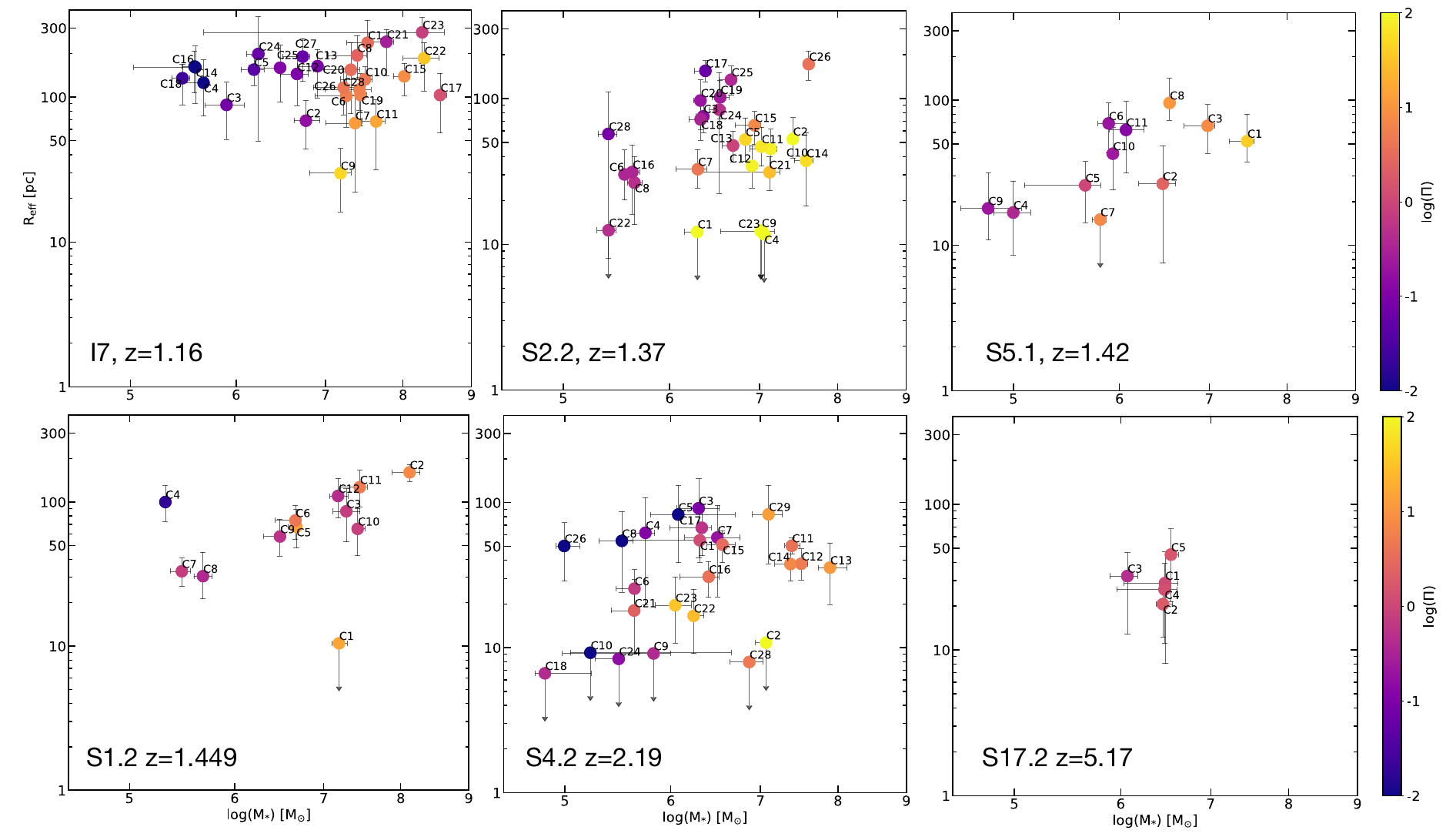}
    \caption{Sizes versus mass plots of the clumps in the six individual galaxies (displayed in the same order as in Fig.~\ref{fig:CCDs}).  We show for each galaxy the clumps detected in the image with the highest number of clumps,  which corresponds in all the cases to the highest magnified image of the galaxy. Size upper-limits are included with arrows. The points are colour-coded according to the estimated $\Pi$ value.}
       \label{fig:bound_ysc}
\end{figure*}

\begin{figure*}
    \includegraphics[width=17cm]{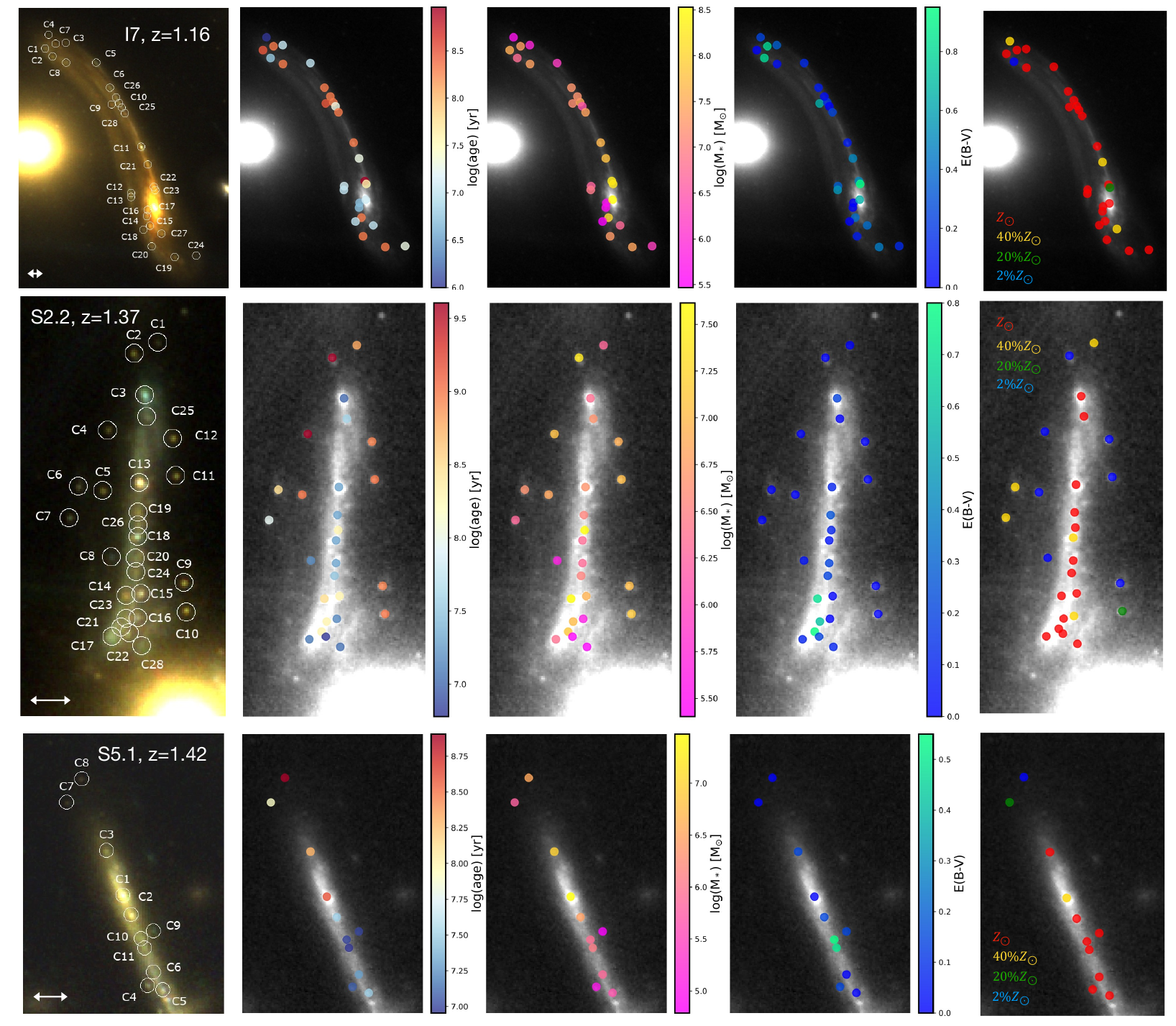}
    \caption{ From left to right: colour image (produced with F090W, F150W and F200W NIRCam filters) and F150W frames of the galaxies I7, S2.2 and S5.1, with the position of detected clumps, colour–coded by their best-fitted age (left), mass (middle left), extinction (middle right) and metallicity (right) measured from the reference SED model (with a continuous star formation for 10 Myr). The white line in each panel is 0.4'' long, which corresponds on average (assuming the mean magnification factor reported in Table~\ref{tab:sample}) to 0.33 kpc for I7, 0.46 kpc for S2.2 and 0.48 kpc for S5.1 in the source plane.}
    \label{fig:fig_pos_clumps}
\end{figure*}

\begin{figure*}
    \includegraphics[width=17cm]{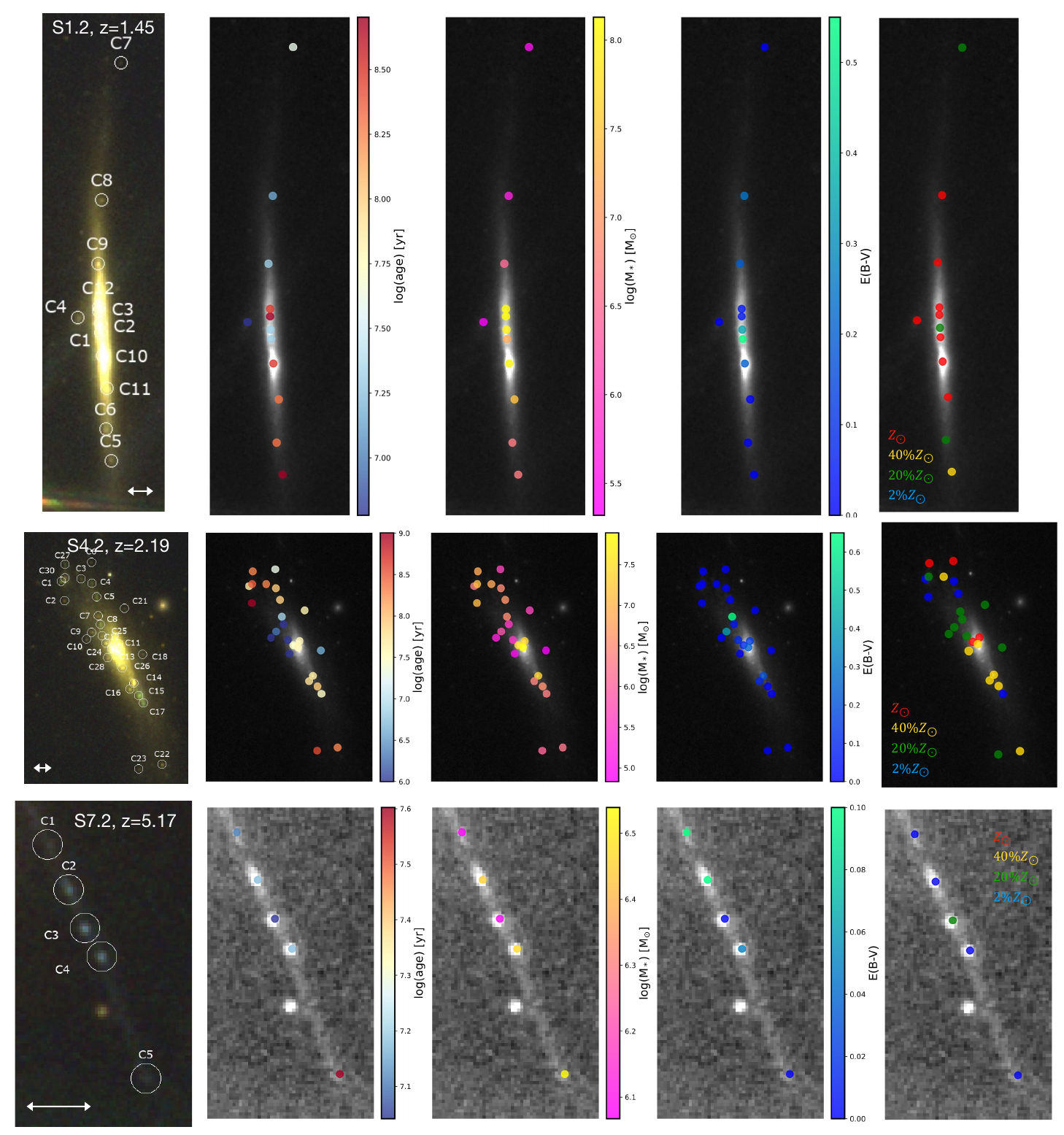}
    \caption{From left to right: colour image (produced with F090W, F150W and F200W NIRCam filters) and  F150W frames of the galaxies S1.2, S4.1 and S17.2, with the position of detected clumps colour–coded by their age (left), mass (middle left), extinction (middle right) and metallicity (right) measured from the reference SED model (with a continuous star formation for 10 Myr). The white line in each panel is 0.4'' long, which corresponds on average (assuming the mean magnification factor reported in Table~\ref{tab:sample}) to 0.48 kpc for S1.2, 0.39 kpc for S4.1 and 0.05 kpc for S17.2 in the source plane.}
    \label{fig:fig_pos_clumps2}
\end{figure*}

In this section we focus on the clump physical properties of some of the most interesting targets in the sample with redshift smaller than 6. From Fig.~\ref{fig:CCDs} to \ref{fig:fig_pos_clumps2} we show the selected targets, ordered by increasing redshift. Measurements and clump physical properties are also listed in Tables~\ref{tab:table_clumps1} to \ref{tab:table_clumps4}.

In Fig.~\ref{fig:CCDs}, we show the colour-colour diagrams of the clumps analysed within each galaxy. In these plots, we include Yggdrasil evolutionary tracks and label the axes with the pseudo rest-frame bands and in brackets the corresponding JWST filters used. We add different sets of evolutionary tracks, to show the reader the main parameter space covered by the models used in this work. In Fig.~\ref{fig:bound_ysc}, we show the mass versus size plot for each galaxy, where the symbols are colour-coded depending on the determined value of $\Pi$. We notice here that in each galaxy we always reach sizes (considering the upper limits) and masses that overlap with massive star clusters. Several of these star cluster candidates appear also gravitationally bound, allowing us to conclude that young star cluster candidates are detected in these systems. In Fig.~\ref{fig:fig_pos_clumps}, we show the positions of the clumps on the galaxy colour-coded accordingly to the age (middle left), mass (centre), extinction (middle right) and metallicity (right).  

\subsection{The Beret galaxy}

Galaxy I7, at redshift $z=1.16$, is a single image system, named the Beret, because of its morphological resemblance to a cap (see green inset in Fig.~\ref{fig:im_cluster}). The lens model presents for this galaxy a grazing critical curve providing high magnifications. 
However, a careful look at the lensing symmetry does not show obvious sign of multiple images. The nearby cluster member galaxy acts as a secondary lens, offering an additional boost of amplification (reaching $\mu=10$ for the most magnified clump).
It hosts a populous cluster/clump population of 28 detected systems with sizes from $\sim100$ down to upper limits of  10 parsec, i.e. between star-forming region and star cluster ranges (see top panel of Fig.~\ref{fig:bound_ysc}). The pseudo $B-V$ versus $V-I$ colour-colour diagram (top panel of Fig.~\ref{fig:CCDs}) show clumps distributed along the evolutionary tracks. Differential extinction can account for the colours distributions, with ages covering a large range from a few to several hundreds of Myr. Masses in the system range between $10^5$ and $10^8$ \msun, with the brightest clump (probably a proto-bulge structure) being the most massive, a few times $10^8$ \msun (see top panels of Fig.~\ref{fig:fig_pos_clumps}). The proto-bulge has a very young age (around 10 Myr) and it has large extinction. This result does not change if we use a longer SFH of 100 Myr (with a three times higher reduced $\chi^2$), reinforcing the finding that this region is still young. Solar metallicity models produce the best-solutions for the majority of the clumps, in agreement with this target being a spiral system.

\subsection{The Sparkler galaxy}

The Sparkler galaxy has attracted large attention because of the numerous visible star clusters surrounding the main body of the galaxy (see Fig.~\ref{fig:fig_pos_clumps}). Using a different analysis approach, consisting of fixed aperture photometry with radii of 0.2'' and SED analysis performed with a code that reconstruct variable and cumulative SFHs, \citet{mowla2022} report for 5 of the star clusters detected in the Sparkler, ages of $\sim4$ Gyr and metallicities between 20 and 70 \% solar. They also report a total mass of $10^9$ \msun\, for the host galaxy and ages of 30 to 300 Myr for the remaining of the star clusters/clumps analysed in their work. 

In our analysis the Sparkler corresponds to system S2.2. Our lens model presents high magnification (around $\mu=9$) and shear for the highest magnified image.  The 4 Gyr old clusters labeled 1, 2, 4, 8, 10 by \citet{mowla2022}, correspond in our sample to C4, C12, C5, C9, C2, respectively. We will refer to this group, hereafter, as globular clusters. In total we analysed 28 clumps/clusters in this system. The recovered $R_{\rm eff}$ range from upper-limits of a few parsec to 100 pc. In particular the globular cluster sample has sizes between 10 and 50 pc, and $\Pi>>1$ values all consistent with being bound star clusters. In Fig.~\ref{fig:fig_pos_clumps}, we show the recovered ages, masses, extinction, and metallicities derived with our reference model assumptions (continuous SFH over 10 Myr and $\rm E(B-V)<1$, except for the clumps in the outskirts where we assume $\rm E(B-V)=0$). We find significant differences in the age ranges of  the clusters surrounding the galaxy and those within the main body. The clusters/clumps within the galaxy have ages between hundreds down to a few Myr, suggesting that star formation has continued in the galaxy at a constant rate. The clusters in the outskirts have ages between 0.1 and 4 Gyr. In general, masses range between $10^5$ and $10^7$ \msun, and extinction are moderately low in clumps within the galaxy (as typically observed for star clusters populations in local galaxies).

For the cluster candidates sitting in the outskirts of the galaxy, our derived ages confirm those reported by \citet{mowla2022}, although some differences are noticeable.  In the top-central panel of Fig.~\ref{fig:CCDs}, we show the pseudo $B-V$ versus $V-I$ colour diagram for the detected clumps. We include the Yggdrasil evolutionary tracks with $\rm E(B-V)=0$ and $Z=0.02$ (blue), $Z=0.004$ (green) and $Z=0.0004$ (red). The latter evolutionary track corresponds to the best fitted extinction value for the old cluster sample. We observe that the colours of the globular clusters in the outskirts of the galaxy are gathered in a very narrow space of the colour-colour diagram, e.g. 0.5 - 4 Gyr are the model steps (red line) closest to that space.
Two clusters in the outskirts, C6 and C7, appear bluer and have colours coinciding with younger ages (80-100 Myr). Overall the recovered age range would indicate that C2, C4 in the outskirts have formed around redshift 9, i.e. around 12.8 Gyr ago. Another fraction of these proto-globular clusters (C5, C10, C11, C12) have ages between 0.9 and 1 Gyr, suggesting they formed at redshift $\sim$1.8 or 9.8 Gyr ago. C1 appears to be the youngest of the globular clusters with an age of 600 Myr (redshift formation 1.6 or 9.5 Gyr ago). The best ages are produced by models with metallicities varying between 2\% for the oldest up to 40\% solar abundances for the younger ones, which would correspond to abundances of [Fe/H] between  $-2$ and $-0.4$ over an age range of 12.8 and 9.5 Gyr and would overlap with the sequence of metal-poor and metal-rich population of globular clusters in the Milky Way \citep{forbes2010, vandenberg2013} and Large Magellanic Cloud \citep{narloch2022}. 

For the interested reader, we present a more detailed and one-to-one comparison for the outskirt clumps with the results presented in \citet{mowla2022} in  Appendix~\ref{sec:comp_sparkler}.

The position of the globular clusters in the outskirts of the Sparkler would suggest either an accretion or merger event. In both cases, dynamical interaction would have brought the globular clusters in their current position. However, in the first scenario (accretion) some of these globular clusters have formed in a different host and later accreted. While in the second scenario (merger) the globular clusters would have formed during the merger event and then be ejected in the outskirts.

\subsection{Arcs S5 and S1}

Arc S5 at redshift 1.42 shows a bright central clump, a proto-bulge candidate with age of 400  Myr and mass of a few times $10^7$ \msun. Younger star-forming clumps are visible in the galaxy which show a small extinction range. For zero extinction the age of the proto-bulge could be as old as 500 Myr. 

On the other hand, clumps in the arc S1 at redshift 1.5 show colours that occupy a very small region in the parameter space, suggesting that a specific star formation episode has led to the formation of these stellar structures. The age range spans 30 to 100s Myr. The brightest central clump is among the youngest structures in the galaxy, opposite to what is observed in S5.1 and the Beret. 

Magnification values in the image S5.1 varies across the arc from 10 to 80 for the clumps closest to the critical line where the shear is stronger. Magnification values in S1.2 varies across the arc from 12 to 24. In both galaxies, several clumps, especially the most compacts, are potentially gravitationally bound ($\Pi\geq1$) systems and could indeed be proto-globular clusters.

\subsection{The Firework, system S4}
At redshift 2.19, system S4, dubbed the firework galaxy, has 27 detected clusters, surrounding and/or located within the galaxy with magnification values varying along the arc between 12 and 65 for the clumps closest to the critical line. Unfortunately, data for this target are nosier than the cluster detected in the Sparkler. However, we notice that several clusters occupy a small region of the colour-colour diagram suggestive of a synchronised star formation event between 0.5 and 1 Gyr ago followed by significant star formation until the present time. Extinction in the galaxy is significantly low. We do not see specific trends in age and position as observed with the Sparkler, although the majority of the clumps located outside the galaxy coincide with older ages. A significant amount of clumps/clusters in the Northern-East side of the target are very young. Again we notice that the most massive region in the target is coincident with the structure that can be considered a proto-bulge.

\subsection{Arc S7}
Located at redshift $z=5.17$ this galaxy appears quite strikingly as a red arc, located above the Beret, in the JWST ERO images. This arc straddles the critical curve, presenting a large magnifications gradient from about 11 to 1000. It hosts 4 well detected clumps with ages between 10 and 40 Myr, and a fifth one with poorer constraints on physical properties but colours compatible with ages of $\sim$10 Myr. Three out of the five clumps have size upper-limits of 30 pc and $\Pi > 1$, suggesting that we are looking at bound star clusters at redshift 5, as similarly found in another arc observed with JWST by \citet{vanzella2022}.

\section{Star cluster detection in galaxies at redshift above 6?}
\label{sec:starclusters}

Recent cosmological simulations that include formation and evolution of star clusters \citep[e.g.,][]{pfeffer2018} find that the bulk of the star clusters more massive than $10^5$ \msun\, that have survived until redshift 0 as globular clusters, form mainly between redshift 6 and 1 \citep{reinacampos2019}. For a long time, globular clusters have been listed among the potential players which contributed to reionisation \citep[e.g.,][]{ricotti2002, trenti2015, sameie2022}. However, their volumetric number density is, so far, model and assumption dependent \citep[e.g][]{BK2018}. As a consequence, predictions of detection of potential proto-globular clusters at redshift beyond 6 can dramatically change \citep[e.g.,][]{renzini2017, kruijssen2019, sameie2022}. To date the highest redshift at which a potential star cluster has been detected is around redshift 6 \citep{Vanzella2019}. JWST observations will be a game changer for this field. 

In this work we have included in total 4 galaxies with spectroscopic confirmed redshift higher than 6, e.g., fully overlapping with the reionisation era. 

\begin{figure*}
    \includegraphics[width=18cm]{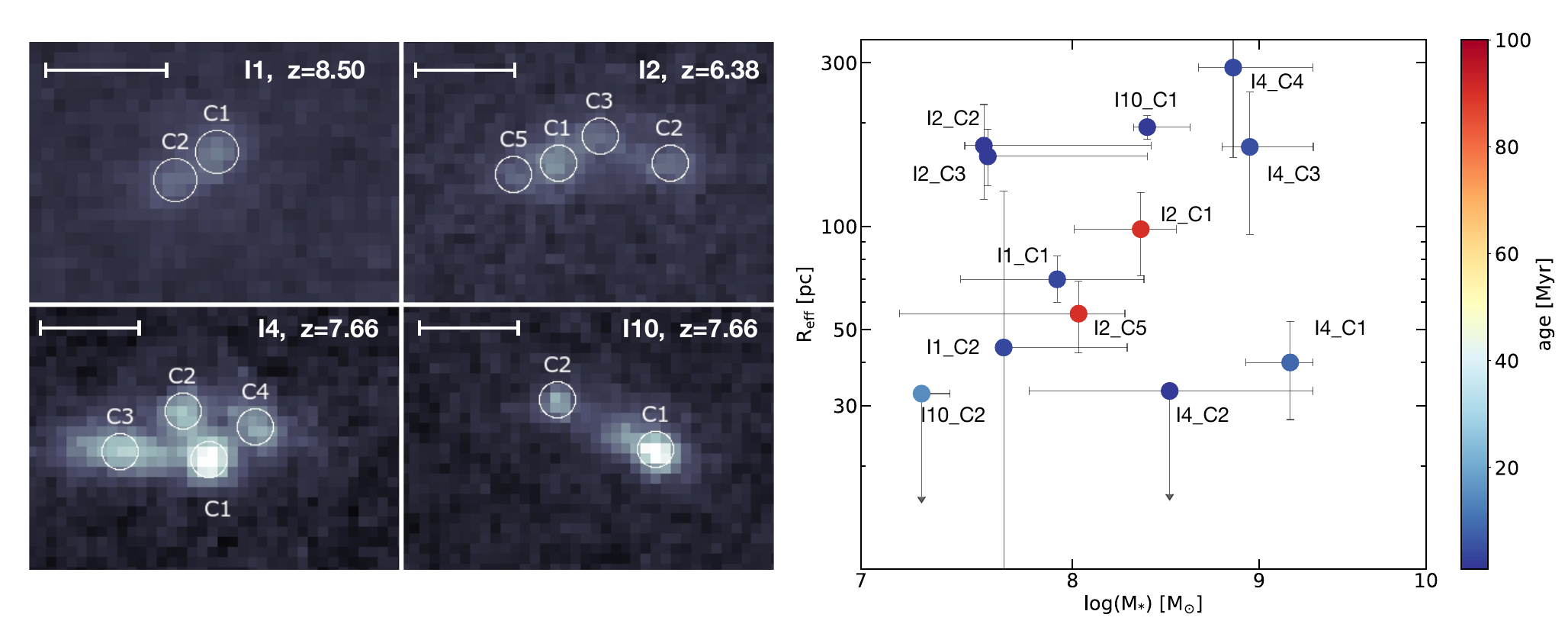}
    \caption{{\bf Left:} F200W images of the 4 highest redshift galaxies (I1, I2, I4 and I10). The pink circles show the clumps detected in each image and the line represent 2 arcseconds. {\bf Right:} Size versus mass diagram of the clumps detected in these 4 galaxies. The points are colour-coded with age and we show the ID of each clump in black.}
    \label{fig:fig_high_z}
\end{figure*}

Three of these galaxies, I1, I4 and I10,  have been recently studied by \citet{Schaerer2022}, who reported $\log(M_*)\times\mu$ of 9, 9.2, 8.9, respectively, the lensing model used in this work, predicts slightly lower magnification for system I1. We apply our mean magnification to estimate the de-lensed masses to be $2.3\times10^8$ \msun, $1.3\times10^9$ \msun, and $5.8\times10^8$ \msun\ for systems I1, I4, I10. These systems are all consistent with being metal poor, dwarf galaxies \citep{Schaerer2022}. Fig.~\ref{fig:fig_high_z} shows the clumps detected within these systems (left), and sizes and masses on the right. In some cases, we reach upper-limits for $R_{\rm eff}$ of 30 pc or better and masses that are all above $10^7$ \msun. We notice that even with the current size upper limits, surface stellar densities in these clumps are similar or larger than in massive star clusters (see star symbols with cyan contours in Fig.~\ref{fig:density_VS_Reff}), suggesting that even if we are not able to resolve the star clusters within these clumps, the latter most likely host them. The three galaxies investigated by \citet{Schaerer2022} host all very young clumps. The two clumps in I1 have ages of 4 Myr. The clumps in I4 have ages from 2 to 9 Myr, while the two clumps in I10 have ages of 2 and 15 Myr. Feedback from these young, massive and compact clumps is most likely the source of ionisation in their host galaxies,  responsible for very high O\three\, equivalent width and high O\three/O\two\, ratio. 

Very interestingly, the fourth galaxy I2, at redshift 6.38, hosts four evolved clumps with ages between 2 and 90 Myr. The two old clumps have  $\Pi > 1$, suggesting that they are long-lived structures in the system and formed about redshift 6.9.  Since this system is detected in O\three\ (see Table~\ref{tab:sample}), we expect the two blue clumps to be fairly young, hinting that star formation can go on in these dwarf galaxies for 10s of Myr.

\section{Discussion and Conclusions}
\label{sec:conclusion}

In this study, we have analysed the optical rest-frame light of stellar clumps in 18 galaxies (for a total of 36 images) using the recently acquired JWST ERO of the galaxy cluster SMACS0723.

The majority of the clumps studied in this work has gone so far undetected at HST wavelengths. First because the RELICS data are not sufficiently deep, and secondly, the signal of point-like structures has been  smeared out at the lower resolution of the WFC3 IR camera.

The detection of clumps has been done in different steps. The initial extraction has been performed via visual inspection on 3-colour images of the targets to overcome contamination from interlopers. Fainter clumps, missed after this first iteration have been included by visually inspecting the galaxy residual images in the reference frames, F150W and F200W. The initial photometric catalogue contains de-lensed photometry (estimates and upper limits) and intrinsic effective radii (or upper limits) for 223 clumps detected in the F150W and F200W filters. Overall, we determine $R_{\rm eff}$ ranging from upper limits of a few parsec up to resolved clumps of a few hundreds parsec. The range of absolute $V$ band magnitudes goes from $-10$ to $-20$ ABmag. A comparison with (rare) massive young star clusters in the local universe, confirm that we are resolving \emph{bona-fide} star clusters in galaxies across a large fraction of the cosmic formation history. As already noticed in recent studies of clump populations in deep HST survey of the Frontier Field lensing fields \citep[][and Claeyssens et al., in prep.]{mestric2022}, stellar clumps are not single entities within their host galaxies, but they progressively resolve into compact systems. We, therefore, refer to stellar clumps not as star formation units but as clustered stellar regions, for which we investigate their nature. When comparing size and FUV magnitude of clumps detected in the Frontier Field regions covered with HST and the clumps analysed in this work we derive significantly smaller sizes and lower magnitudes, as a result of the superior spatial sampling and sensitivity of JWST observations.  

We derive ages, masses, extinctions, metallicities, and $R_{\rm eff}$ for a population of 221 clumps in the redshift range 1 to 8.5. Even with available optical SEDs, we still notice unbreakable degeneracies such as SFH, age-extinction, and metallicities. We tested models with different SFH assumptions, i.e. from instantaneous burst (single stellar population, typical of star clusters in the local universe) to 10 Myr and 100 Myr continuous star formation followed by passive stellar evolution. These three models produce similarly good reduced $\chi^2$. However, the resulting age distribution changes significantly,  with  peak ages shifting towards older values (5-10 Myr for  IB, 10-20 Myr for 10Myr, 50-200 Myr for 100Myr). Recovered ranges of masses and extinction do not change significantly.

By learning from studies of resolved star-forming regions in the local universe, we adopt as reference model the one with continuous SFR over 10 Myr (to account that over regions of 10s of parsec star formation time scales are longer than instantaneous).

We analyse the sample of 221 clumps, by dividing it into three different redshift bins that bracket the cosmic dawn ($z>5$), cosmic noon ($1.5<z<3$) and cosmic afternoon ($z<1.5$). With the current statistics, we do not find a significant change in the age distributions of the three samples that would suggest survival timescales longer than several hundreds of Myr. If we account for the average age, there is roughly 1 Gyr time range between the two highest redshift bins and $\sim3$ Gyr between the highest redshift sample and the lowest one. If clumps forming at redshift 5 and above survive until redshift 1 we should recover clumps with ages larger than 1 Gyr.
We notice however, that thanks to JWST we are able to detect globular cluster candidates around high-redshift galaxies (see the Sparkler galaxy as an example). The detection of proto-globular clusters at high-redshift might point toward the fact that while clumps dissolve, the globular clusters that formed within them might survive for much longer times.   

To build a more complete picture of the nature of clumps and their formation and evolution, we also derive the mass--radius relation, clump surface densities and dynamical timescales. We find that clumps follow a shallower mass--size relation (recovered slope is $0.24\pm0.10$) than commonly found for GMCs in the local and high-redshift universe (around $\sim$0.5). However, the recovered slope is very similar to the relation derived for gravitationally bound star clusters in local galaxies ($0.24$, but with different normalisation). The deviation from the GMC mass--size relation could be understood if, thanks to lensing magnification, we can better resolve the denser structures of clumps, resulting in smaller sizes than their parent GMCs. Indeed, we find that clump surface densities are quite high, especially so for the higher redshift bin, as predicted by \citet{Livermore2015}, but here extrapolated to redshift higher than 5. The evolution of clump densities is also important if we think that these are the regions where globular cluster progenitors have formed. If the average densities of clumps evolve with redshift, then the average conditions for cluster formation have also evolved. These changes reflect in the average physical properties of the populations of star clusters which in local galaxies rarely reach masses above $10^5$ \msun, except in specific events such as accretion of gas, interactions and mergers \citep[e.g.,][]{adamo2020a}. 

We estimate the boundness of the stellar clumps by comparing their crossing time to the their ages. We find that for the reference model about 53 \% of clumps appear to be gravitationally bound. This fraction changes between 45 and 60 \% if different SFH are assumed (IB and 100Myr, respectively). 

Combining all these results together, we conclude that at a resolution from a few to 100 pc, clumps appear as dense stellar structures that share many similar properties to local massive gravitationally bound star clusters in the local universe. Their survival timescales are not longer than 1 Gyr, probably due to the fact that they form in thick and gravitationally unstable disks with elevated gas fractions, meaning that they are dissolved by large shear effects, on time scales of hundreds of Myr.

By focusing on single galaxies, we witness a rich variation of star formation events taking place in these systems and detect several clumps, that are indeed star clusters in the redshift range 1 to 5. Two clump/cluster rich systems, the Sparkler and the Firework, have clumps with colours and derived physical properties that would suggest the presence of proto-globular clusters. Their location in the outskirts of the host galaxies would suggest that they have been dynamically ejected, either via accretion or merger events. Both systems have continued forming stars until now, as showed by the distribution of colours and ages of the younger clumps within their host galaxies. The other galaxies, at redshift smaller than 6, show star formation events that have formed clusters/clumps during the last $\sim$100 Myr until now. In all these systems, we always detect a fraction of clumps that are consistent with being gravitationally bound, i.e. with star cluster candidates.

Four of our galaxies have redshifts higher than 6, making them potentially interesting to study as reionisation-era galaxies. Spectroscopic analyses of three of these systems have revealed low mass, low metallicity galaxies with ionised gas properties consistent with Lyman continuum leakers. We detect between 2 and 4 clumps in each of these systems. The majority of these clumps have sizes or upper-limits smaller than 50 pc and masses above $10^7$ \msun. Their stellar surface densities are comparable or higher that massive star clusters, suggesting that even if we do not resolve them, they must host star clusters. All the galaxies host very young clumps and, in some cases, also older clumps suggesting that star formation has been ongoing for several Myr.

To conclude, JWST observations of a poorly known galaxy cluster region have shown the incredible leap forward that we are taking in studying star formation within rapidly evolving galaxies. The unprecedented sensitivity and high resolution power at IR wavelengths is a key factor in the detection and analysis of compact stellar structures of these young galaxies, previously only detected at the FUV restframe. Our initial analyses reveal that star cluster candidates can be detected and studied in these systems and can be used to trace the recent star formation history of their host galaxies, as well as, under what conditions star formation is taking place. The study presented here can only be undertaken thanks to gravitational lensing. As such it is also related to our ability to reconstruct a reliable model of the underlying lens. Our results are, hence, related to the quality of the lens model and can be improved. 

Finally, many of the clumps in these galaxies are too faint to be studied with 3D spectroscopy. The latter will be fundamental to evaluate the effect of stellar feedback originating from the clumps into the interstellar medium of their host galaxies. Therefore, another important improvement to our analysis can be made by including deeper HST observations in the 0.4--0.8 $\mu$m wavelength range as well as a better imaging coverage between 0.9 and 4 $\mu$m, to better sample key features in the optical rest-frame of the clumps (Balmer break, strong emission lines) and therefore better break age-extinction-SFH degeneracies.

\section*{Acknowledgements}

We are indebted to G. Brammer for making publicly accessible the final reduced data of SMACS0723. We thank D. Coe, F. Calura, U. Me{\v{s}}tric, C. Usher, E. Vanzella, and the anonymous referee for insightful comments and discussions which have helped to improve the manuscript. 
A.A. and M.M. acknowledge support from the Swedish Research Council (Vetenskapsr\aa{}det project grants 2021-05559 and 2019-00502, respectively). G.M. acknowledges funding from the European Union’s Horizon 2020 research and innovation program under the Marie Skłodowska-Curie grant agreement No MARACHAS - DLV-896778.

Based on observations made with the NASA/ESA \textit{Hubble Space Telescope}, obtained at the Space Telescope Science  Institute, which is operated by the Association of Universities for Research in Astronomy, Inc., under NASA contract NAS 5-26555. These observations are associated with programs GO-11103, GO-12166, GO-12884, GO-14096.
This work is based on observations made with the NASA/ESA/CSA \textit{JWST}. The data were obtained from the Mikulski Archive for Space Telescopes at the Space Telescope Science Institute, which is operated by the Association of Universities for Research in Astronomy, Inc., under NASA contract NAS 5-03127 for \textit{JWST}. These observations are associated with program \#2736.

\section*{Data Availability}

Data are publicly accessible via the MAST archive. Higher-level quantities used in this manuscript are included in the Appendix.



\bibliographystyle{mnras}
\bibliography{biblio} 




\appendix
\section{Clumps selection}
\begin{figure}
    \includegraphics[width=9cm]{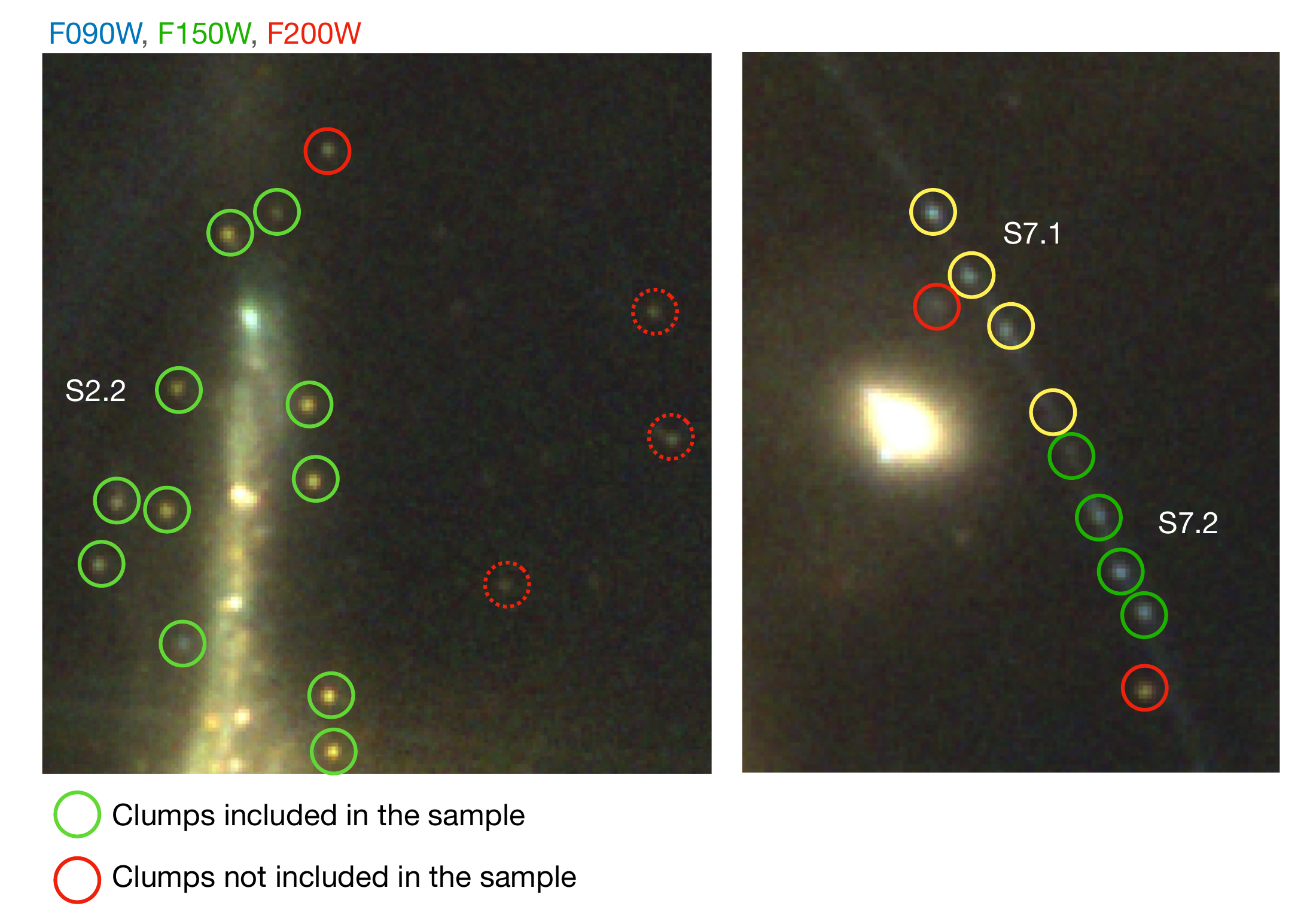}
    \caption{\textbf{Left}: NIRCam 3-colours image (with F090W, F150W and F200W filters) of the Sparkler galaxy. The green circle represent the outskirt clumps detected in the galaxy.the solid red circle shows a clump which was not selected as a galaxy clump based on his bluer colour very similar to the intra-cluster globular clusters (dashed red circles). \textbf{Right}: NIRCam colour image of the system S7. The green/yellow circles represent the clumps detected in these two multiple images. The red circles represents the clumps which were not included in the catalogue based on both their colour and the fact that they are not detected in the two multiple images.}
    \label{fig:selection_clumps}
\end{figure}

In this section we briefly illustrate the clump selection based on colour images. As explained in the Sect.~\ref{sec:cl_identification}, we used 3-colours images in order to discriminate which compact clump belongs to the selected galaxies or not. In Fig.~\ref{fig:selection_clumps}  we show this colour identification for two galaxies. On the left side, we show the outskirts clumps detected in the Sparkler (S2.2) in green and one clump rejected from the final sample in red. We rejected this clump because its colour is very similar to other compact globular clusters likely belonging to the galaxy clusters (highlighted with the dashed red circles). On the right panel, we show two images of the system S7 (S7.1 and S7.2) and the detected clumps in white. The two red circles present non-selected clumps based on their colour and the fact that they are not observed in both multiple images of this galaxy.

\section{Complete clumps catalogues}
In this section we include a complete list of the JWST/NIRCam photometry (Tables~\ref{tab:table_clumps_phot1} to~\ref{tab:table_clumps_phot4})  and physical properties (Tables~\ref{tab:table_clumps1} to ~\ref{tab:table_clumps4}) measured and derived for the 223 clumps detected in the two reference NIRCam bands, F150W and F200W. See Sect.~\ref{sec:cl_analysis} for a detailed description of the analysis.

\begin{table*} 
\centering 
\resizebox{\textwidth}{!}{
\begin{tabular}{llllllllllll} 
Image ID & Clump ID & z & RA & DEC & $\mu$  & F090W & F150W & F200W & F277W & F356W & F444W  \\ 
 &  &  &  &    & &  [ABmag] & [ABmag] & [ABmag] & [ABmag] & [ABmag] & [ABmag]  \\ 
(1) & (2) &  (3) & (4)  & (5)   &  (6)& (7) & (8) & (9) & (10) & (11) & (12) \\ 
\hline 
\hline 
S1.1 & C1.1 & 1.449 & 110.84069 & -73.451065 & 6.12$^{+0.57}_{-0.46}$ & 25.09$\pm$ 0.29 & 24.39 $\pm$ 0.25 & 24.38 $\pm$ 0.17  &23.85$\pm$ 0.2 &23.7$\pm$ 0.18& 23.49 $\pm$0.19\\  
S1.1 & C2.1 & 1.449 & 110.840775 & -73.451126 & 6.38$^{+0.69}_{-0.46}$ & 25.69$\pm$ 0.37 & 25.0 $\pm$ 0.36 & 24.6 $\pm$ 0.42  &23.96$\pm$ 0.24 &23.75$\pm$ 0.22& 23.6 $\pm$0.2\\  
S1.1 & C3.1 & 1.449 & 110.84103 & -73.451416 & 8.05$^{+0.93}_{-0.58}$ & 30.17$\pm$ 0.87 & 28.77 $\pm$ 0.4 & 28.35 $\pm$ 0.25  &27.89$\pm$ 0.16 &27.67$\pm$ 0.12& 27.64 $\pm$0.13\\  
S1.1 & C4.1 & 1.449 & 110.84191 & -73.452255 & 26.63$^{+6.9}_{-2.51}$ & 31.16$\pm$ 0.72 & 29.15 $\pm$ 0.38 & 29.04 $\pm$ 0.13  &28.78$\pm$ 0.21 &28.67$\pm$ 0.08& 28.46 $\pm$0.13\\  
S1.1 & C5.1 & 1.449 & 110.84136 & -73.45193 & 14.64$^{+2.5}_{-1.13}$ & 31.7$\pm$ 1.01 & 28.96 $\pm$ 0.31 & 28.81 $\pm$ 0.17  &28.81$\pm$ 0.31 &28.73$\pm$ 0.29& 28.63 $\pm$0.09\\
\end{tabular}} 
\caption{NIRCam photometry of the clumps. (1) and (2) images and clumps ID; (3) redshifts; (4) and (5) RA and Dec coordinates; (6) lensing magnification; (7)-(12) observed AB magnitudes}
\label{tab:table_clumps_phot1} 
\end{table*}

\begin{table*} 
\centering 
\resizebox{\textwidth}{!}{\begin{tabular}{llllllllllll} 
Image ID & Clump ID & z & RA & DEC & $\mu$  & F090W & F150W & F200W & F277W & F356W & F444W  \\ 
 &  &  &  &    & &  [ABmag] & [ABmag] & [ABmag] & [ABmag] & [ABmag] & [ABmag]  \\ 
(1) & (2) &  (3) & (4)  & (5)   &  (6)& (7) & (8) & (9) & (10) & (11) & (12) \\ 
\hline 
\hline 
S3.3 & C5.1 & 1.9914 & 110.82506 & -73.45966 & 10.77$^{+2.39}_{-1.06}$ & 27.03$\pm$ 0.37 & 26.08 $\pm$ 0.29 & 26.08 $\pm$ 0.17  &25.29$\pm$ 0.16 &25.04$\pm$ 0.14& 24.9 $\pm$0.12\\  
S3.2 & C1.1 & 1.9914 & 110.831985 & -73.4552 & 3.56$^{+0.59}_{-0.24}$ & 26.61$\pm$ 0.58 & 24.97 $\pm$ 0.51 & 25.51 $\pm$ 0.45  &26.55$\pm$ 0.3 &26.05$\pm$ 0.17& 26.83 $\pm$0.27\\  
S3.2 & C2.1 & 1.9914 & 110.83149 & -73.45522 & 3.4$^{+0.61}_{-0.36}$ & 27.92$\pm$ 0.17 & 26.16 $\pm$ 0.11 & 26.56 $\pm$ 0.11  &26.32$\pm$ 0.16 &26.06$\pm$ 0.14& 26.15 $\pm$0.18\\  
S3.2 & C3.1 & 1.9914 & 110.831764 & -73.45524 & 3.53$^{+0.61}_{-0.24}$ & 27.75$\pm$ 1.12 & 26.92 $\pm$ 0.59 & 27.5 $\pm$ 1.13  &26.04$\pm$ 0.2 &25.83$\pm$ 0.14& 25.8 $\pm$0.2\\  
S3.2 & C4.1 & 1.9914 & 110.8317 & -73.45524 & 3.5$^{+0.61}_{-0.24}$ & 25.11$\pm$ 0.51 & 24.35 $\pm$ 0.54 & 24.63 $\pm$ 0.58  &24.69$\pm$ 0.39 &24.65$\pm$ 0.37& 24.69 $\pm$0.34\\
\end{tabular}} 
\caption{Same as Table~\ref{tab:table_clumps_phot1}}
\label{tab:table_clumps_phot2} 
\end{table*}

\begin{table*} 
\centering 
\resizebox{\textwidth}{!}{\begin{tabular}{llllllllllll} 
Image ID & Clump ID & z & RA & DEC & $\mu$  & F090W & F150W & F200W & F277W & F356W & F444W  \\ 
 &  &  &  &    & &  [ABmag] & [ABmag] & [ABmag] & [ABmag] & [ABmag] & [ABmag]  \\ 
(1) & (2) &  (3) & (4)  & (5)   &  (6)& (7) & (8) & (9) & (10) & (11) & (12) \\ 
\hline 
\hline 
S5.1 & C3.1 & 1.42 & 110.824104 & -73.45171 & 16.12$^{+1.97}_{-1.86}$ & 28.18$\pm$ 0.4 & 27.37 $\pm$ 0.3 & 27.39 $\pm$ 0.24  &26.59$\pm$ 0.25 &26.51$\pm$ 0.26& 26.4 $\pm$0.31\\  
S5.1 & C4.1 & 1.42 & 110.823555 & -73.45222 & 68.6$^{+25.68}_{-10.65}$ & 28.21$\pm$ 0.54 & 27.87 $\pm$ 0.4 & 27.9 $\pm$ 0.19  &27.64$\pm$ 0.5 &27.5$\pm$ 0.64& 27.3 $\pm$0.69\\  
S5.1 & C5.1 & 1.42 & 110.82334 & -73.45224 & 92.51$^{+43.15}_{-16.06}$ & 27.22$\pm$ 0.53 & 27.01 $\pm$ 0.41 & 26.83 $\pm$ 0.37  &26.52$\pm$ 0.62 &26.27$\pm$ 0.69& 26.03 $\pm$0.72\\  
S5.1 & C6.1 & 1.42 & 110.82347 & -73.45217 & 56.21$^{+17.73}_{-8.2}$ & 26.31$\pm$ 0.57 & 26.37 $\pm$ 0.56 & 26.32 $\pm$ 0.65  &25.73$\pm$ 0.54 &25.64$\pm$ 0.53& 25.57 $\pm$0.53\\  
S5.1 & C7.1 & 1.42 & 110.82462 & -73.45153 & 12.3$^{+1.25}_{-1.48}$ & 30.34$\pm$ 0.47 & 29.11 $\pm$ 0.25 & 28.71 $\pm$ 0.1  &28.85$\pm$ 0.37 &29.42$\pm$ 0.1& 29.88 $\pm$0.29\\
\end{tabular}} 
\caption{Same as Table~\ref{tab:table_clumps_phot1}}
\label{tab:table_clumps_phot3} 
\end{table*}

\begin{table*} 
\centering 
\resizebox{\textwidth}{!}{\begin{tabular}{llllllllllll} 
Image ID & Clump ID & z & RA & DEC & $\mu$  & F090W & F150W & F200W & F277W & F356W & F444W  \\ 
 &  &  &  &    & &  [ABmag] & [ABmag] & [ABmag] & [ABmag] & [ABmag] & [ABmag]  \\ 
(1) & (2) &  (3) & (4)  & (5)   &  (6)& (7) & (8) & (9) & (10) & (11) & (12) \\ 
\hline 
\hline 
I7.1 & C6.1 & 1.16 & 110.79874 & -73.44817 & 7.68$^{+1.59}_{-1.25}$ & 27.02$\pm$ 0.46 & 26.84 $\pm$ 0.37 & 27.07 $\pm$ 0.42  &26.2$\pm$ 0.32 &26.03$\pm$ 0.33& 25.9 $\pm$0.32\\  
I7.1 & C7.1 & 1.16 & 110.800125 & -73.44785 & 4.55$^{+0.54}_{-0.33}$ & 29.42$\pm$ 0.54 & 28.93 $\pm$ 0.57 & 28.41 $\pm$ 0.69  &27.35$\pm$ 0.33 &27.13$\pm$ 0.27& 26.81 $\pm$0.21\\  
I7.1 & C8.1 & 1.16 & 110.799866 & -73.44799 & 5.61$^{+0.78}_{-0.78}$ & 26.99$\pm$ 0.35 & 26.75 $\pm$ 0.36 & 26.91 $\pm$ 0.43  &26.34$\pm$ 0.27 &26.2$\pm$ 0.27& 26.02 $\pm$0.26\\  
I7.1 & C9.1 & 1.16 & 110.798706 & -73.448296 & 8.98$^{+2.55}_{-1.11}$ & 30.57$\pm$ 0.68 & 28.77 $\pm$ 0.48 & 28.08 $\pm$ 0.36  &27.3$\pm$ 0.41 &26.99$\pm$ 0.43& 26.85 $\pm$0.5\\  
I7.1 & C10.1 & 1.16 & 110.79851 & -73.44829 & 8.65$^{+2.42}_{-1.1}$ & 26.81$\pm$ 0.31 & 26.44 $\pm$ 0.26 & 26.39 $\pm$ 0.24  &25.62$\pm$ 0.26 &25.37$\pm$ 0.26& 25.28 $\pm$0.22\\
\end{tabular}} 
\caption{Same as Table~\ref{tab:table_clumps_phot1}}
\label{tab:table_clumps_phot4} 
\end{table*}

\begin{table*} 
\centering 
\begin{tabular}{llllllllllll} 
Image ID & Clump ID & z & RA & Dec & $\mu$  & $R_{\rm eff}$ & F150W & M$_{*}$ & Age & $\rm E(B-V)$ & Metallicity \\ 
 &  &  &  &    & & [pc] & [AB mag] & log([$M_{\odot}$]) & [Myr] &   \\ 
(1) & (2) &  (3) & (4)  & (5)   &  (6)& (7) & (8) & (9) & (10) & (11) & (12) \\ 
\hline 
\hline 
S1.1 & C1.1 & 1.449 & 110.84069 & -73.451065 & 6.12$^{+0.57}_{-0.46}$ & 108$_{-30}^{+32}$ & -17.88$\pm$0.33 & 7.7$_{-0.0}^{+0.1}$ & 11$_{-1}^{+1}$ & 0.3$_{-0.2}^{+0.1}$  &0.02   \\ 
S1.1 & C2.1 & 1.449 & 110.840775 & -73.451126 & 6.38$^{+0.69}_{-0.46}$ & 171$_{-45}^{+46}$ & -17.22$\pm$0.45 & 7.7$_{-0.1}^{+0.1}$ & 20$_{-9}^{+10}$ & 0.3$_{-0.1}^{+0.25}$  &0.02   \\ 
S1.1 & C3.1 & 1.449 & 110.84103 & -73.451416 & 8.05$^{+0.93}_{-0.58}$ & <15 & -13.2$\pm$0.49 & 6.9$_{-0.6}^{+0.2}$ & 80$_{-73}^{+1933}$ & 0.45$_{-0.45}^{+0.45}$ & 0.004   \\ 
S1.1 & C4.1 & 1.449 & 110.84191 & -73.452255 & 26.63$^{+6.9}_{-2.51}$ & 23$_{-7}^{+4}$ & -11.52$\pm$0.55 & 6.2$_{-0.1}^{+0.1}$ & 802$_{-299}^{+204}$ & 0.0$^*$  &0.02   \\ 
S1.1 & C5.1 & 1.449 & 110.84136 & -73.45193 & 14.64$^{+2.5}_{-1.13}$ & <8 & -12.36$\pm$0.44 & 6.8$_{-0.2}^{+0.0}$ & 4029$_{-993}^{+1000}$ & 0.0$^*$& 0.0004   \\
\end{tabular} 
\caption{Main properties of the clumps. (1) and (2) images and clumps ID; (3) redshifts; (4) and (5) RA and Dec coordinates; (6) lensing magnification; (7) intrinsic effective radius measured on NIRCam/F150W; (8) NIRCam/F150W absolute magnitude; (9)-(12) SED properties measured with the reference model (a continuum star fromation for 10 Myr). Upper limits in size are indicated with <. The clumps with an $E(B-V)$ value at 0$^*$ have been fitted with no extinction.}
\label{tab:table_clumps1} 
\end{table*}

\begin{table*} 
\centering 
\begin{tabular}{llllllllllll} 
Image ID & Clump ID & z & RA & Dec & $\mu$  & $R_{\rm eff}$ & F150W & M$_{*}$ & Age & $\rm E(B-V)$ & Metallicity \\ 
 &  &  &  &    & & [pc] & [AB mag] & log([$M_{\odot}$]) & [Myr] &   \\ 
(1) & (2) &  (3) & (4)  & (5)   &  (6)& (7) & (8) & (9) & (10) & (11) & (12) \\ 
\hline 
\hline 
S2.3 & C1.1 & 1.3779 & 110.83601 & -73.45887 & 5.39$^{+0.33}_{-0.43}$ & 142$_{-18}^{+28}$ & -13.57$\pm$0.82 & 7.4$_{-0.3}^{+0.1}$ & 3036$_{-2233}^{+993}$ & 0.0$^*$  &0.02   \\ 
S2.3 & C2.1 & 1.3779 & 110.83632 & -73.45887 & 5.4$^{+0.33}_{-0.43}$ & 133$_{-17}^{+20}$ & -15.8$\pm$1.03 & 6.6$_{-0.2}^{+0.2}$ & 8$_{-2}^{+4}$ & 0.2$_{-0.2}^{+0.35}$  &0.02   \\ 
S2.3 & C3.1 & 1.3779 & 110.83649 & -73.45881 & 5.57$^{+0.33}_{-0.44}$ & 111$_{-31}^{+35}$ & -15.46$\pm$0.43 & 7.0$_{-0.2}^{+0.1}$ & 30$_{-22}^{+9}$ & 0.1$_{-0.1}^{+0.25}$  &0.008   \\ 
S2.3 & C4.1 & 1.3779 & 110.83667 & -73.4587 & 5.88$^{+0.44}_{-0.44}$ & 44$_{-12}^{+13}$ & -14.34$\pm$0.3 & 5.8$_{-0.0}^{+0.0}$ & 4$_{-1}^{+2}$ & 0.25$_{-0.05}^{+0.05}$  &0.008   \\ 
S2.3 & C5.1 & 1.3779 & 110.83641 & -73.45889 & 5.35$^{+0.33}_{-0.44}$ & <23 & -13.22$\pm$1.08 & -- & -- & --& --  \\
\end{tabular} 
\caption{Same as Table~\ref{tab:table_clumps1}}
\label{tab:table_clumps2} 
\end{table*}

\begin{table*} 
\centering 
\begin{tabular}{llllllllllll} 
Image ID & Clump ID & z & RA & Dec & $\mu$  & $R_{\rm eff}$ & F150W& M$_{*}$ & Age & $\rm E(B-V)$ & Metallicity \\ 
 &  &  &  &    & & [pc] & [AB mag] & log([$M_{\odot}$]) & [Myr] &   \\ 
(1) & (2) &  (3) & (4)  & (5)   &  (6)& (7) & (8) & (9) & (10) & (11) & (12) \\ 
\hline 
\hline 
S4.1 & C5.1 & 2.19 & 110.807 & -73.45843 & 8.04$^{+0.7}_{-0.5}$ & <15 & -16.69$\pm$0.33 & 8.1$_{-0.1}^{+0.1}$ & 302$_{-282}^{+100}$ & 0.05$_{-0.05}^{+0.5}$ & 0.004   \\ 
S4.1 & C6.1 & 2.19 & 110.80713 & -73.45845 & 7.92$^{+0.71}_{-0.5}$ & 93$_{-16}^{+16}$ & -16.98$\pm$0.4 & 7.7$_{-0.2}^{+0.1}$ & 30$_{-10}^{+9}$ & 0.15$_{-0.05}^{+0.1}$  &0.008   \\ 
S4.1 & C7.1 & 2.19 & 110.8083 & -73.45878 & 6.63$^{+0.51}_{-0.41}$ & <19 & -13.67$\pm$0.4 & 7.0$_{-0.4}^{+0.1}$ & 802$_{-500}^{+204}$ & 0.0$^*$& 0.0004   \\ 
S4.1 & C8.1 & 2.19 & 110.80833 & -73.45885 & 6.43$^{+0.51}_{-0.41}$ & 118$_{-27}^{+31}$ & -14.82$\pm$0.45 & 7.1$_{-0.1}^{+0.2}$ & 200$_{-150}^{+101}$ & 0.0$^*$  &0.02   \\ 
S4.1 & C9.1 & 2.19 & 110.80728 & -73.45848 & 7.78$^{+0.71}_{-0.51}$ & 44$_{-14}^{+15}$ & -14.17$\pm$0.37 & 6.7$_{-0.2}^{+0.0}$ & 20$_{-6}^{+10}$ & 0.3$_{-0.1}^{+0.05}$  &0.02   \\
\end{tabular} 
\caption{Same as Table~\ref{tab:table_clumps1}}
\label{tab:table_clumps3} 
\end{table*}

\begin{table*} 
\centering 
\begin{tabular}{llllllllllll} 
Image ID & Clump ID & z & RA & Dec & $\mu$  & $R_{\rm eff}$ & F150W  & M$_{*}$ & Age & $\rm E(B-V)$ & Metallicity \\ 
 &  &  &  &    & & [pc] & [AB mag] & log([$M_{\odot}$]) & [Myr] &   \\ 
(1) & (2) &  (3) & (4)  & (5)   &  (6)& (7) & (8) & (9) & (10) & (11) & (12) \\ 
\hline 
\hline 

S7.2 & C1.1 & 5.17 & 110.794975 & -73.448944 & 39.28$^{+56.47}_{-4.24}$ & 28$_{-20}^{+18}$ & -14.16$\pm$1.29 & 6.5$_{-0.4}^{+0.1}$ & 15$_{-11}^{+5}$ & 0.2$_{-0.15}^{+0.05}$  &0.0004   \\ 
S7.2 & C2.1 & 5.17 & 110.794846 & -73.449036 & 28.7$^{+24.88}_{-2.64}$ & 20$_{-8}^{+5}$ & -14.82$\pm$0.69 & 6.4$_{-0.1}^{+0.1}$ & 15$_{-1}^{+5}$ & 0.1$_{-0.1}^{+0.05}$  &0.0004   \\ 
S7.2 & C3.1 & 5.17 & 110.79474 & -73.4491 & 24.08$^{+15.36}_{-3.09}$ & 32$_{-19}^{+14}$ & -15.17$\pm$0.7 & 6.1$_{-0.2}^{+0.1}$ & 11$_{-6}^{+4}$ & 0.0$_{-0.0}^{+0.05}$  &0.004   \\ 
S7.2 & C4.1 & 5.17 & 110.79463 & -73.44915 & 21.42$^{+11.84}_{-2.94}$ & 26$_{-14}^{+13}$ & -15.16$\pm$0.66 & 6.5$_{-0.5}^{+0.1}$ & 15$_{-12}^{+5}$ & 0.05$_{-0.05}^{+0.05}$  &0.0004   \\ 
S7.2 & C5.1 & 5.17 & 110.794334 & -73.44937 & 16.57$^{+6.49}_{-1.92}$ & 45$_{-23}^{+22}$ & -14.18$\pm$0.66 & 6.5$_{-0.1}^{+0.1}$ & 40$_{-19}^{+10}$ & 0.0$_{-0.0}^{+0.1}$  &0.0004   \\
\end{tabular} 
\caption{Same as Table~\ref{tab:table_clumps1}}
\label{tab:table_clumps4} 
\end{table*}

\section{NIRCam PSF}
\label{sec:comp_psf}
We compare the JWST model PSF, webbpsf, to the one create using not-saturated, isolated stars in the field of view of the galaxy cluster in the reference filter F150W. Overall, the two PSFs are in very good agreement, with variations limited to few percents.

\begin{figure*}
    \includegraphics[width=18cm]{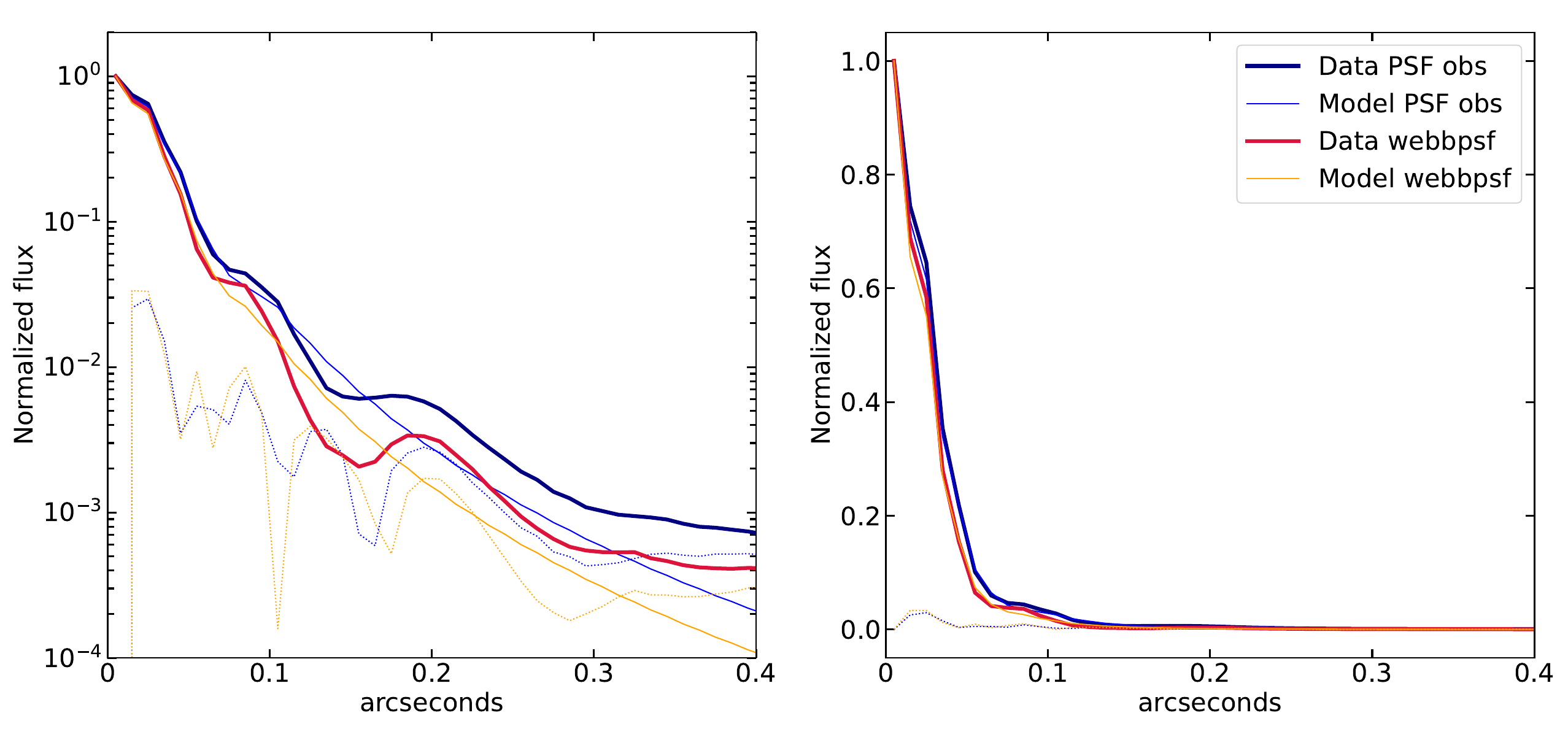}
    \caption{Comparison between the observed PSF (built with a stack of 3 non saturated stars) in blue and the webbpsf PSF for the F150W filter in red/orange. We show the light profile of the PSF in thick solid lines and the results of the fit with the thin line. The residuals between the fit and the data for each PSF are represented in dotted lines. The left panel shows the light profile in log scale and the right panel the profile in linear scale. }
    \label{fig:comp_psf}
\end{figure*}

\section{Clump size measurements}
\label{sec:comp_sizes}
\begin{figure}
    \includegraphics[width=8cm]{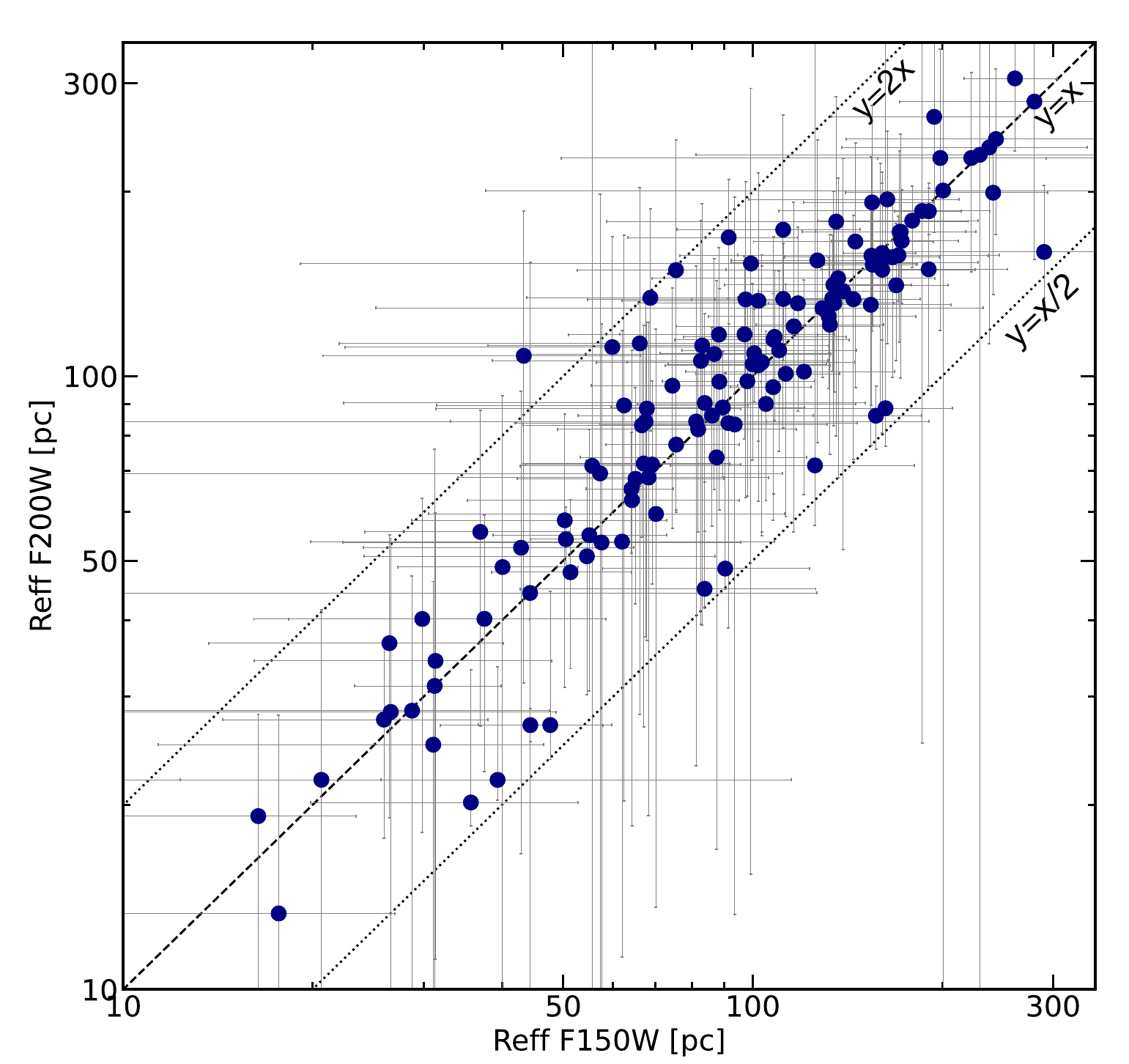}
    \caption{Clump sizes measured on F200W versus sizes measured on F150W. We show here only the resolved clumps larger than 0.4 pixels in both direction. The dashed / dotted black lines represent $y=x$, $y=x/2$ and $y=2x$.}
    \label{fig:comp_sizes}
\end{figure}

During testing, we measured clumps sizes independently on the two NIRCam reference filters: F150W and F200W. The two datasets are presented in Fig.~\ref{fig:comp_sizes} and produce consistent results. The PSF built from observed stars in the field of view of the galaxy cluster has been used to extract clump sizes and fluxes (see Sect.~\ref{sec:cl_model}).

\section{Residual images after extracting clump photometry}
\label{sec:clump_models}
\begin{figure*}
    \includegraphics[width=11cm]{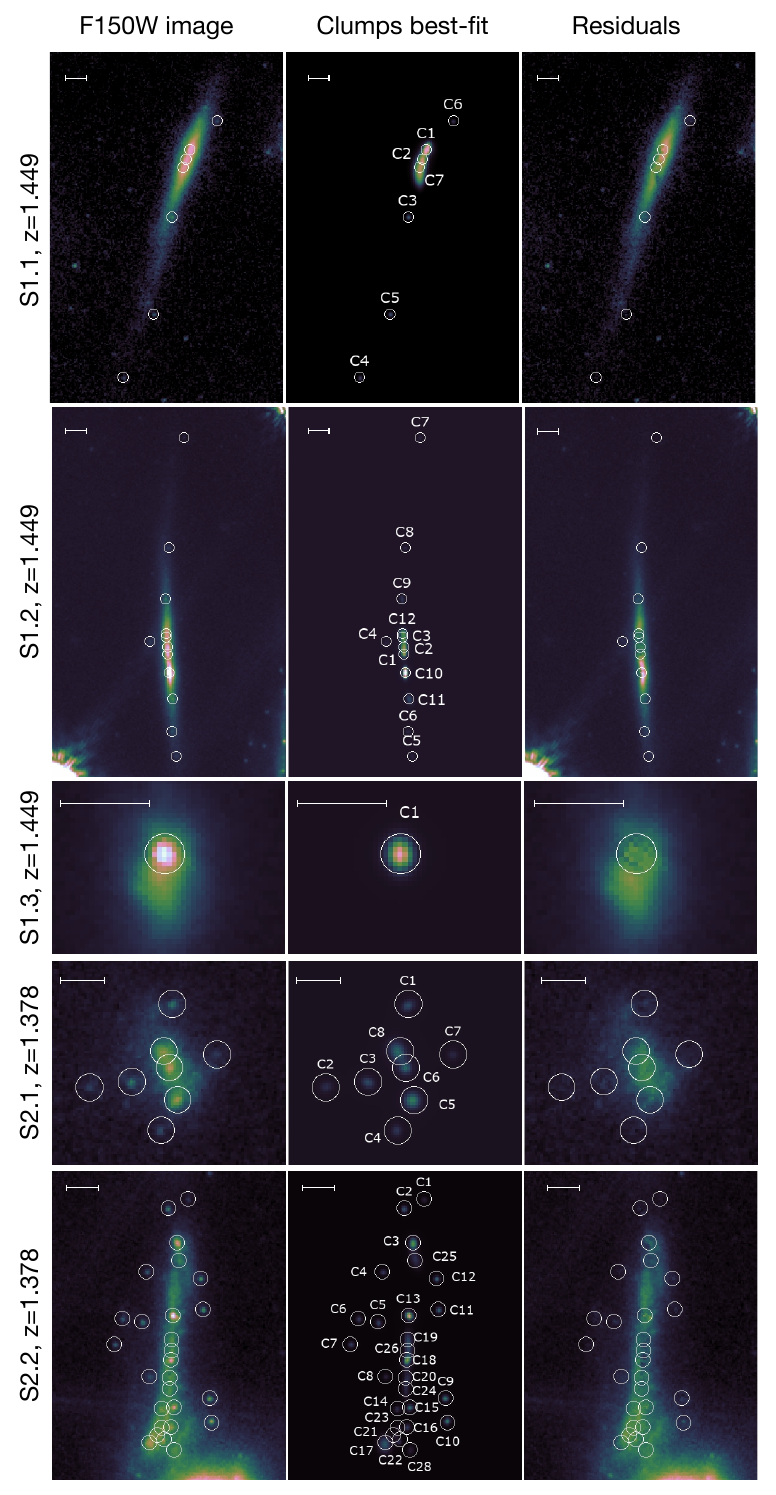}
    \caption{F150W image, clumps' best-fit model and residuals for each image of the sample. The selected clumps are shown in white. Images of the same object are at the same scale. The white line is 0.4'' long. }
    \label{fig:residuals1}
\end{figure*}

\begin{figure*}
    \includegraphics[width=14cm]{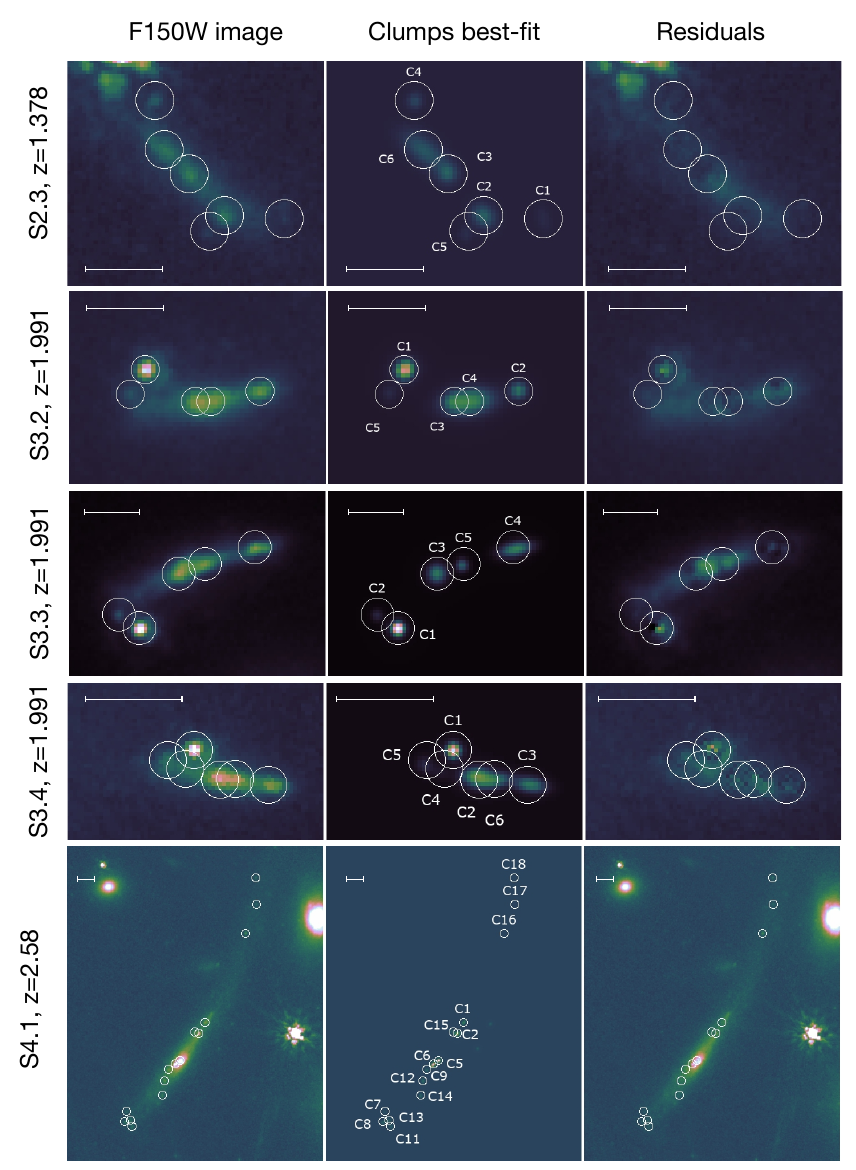}
    \caption{F150W image, clumps' best-fit model and residuals for each image of the sample. The selected clumps are shown in white. Images of the same object are at the same scale. The white line is 0.4'' long. }
    \label{fig:residuals2}
\end{figure*}

\begin{figure*}
    \includegraphics[width=11cm]{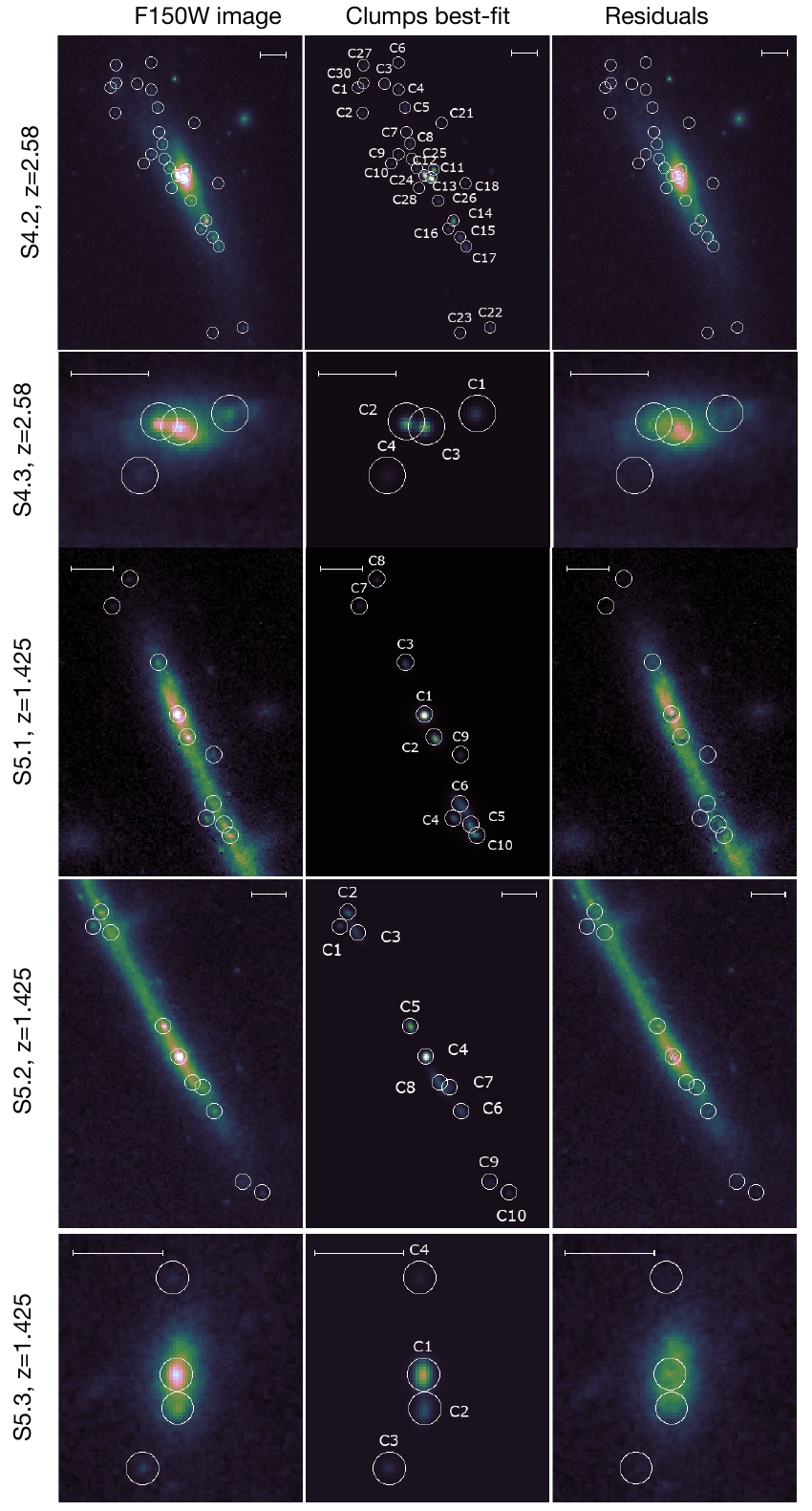}
    \caption{F150W image, clumps' best-fit model and residuals for each image of the sample. The selected clumps are shown in white. Images of the same object are at the same scale. The white line is 0.4'' long. }
    \label{fig:residuals3}
\end{figure*}

\begin{figure*}
    \includegraphics[width=11cm]{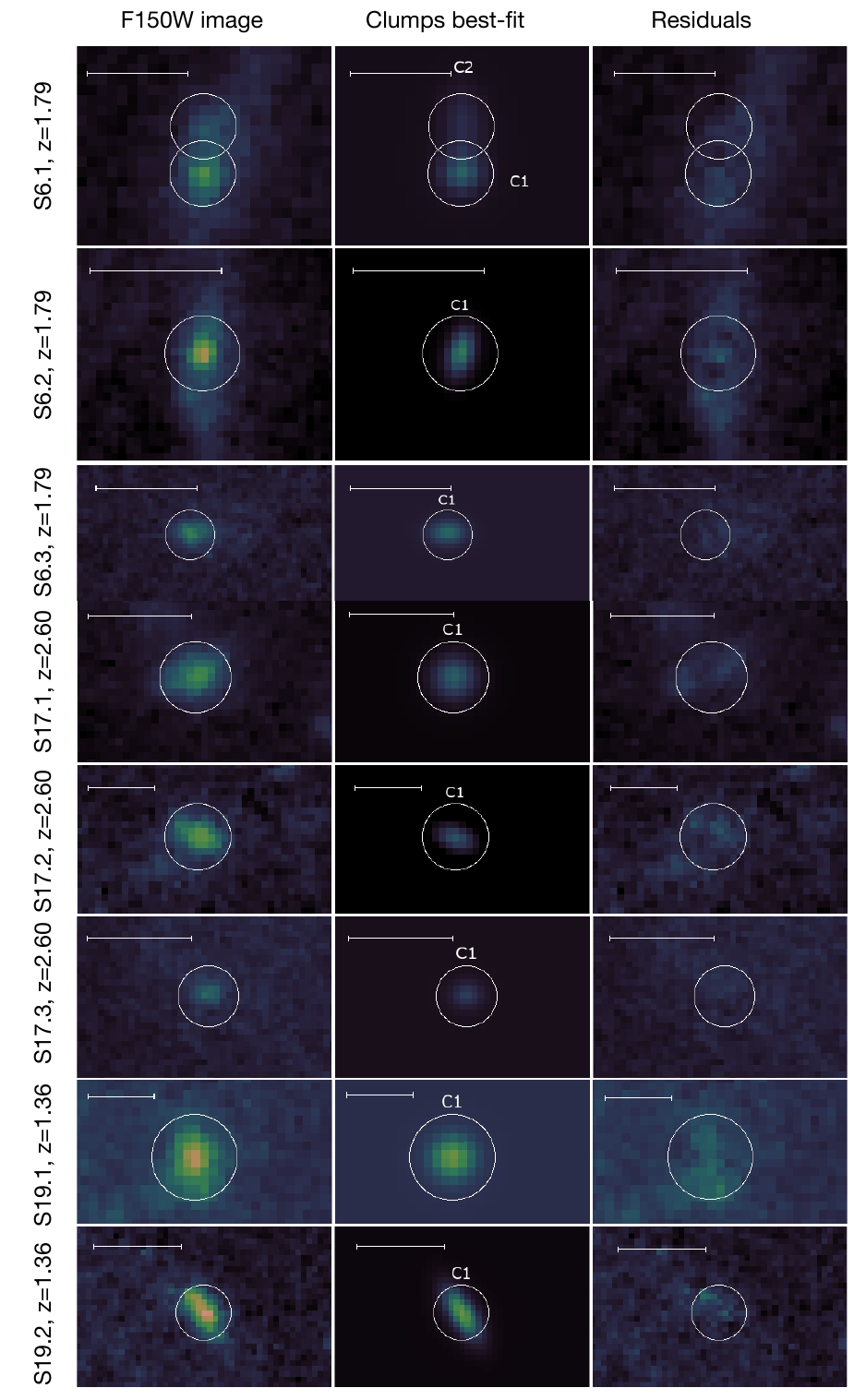}
    \caption{F150W image, clumps' best-fit model and residuals for each image of the sample. The selected clumps are shown in white. Images of the same object are at the same scale. The white line is 0.4'' long. }
    \label{fig:residuals4}
\end{figure*}

\begin{figure*}
    \includegraphics[width=11cm]{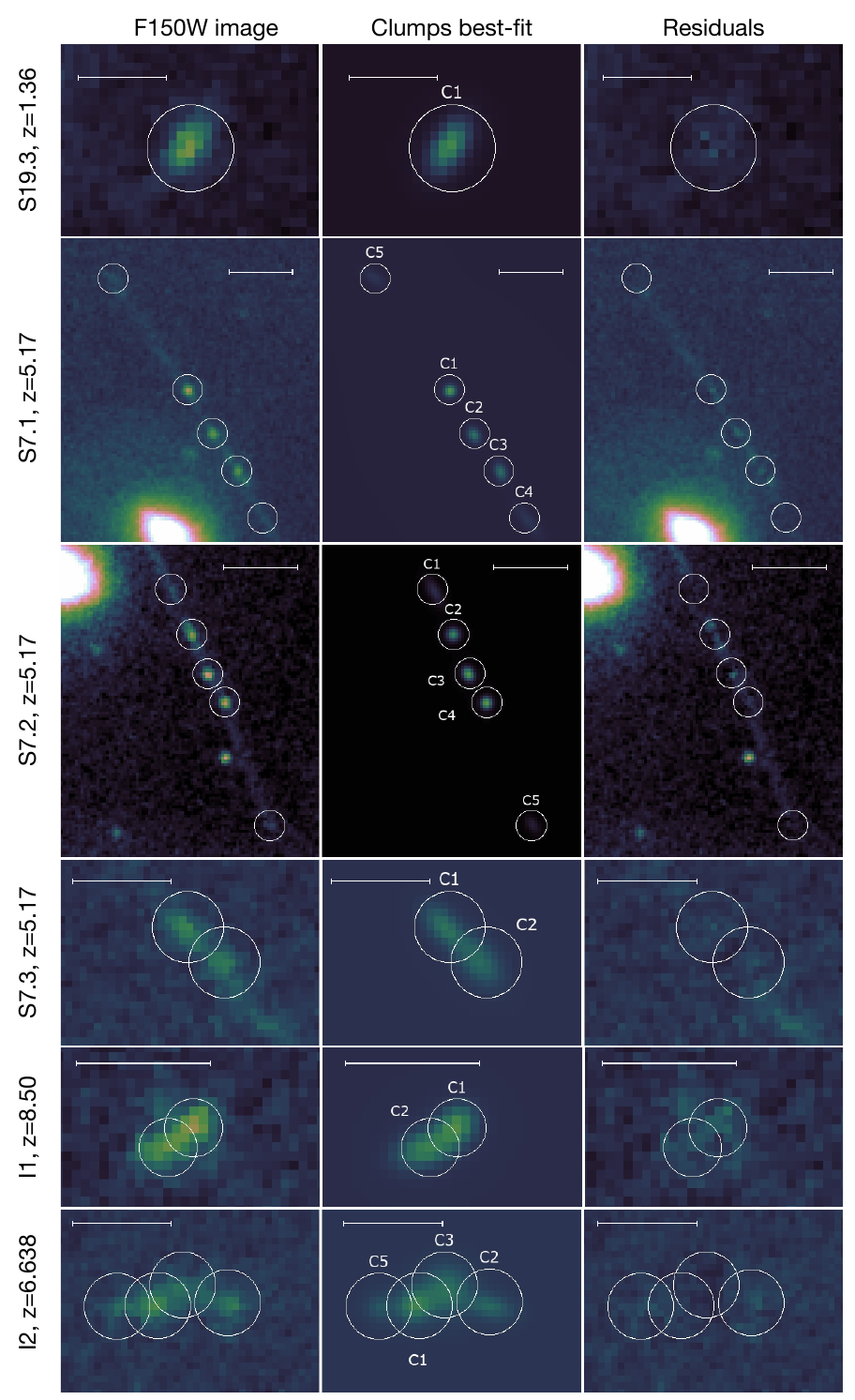}
    \caption{F150W image, clumps' best-fit model and residuals for each image of the sample. The selected clumps are shown in white. Images of the same object are at the same scale. The white line is 0.4'' long. }
    \label{fig:residuals5}
\end{figure*}

\begin{figure*}
    \includegraphics[width=11cm]{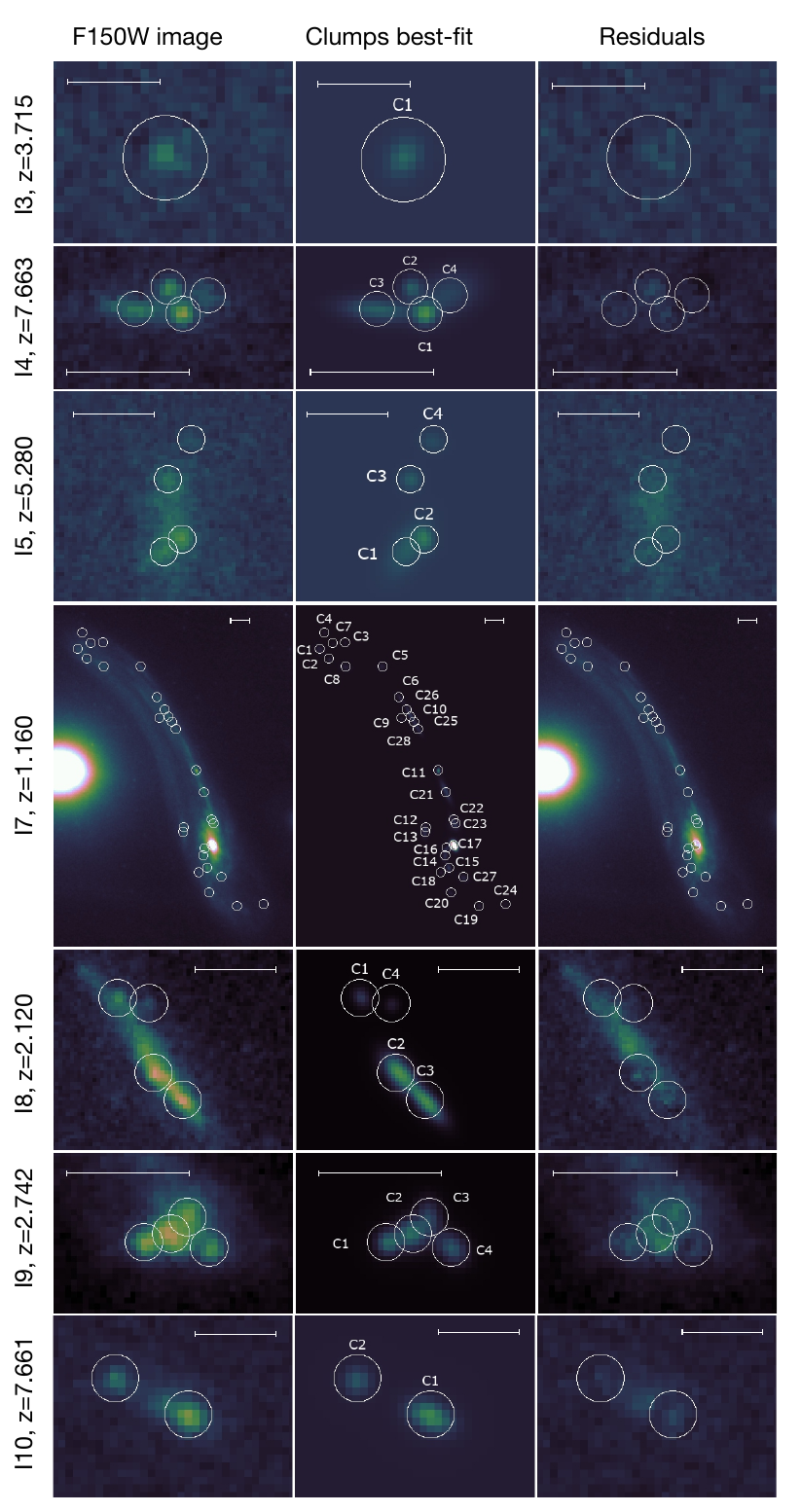}
    \caption{F150W image, clumps' best-fit model and residuals for each image of the sample. The selected clumps are shown in white. Images of the same object are at the same scale. The white line is 0.4'' long. }
    \label{fig:residuals6}
\end{figure*}

We list here the observed image of the galaxy including the clump positions, the modelled clumps and the galaxy residual image after the modelled clumps have been subtracted (Fig.~\ref{fig:residuals1} to ~\ref{fig:residuals6}). See Sect.~\ref{sec:cl_model} and Fig.~\ref{fig:im_residuals} of the main text as reference.

\section{The Sparkler}
\label{sec:comp_sparkler}
\begin{figure*}
    \includegraphics[width=18cm]{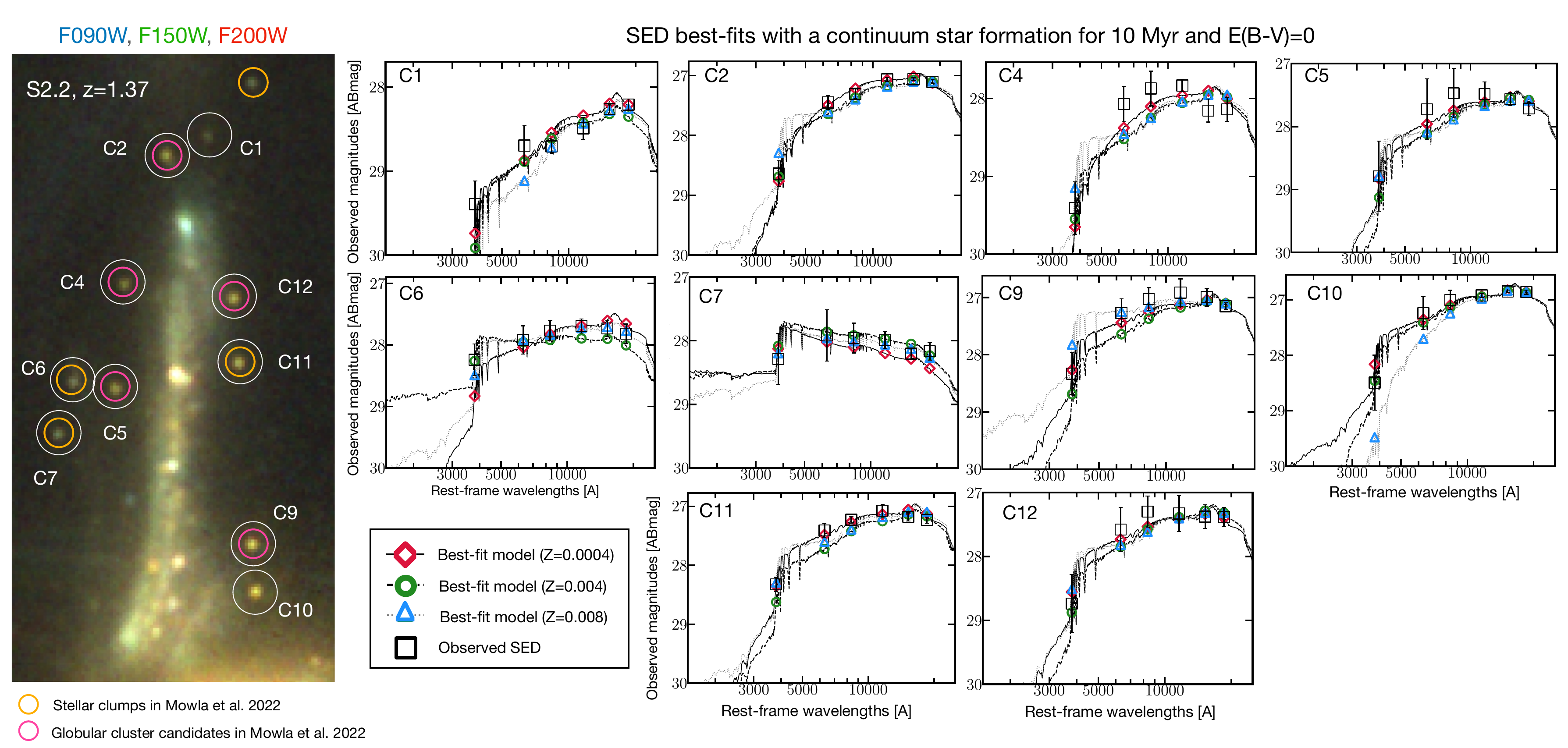}
    \caption{Comparison with the results from \citet{mowla2022} for the 10 outskirts clumps of the Sparkler galaxy. \textbf{Left}: NIRCam colour image (using F090W, F150W and F200W filters) of the Sparkler galaxy (S2.2 at $z=1.37$). The white circles show the 10 outskirts clumps we detected in this work. The orange and pink circles represent the stellar clumps and globular clusters candidates respectively, studies in \citet{mowla2022}. We indicate the ID of each clump in white. \textbf{Right}: SED best-fits for each clump based on a model with a continuous star formation for 10 Myr and no extinction (i.e., $\rm E(B-V)=0$). We present for each clump the best model with $Z=0.008$ (in dotted grey line and blue triangles), with $Z=0.004$ (in dashed line and green circles) and with $Z=0.0004$ (in solid line and red diamonds). The black squares represent the observed SED and associated error-bars.} 
    \label{fig:comp_sparkler}
\end{figure*}

\begin{figure*}
    \includegraphics[width=18cm]{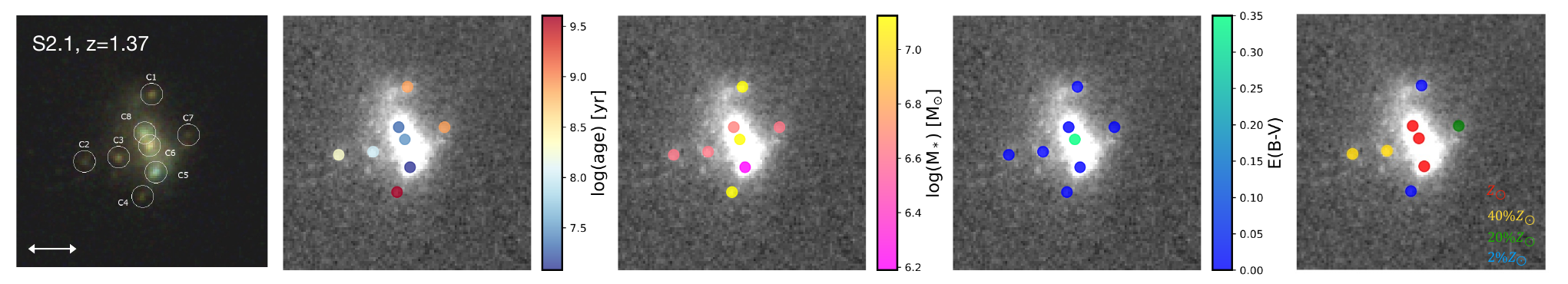}
    \caption{From left to right: colour image (produced with F090W, F150W and F200W NIRCam filters) and F150W frames of the counter-image of the Sparkler galaxy (S2.1), with the position of detected clumps colour-coded by their age (left), mass (middle left), extinction (middle right), and metallicity (right) obtained from the reference SED model (with a continuous star formation for 10 Myr). The white line is 0.4'' long.} 
    \label{fig:comp_sparkler_CI}
\end{figure*}

In this section, we present a more detailed comparison between this study and the results presented in \citet{mowla2022} for the outskirt clusters of the Sparkler galaxy. In Fig.~\ref{fig:comp_sparkler}, we present the 10 globular clusters included in our analysis (outlined with white circles) and the ones studied by \citet{mowla2022} (in orange and pink). We included in our analysis 2 extra cluster candidates (C1 and C10) which should belong to the galaxy based on their location and their colour very similar to the closest clusters. However, we did not include the top right cluster (number 9 in \citet{mowla2022}) which is located farther from the host and has a colour that is closer to the intracluster globular clusters (see Fig.~\ref{fig:selection_clumps}). 
In Fig.~\ref{fig:comp_sparkler}, we also present the best-SED fit (using three different metallicities Z=0.008, Z=0.004 and Z=0.0004) for these globualr clusters. Giving the fact that we detect these clusters only in the NIRCam images, the SED fit is based only on 6 filters covering the rest-frame 3600-19000 \AA\ wavelength range, which is not enough to break the age-extinction degeneracy. If we let the extinction parameter free, we find best-fit solutions with young ages (around 100 Myr) but high extinction values (between 0.3 and 0.5 mag) which is not realistic for clumps/clusters located outside of the main body of their host galaxy. By fixing $\rm E(B-V)=0$ we measure older ages (few Gyr). The reduced $\chi^2$ of these two sets of solutions (free extinction vs. $\rm E(B-V)=0$) are similar. We also notice a degeneracy with metallicity for some of these clusters resulting in similar $\chi^2$ (the best-fit parameters for these clusters are resumed in the Table~\ref{tab:table_sparkler}). The solution obtained with $Z=0.0004$ are on average older than the ones measured with $Z=0.008$. Finally, in Fig.~\ref{fig:comp_sparkler_CI}, we present the SED best-fit parameters of the clusters detected in the counter-image (i.e. less magnified) of the Sparkler galaxy. In this image, we detect only 8 clusters (presented in the left panel of the figure). Using the same assumptions, and fixing $\rm E(B-V)=0$ for the outskirts clumps (C1, C2, C3, C4 and C7), we derived consistent results with the Sparkler images.

\begin{table*} 
 
\begin{tabular}{lllllllllll} 
& &   & $Z=0.008$ & & & $Z=0.004$ &  &  & $Z=0.0004$ &  \\
 ID &  ID & Age & Mass & $\chi^2_{\rm red}$ &Age & Mass & $\chi^2_{\rm red}$ & Age & Mass & $\chi^2_{\rm red}$\\ 
 This work &  \citet{mowla2022} & [Myr] & [log($\rm M_{\odot}$)] & & [Myr] & [log($\rm M_{\odot}$)] &  &  [Myr] & [log($\rm M_{\odot}$)] & \\ 

\hline 
\hline 

C1 & -- & $604^{+200}_{-500}$ &$6.29^{+0.01}_{-0.14}$ & 1.13&$912^{+100}_{-300}$ & $6.51^{+0.05}_{-0.13} $ & 1.41 &$605^{+100}_{-100}$ & $6.28^{+0.01}_{-0.13}$& 1.38  \\

C2 & 10 & $401^{+100}_{-100}$&$6.86^{+0.06}_{-0.02}$ & 2.72 &$1007^{+1000}_{-100}$ &$7.03^{+0.01}_{-0.04} $ & 1.63 &$4030^{+1000}_{-1000}$ &$7.41^{+0.01}_{-0.01}$ & 0.70 \\

C4 & 1 & $401^{+100}_{-100}$& $6.51^{+0.04}_{-0.16}$& 1.93 &$912^{+100}_{-400}$ & $6.65^{+0.10}_{-0.16} $ & 1.93 & $4030^{+1000}_{-1000}$& $7.05^{+0.01}_{-0.05} $ & 1.25 \\

C5 & 4 & $401^{+200}_{-100}$& $6.66^{+0.04}_{-0.10}$ & 0.48 &$800^{+1200}_{-700}$ &$6.76^{+0.13}_{-0.29}$ & 0.53 &$1007^{+1000}_{-200}$ & $6.83^{+0.03}_{-0.06}$ & 0.31 \\

C6 & 5 &$302^{+100}_{-100}$ & $6.66^{+0.01}_{-0.13} $& 0.22 & $100^{+100}_{-60}$& $6.3^{+0.01}_{-0.13}$ & 0.77 & $800^{+100}_{-100}$& $6.75^{+0.07}_{-0.02}$ &  0.95\\

C7 & 6 & $90^{+110}_{-50}$&$6.29^{+0.10}_{-0.23}$ &  0.22 &$80^{+420}_{-30}$ & $6.30^{+0.09}_{-0.23} $ &0.55 & $90^{+10}_{-10}$& $6.29^{+0.09}_{-0.24} $& 0.78\\

C9 & 8 & $302^{+100}_{-100}$& $6.93^{+0.01}_{-0.18} $& 0.79 &$1007^{+1000}_{-200}$ & $7.03^{+0.01}_{-0.10} $& 0.90 &$912^{+100}_{-200}$ &$7.02^{+0.01}_{-0.05} $ & 0.36 \\

C10 & -- & $3036^{+1000}_{-1000}$&$7.40^{+0.01}_{-0.14} $ &1.20 & $1007^{+1000}_{-500}$& $7.13^{+0.07}_{-0.25}$& 0.20 & $706^{+200}_{-100}$& $6.98^{+0.12}_{-0.16}$ & 0.30 \\

C11 & 3 & $401^{+100}_{-100}$& $6.85^{+0.01}_{-0.07}$& 2.71 &$503^{+100}_{-100}$ &$6.76^{+0.03}_{-0.01} $ & 3.54 &$1007^{+1000}_{-100}$ &$7.02^{+0.01}_{-0.07} $ & 1.28 \\

C12 & 2 & $401^{+100}_{-100}$& $6.77^{+0.02}_{-0.09} $ & 0.34 & $912^{+1100}_{-500}$&$6.92^{+0.12}_{-0.22} $ &0.29 &$912^{+100}_{-300}$ & $6.91^{+0.07}_{-0.05}$ & 0.18\\
 
\end{tabular} 
\caption{SED best-fit parameters for the 10 outskirt clumps from the Sparkler galaxy using a continuum star formation for 10 Myr and no extinction (i.e., $\rm E(B-V)=0$). From left to right: clumps ID in this work, corresponding clumps ID in \citet{mowla2022} (when existing), age, mass and reduced $\chi^2$ for $Z=0.008$, age, mass and reduced $\chi^2$ for $Z=0.004$ and age, mass and reduced $\chi^2$ for $Z=0.0004$. }
\label{tab:table_sparkler} 
\end{table*}

\bsp	
\label{lastpage}
\end{document}